\documentclass[aps,prd,
10pt,nofootinbib,floatfix,preprintnumbers,showpacs]{revtex4}
\usepackage{amssymb}
\usepackage{amsmath}
\usepackage{comment}
\usepackage[dvips]{color}
\usepackage[dvips]{graphicx}
\usepackage{epsfig}
\usepackage{multirow}
\newcommand{\err}[2]{%
{{\renewcommand{\arraystretch}{0.4}%
\ensuremath{\mathop{\raisebox{0.1\height}{\scriptsize
$\begin{array}{@{}c@{}}+\\-\end{array}$}}%
\raisebox{0.1\height}{\scriptsize
$\begin{array}{@{}r@{}}#1\\#2\end{array}$}}}}%
}

\def\a{{\alpha}}
\def\b{{\beta}}
\def\d{{\delta}}
\def\D{{\Delta}}
\def\e{{\epsilon}}
\def\g{{\gamma}}
\def\G{{\Gamma}}

\def\l{{\lambda}}
\def\L{{\Lambda}}

\def\w{{\omega}}
\def\O{{\Omega}}
\def\S{{\Sigma}}
\def\s{{\sigma}}

\def\ol#1{{\overline{#1}}}

\def\str{\text{str}}

\def\mc#1{{\mathcal #1}}

\def\eqref#1{{(\ref{#1})}}

\preprint{JLAB-THY-08-838}
\preprint{MIT-CTP 3954}
\preprint{TUM-T39-08-11}
\preprint{UMD-40762-412}

\newcommand{\mud}{a m_{u/d}^{\rm asqtad}}
\newcommand{\muddwf}{a m_{u/d}^{\rm DWF}}
\newcommand{\msdwf}{a m_s^{\rm DWF}}
\begin{document}
\bibliographystyle{apsrev}
\title{Light hadron spectroscopy using domain wall valence quarks on an Asqtad sea}

\author{A. Walker-Loud}
\affiliation{Maryland Center for Fundamental Physics, Department of Physics, University of Maryland, 
College Park, MD 20742-4111, USA}

\author{H.-W. Lin}
\affiliation{Thomas Jefferson National Accelerator Facility, Newport News, VA 23606, USA}

\author{D.G. Richards}
\affiliation{Thomas Jefferson National Accelerator Facility, Newport News, VA 23606, USA}

\author{R.G. Edwards}
\affiliation{Thomas Jefferson National Accelerator Facility, Newport News, VA 23606, USA}

\author{M. Engelhardt}
\affiliation{Physics Department, New Mexico State University, Las Cruces, NM 88003-8001, USA}

\author{G.T. Fleming}
\affiliation{Sloane Physics Laboratory, Yale University, New Haven, CT 06520, USA}

\author{Ph. H\"{a}gler}
\affiliation{Institut f¬ur Theoretische Physik T39, Physik-Department der TU M\"{u}nchen, James-Franck-Strasse, D-85747 Garching, Germany}

\author{B. Musch}
\affiliation{Institut f¬ur Theoretische Physik T39, Physik-Department der TU M\"{u}nchen, James-Franck-Strasse, D-85747 Garching, Germany}


\author{M.F. Lin}
\affiliation{Center for Theoretical Physics, Massachusetts Institute of Technology, Cambridge, MA 02139, USA}

\author{H. Meyer}
\affiliation{Center for Theoretical Physics, Massachusetts Institute of Technology, Cambridge, MA 02139, USA}

\author{J.W. Negele}
\affiliation{Center for Theoretical Physics, Massachusetts Institute of Technology, Cambridge, MA 02139, USA}

\author{A.V. Pochinsky}
\affiliation{Center for Theoretical Physics, Massachusetts Institute of Technology, Cambridge, MA 02139, USA}

\author{M. Procura}
\affiliation{Center for Theoretical Physics, Massachusetts Institute of Technology, Cambridge, MA 02139, USA}

\author{S. Syritsyn}
\affiliation{Center for Theoretical Physics, Massachusetts Institute of Technology, Cambridge, MA 02139, USA}

\author{C.J. Morningstar}
\affiliation{Department of Physics, Carnegie Mellon University, Pittsburgh, PA 15213, USA}

\author{K. Orginos}
\affiliation{Department of Physics, College of William and Mary, Williamsburg VA 23187-8795}
\affiliation{Thomas Jefferson National Accelerator Facility, Newport News, VA 23606, USA}

\author{D.B. Renner}
\affiliation{DESY Zeuthen, Theory Group, Platanenallee 6, D-15738 Zeuthen, Germany}

\author{W. Schroers}
\affiliation{Department of Physics, Center for Theoretical Sciences, National Taiwan University, Taipei 10617, Taiwan}

\collaboration{ LHP Collaboration }
\noaffiliation

\date{\today}
\begin{abstract}
We calculate the light hadron spectrum in full QCD using two plus one flavor Asqtad sea quarks and domain wall valence quarks.  Meson and baryon masses are calculated on a lattice of spatial size $L \approx 2.5$~\texttt{fm}, and a lattice spacing of $a \approx 0.124$~\texttt{fm}, for pion masses as light as $m_\pi \approx 300$~\texttt{MeV}, and compared with the results by the MILC collaboration with Asqtad valence quarks at the same lattice spacing.  Two- and three-flavor chiral extrapolations of the baryon masses are performed using both continuum and mixed-action heavy baryon chiral perturbation theory.  Both the three-flavor and two-flavor functional forms describe our lattice results, although the low-energy constants from the next-to-leading order $SU(3)$ fits are inconsistent with their phenomenological values.  Next-to-next-to-leading order $SU(2)$ continuum formulae provide a good fit to the data and yield and extrapolated nucleon mass consistent with experiment, but the convergence pattern indicates that even our lightest pion mass may be at the upper end of the chiral regime.  Surprisingly, our nucleon masses are essentially lineaer in $m_\pi$ over our full range of pion masses, and we show this feature is common to all recent dynamical calculations of the nucleon mass.  The origin of this linearity is not presently understood, and lighter pion masses and increased control of systematic errors will be needed to resolve this puzzling behavior.
\end{abstract}

\maketitle
\newpage
\section{Introduction}
A precise calculation of the hadron spectrum is an important achievement of lattice QCD, providing a comparison of the experimentally known states to a first principles prediction from QCD.  These calculations additionally provide insight into presently unexplored states, help to guide experiment, test predictions from QCD-inspired models of hadrons, and set a foundation for more challenging lattice calculations.  A detailed map of the hadron spectrum, including low lying excited states, is a challenging task for lattice QCD which is now well underway.

In this work, we focus on the lowest lying hadrons for a given set of quantum numbers.  This is part of an extensive program by the LHP Collaboration to study hadron structure using a mixed action that exploits the lattice chiral symmetry provided by domain wall valence quarks and ensembles of computationally economical improved staggered sea quark configurations generated by the MILC Collaboration.  Since the MILC Collaboration has performed an extensive investigation of the spectrum using Asqtad valence quarks~\cite{Bernard:2001av,Aubin:2004wf}, this study provides an important test of the systematics of the mixed action formulation, as well as a comparison with the spectrum determined from staggered fermions.  In addition, because of the technical difficulties in the construction of baryon interpolating operators using the staggered quarks, this work also provides a more comprehensive study of the baryon spectrum than has been performed using staggered quarks alone.  Since the range of pion masses treated in this work extends down to $m_\pi \sim 300$~\texttt{MeV}, our lattice calculations provide a fruitful opportunity to explore the applicability of chiral perturbation theory for the extrapolation of hadron masses in this pion mass regime.  

In addition to this work, there now exists a large number of lattice calculations that utilize this specific mixed action framework.  There is other work by the LHP Collaboration exploring hadron structure~\cite{Renner:2004ck,Edwards:2005kw,Edwards:2005ym,Edwards:2006qx,Hagler:2007xi}, as well as related studies of the electromagnetic structure of hadrons~\cite{Alexandrou:2006mc,Alexandrou:2007zz,Alexandrou:2007dt,Engelhardt:2007ub}.  The NPLQCD Collaboration has published a number of calculations focussed on hadron interactions on the lattice~\cite{Beane:2005rj,Beane:2006mx,Beane:2006pt,Beane:2006fk,Beane:2006kx,Beane:2006gj,Beane:2006gf,Beane:2007xs,Beane:2007uh,Beane:2007es,Detmold:2008fn,Beane:2008dv}.  The first calculation of the hyperon axial matrix elements of the octet baryons was recently performed~\cite{Lin:2007ap} and there is also a group in the process of calculating the kaon bag parameter~\cite{Aubin:2007pt}.  Simultaneously with the development of these mixed action calculations, extensive effort produced the mixed action effective field theory (EFT)~\cite{Bar:2002nr,Bar:2003mh,Tiburzi:2005vy,Bar:2005tu,Golterman:2005xa,Tiburzi:2005is,Chen:2005ab,Bunton:2006va,Chen:2006wf,Aubin:2006hg,Orginos:2007tw,Jiang:2007sn,Chen:2007ug}, an extension of chiral perturbation theory $\chi$PT~\cite{Weinberg:1978kz,Gasser:1983yg,Gasser:1984gg,Jenkins:1990jv}, which includes the relevant lattice spacing artifacts for these mixed action lattice calculations.  This allows for a simultaneous extrapolation in the quark mass and the lattice spacing.  By performing the chiral extrapolations of the calculated hadron masses, we will be able to determine the numerical values of the low energy constants (LECs) that appear as coefficients of the operators in the EFT, most importantly the physical LECs that contribute to hadronic processes in the continuum limit.  The extracted values can then be compared with phenomenological determinations, where they exist, as well as provide predictions for those LECs which are notoriously difficult to determine without the ability to vary the quark masses.  The real predictive power then comes when one has confidence in the numerical values of these universal coefficients.  Once determined from one set of hadronic observables, they can then be used to make predictions about other observables, allowing a precision comparison with our experimental knowledge as well as predictions of experimentally unexplored or inaccessible physical processes.

The rest of this paper is organized in four additional sections.  In Section~\ref{sec:computation_details} we present the details of our computation including the tuning of the valence quark masses as well as the determination the residual chiral symmetry breaking inherent in finite fifth-dimensional domain-wall fermion lattice actions.  The hadron spectrum is presented in Section~\ref{sec:spectrum} beginning with the meson masses and decay constants.  Section~\ref{sec:baryons} describes the lattice calculation of the baryon spectrum, including the octet baryons, the nucleon, $\L$, $\S$ and $\Xi$, as well as the decuplet baryons, the $\D$, $\S^*$, $\Xi^*$ and the $\O^-$.  In Section~\ref{sec:ChExtrap}, we present the details of our chiral extrapolations.
A noteworthy result of this analysis is the finding of unexpected and unexplained 
non-analytic quark mass dependence 
in the baryon spectrum.  The behavior is most striking in the resulting nucleon mass calculations which are well described by a linear function in the pion mass.  Comparing our numerical results to those of other lattice collaborations, we find this unexpected chiral non-analytic behavior is also present in their lattice-data sets, suggesting this behavior is not simply the result of lattice artifacts. 
Additionally, we describe and apply the continuum and mixed action versions of heavy baryon chiral perturbation theory, for both two and three flavors, to our lattice calculations.  All the chiral extrapolation formulae used have been determined previously in the literature and we refer the interested reader to the references within the text for a complete understanding of the derivation of these formulae.  
We present our conclusions in Section~\ref{sec:conclusions}.

\section{Computational Details}\label{sec:compute}
\label{sec:computation_details}
We employ the gauge configurations generated by the MILC
collaboration, using $2 + 1$ flavors of improved staggered
quarks using the so-call Asqtad
action~\cite{Orginos:1999cr,Orginos:1998ue}.%
\footnote{Our results depend upon the validity of the rooting procedure used in the generation of the staggered ensembles.  For recent discussion on this topic in the literature, we refer the reader to Refs.~\cite{Durr:2004ta,Durr:2006ze,Shamir:2006nj,Bernard:2006ee,Bernard:2007ma,Bernard:2007eh}.  For the rest of this work, we assume this method to be valid.} 
For this study we
use the $20^3 \times 64$ lattices generated at six values of the
light-quark masses, with the heaviest corresponding to the case of
three degenerate quark masses approximately at the strange-quark mass and the lightest corresponding to a pion mass of $m_\pi\sim 290$~\texttt{MeV}.  The scale is
set by the lattice spacing $ a = 0.12406$ fm determined from heavy
quark spectroscopy\cite{Aubin:2004wf} with an uncertainty of 2\%,
yielding a spatial volume $V = (2.5\,{\rm fm} )^3 $.

For the valence quarks, we adopt domain-wall fermions
(DWF)\cite{Kaplan:1992bt,Furman:1994ky}; these have the arbitrarily
precise realization of chiral symmetry required for our study of
nucleon structure, and admit a straightforward interpretation of
baryon quantum numbers, in particular parity.  In order to reduce the
density of small eigenvalues, the lattices are first
HYP-smeared~\cite{Hasenfratz:2001hp}. 
For computational economy, the majority of the valence propagators are computed on lattices of temporal extent $32$ rather than $64$, with Dirichlet boundary conditions imposed in the temporal
direction.  We refer to these as \textit{chopped} propagators and to the those which are computed with the full length in time as \textit{unchopped}.  We improve statistics by making several computations of the spectrum for each configuration, using different temporal ranges.

The physical quark
fields, $q(\vec x, t)$, reside on the 4-dimensional boundaries with
fifth coordinate 1 and $L_s$ and have bare quark parameter $(a
m)^{DWF}_q$. The wall height $M_5 = 1.7$ was chosen on the basis of
spectral flow analyses to optimize the evaluation of domain wall
propagators, and $L_s = 16$ was chosen\cite{Renner:2004ck} to ensure
that the residual mass characterizing residual chiral symmetry
breaking is always less than 20\% of the physical quark mass as
discussed later.

\subsection{Tuning of the valence-quark mass\label{sec:tuning}}

We set the valence strange-quark mass $\msdwf$ using the $N_f = 3$ ensemble corresponding to three degenerate sea quarks with $\mud$ by imposing that the valence pseudoscalar mass equal the mass of the Goldstone boson constructed using staggered quarks within the calculational precision of a few percent.  This yields $\msdwf = 0.081$.  We maintain the same $\msdwf$ on the $N_f = 2 + 1$ ensembles.  Similarly, the valence quark mass is set by tuning the DWF pion mass to the taste-5 staggered Goldstone-boson pion to within the calculational precision of a few percent.  See Table~\ref{tab:mpi} for more details.  There is a freedom in the tuning choice, and for any choice, the mixed action will violate unitarity at finite lattice spacing.  The role of the mixed action EFT analysis is to recover the proper continuum QCD behavior for any tuning choice~\cite{Bar:2005tu}, and our particular tuning choice provides for the most chiral valence fermions, which recover the QCD point in the continuum limit.  For further details, see Ref.~\cite{Renner:2004ck}.  Throughout this work, we will refer to a lattice ensemble generated with a light Asqtad quark mass of $am^{asqtad}_{u/d} = 0.007$ as the $m007$ ensemble, and similarly for the other light quark masses.

\begin{table}
\caption{\label{tab:mpi} The values of the domain-wall (DW) quark masses, the resulting pion mass obtained using the mixed action (MA) DW valence quarks on the asqtad MILC ensembles, together with the corresponding lightest Goldstone-boson pion mass obtained by the MILC collaboration to which the MA pion masses were tuned.}
\begin{ruledtabular}
\begin{tabular}{ccccccc}
&\multicolumn{6}{c}{$\mud$} \\
& 0.007 & 0.010 & 0.020 & 0.030 & 0.040 & 0.050 \\
\hline
\# props & 3085 & 3729 & 3389 & 2255 & 324 & 425 \\
$\muddwf$ & 0.0081 & 0.0138 & 0.0313 & 0.0478 & 0.0644 & 0.081 \\
$m_\pi $ & 0.1847(7) & 0.2242(5) & 0.3124(4) & 0.3761(5) & 0.4338(12) & 0.4774(10) \\
$m_\pi$ (MILC) & 0.1888(2) & 0.2245(2)& 0.3113(2) & 0.3779(2) & 0.4351(03) & 0.4842(02)\\
\end{tabular}
\end{ruledtabular}
\end{table}

\subsection{Determination of $m_{\rm res}$}
\newcommand{\Pp}{\frac{1+\gamma_5}{2}}
\newcommand{\Pm}{\frac{1-\gamma_5}{2}}

Domain wall fermions in the infinite $L_s$ limit possess  an exact chiral symmetry for vanishing $m_q$. The corresponding symmetry transformation is
\begin{eqnarray}
  \Psi(x,s)  &\rightarrow& e^{i\Gamma_5(s) \theta(x)} \Psi(x,s) \\
  \bar\Psi(x,s)  &\rightarrow& \bar \Psi(x,s) e^{-i\Gamma_5(s) \theta(x)}
\end{eqnarray}
where $\Gamma_5(s) = {\rm sign}( \frac{L_s - 1}{2} - s ) $.
However, at finite $L_s$ this chiral symmetry is explicitly broken by
the coupling of left handed and right handed modes in the middle of
the 5th dimension. As a result one can construct the following
partially conserved axial vector current
\begin{equation}
  {\cal A}_\mu(x) = -\sum_{s=0}^{L_s-1} \Gamma_5(s) j_\mu(x,s)\, ,
\end{equation}
where 
\begin{equation}
j_\mu(x,s) = \frac{1}{2}\left[
\bar{\Psi}(x+\hat{\mu},s)(1+\gamma_\mu)U^\dag_{x+\hat{\mu},\mu}\Psi(x,s) -
\bar{\Psi}(x,s)(1-\gamma_\mu)U_{x,\mu}\Psi(x+\hat{\mu},s)
\right],
\end{equation}
 is the four dimensional conserved vector current that corresponds to the 4D Wilson fermion action. 
This current satisfies a Ward-Takahashi identity which in the flavor non-singlet case
takes the form~\cite{Furman:1994ky}:
\begin{eqnarray}
  \Delta_\mu \langle {\cal A}^a_\mu(x) O(y) \rangle = 
        2m_q \langle \bar q(x)\tau^a \gamma_5 q(x) O(y) \rangle
        + 2 \langle  \bar q_{mp}(x)\tau^a \gamma_5 q_{mp}(x) O(y)
        \rangle + i \langle \delta^a O(y) \rangle \, ,
\label{eq:ward_tak_id}
\end{eqnarray}
where 
\begin{eqnarray}
  q(x) &=& \Pm\Psi(x,0)  +  \Pp\Psi(x,L_s-1) \nonumber \\
  \bar{q}(x) &=&\bar{\Psi}(x,L_s-1)\Pm + \bar{\Psi}(x,0)\Pp \;.
\end{eqnarray}
 are the physical 4D quark degrees of freedom localized at the
 boundaries of the 5th dimension and
\begin{eqnarray}
  q_{mp}(x) &=& \Pm\Psi(x,\frac{L_s}{2})  +  
                \Pp\Psi(x,\frac{L_s}{2}-1)
 \nonumber \\
 \bar{q}_{mp}(x) &=& \bar{\Psi}(x,\frac{L_s}{2}-1)\Pm+
                    \bar{\Psi}(x,\frac{L_s}{2})\Pp 
\end{eqnarray}
\begin{figure}
\begin{tabular}{cc}
\includegraphics[width=0.4\textwidth]{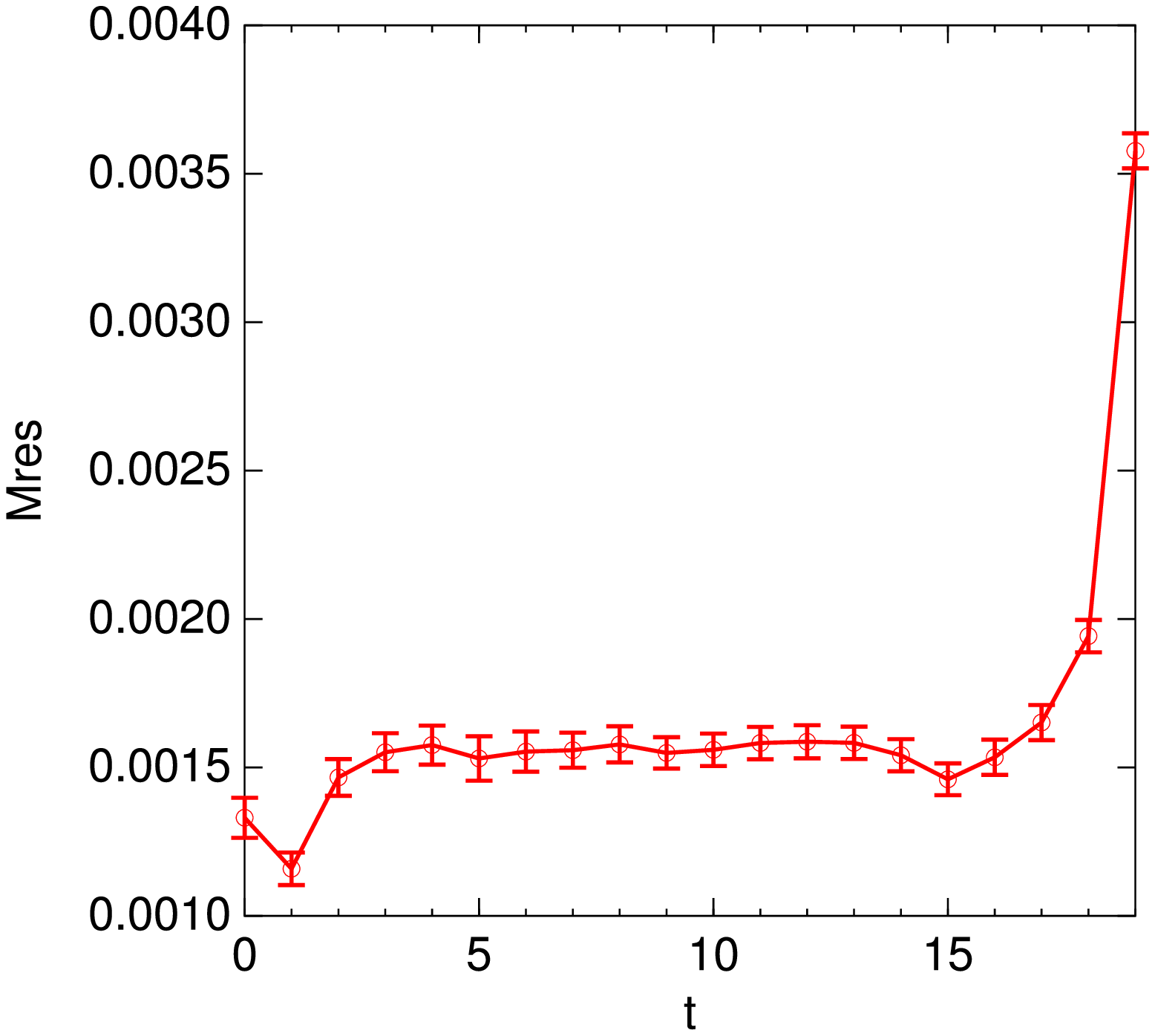}
&\includegraphics[width=0.4\textwidth]{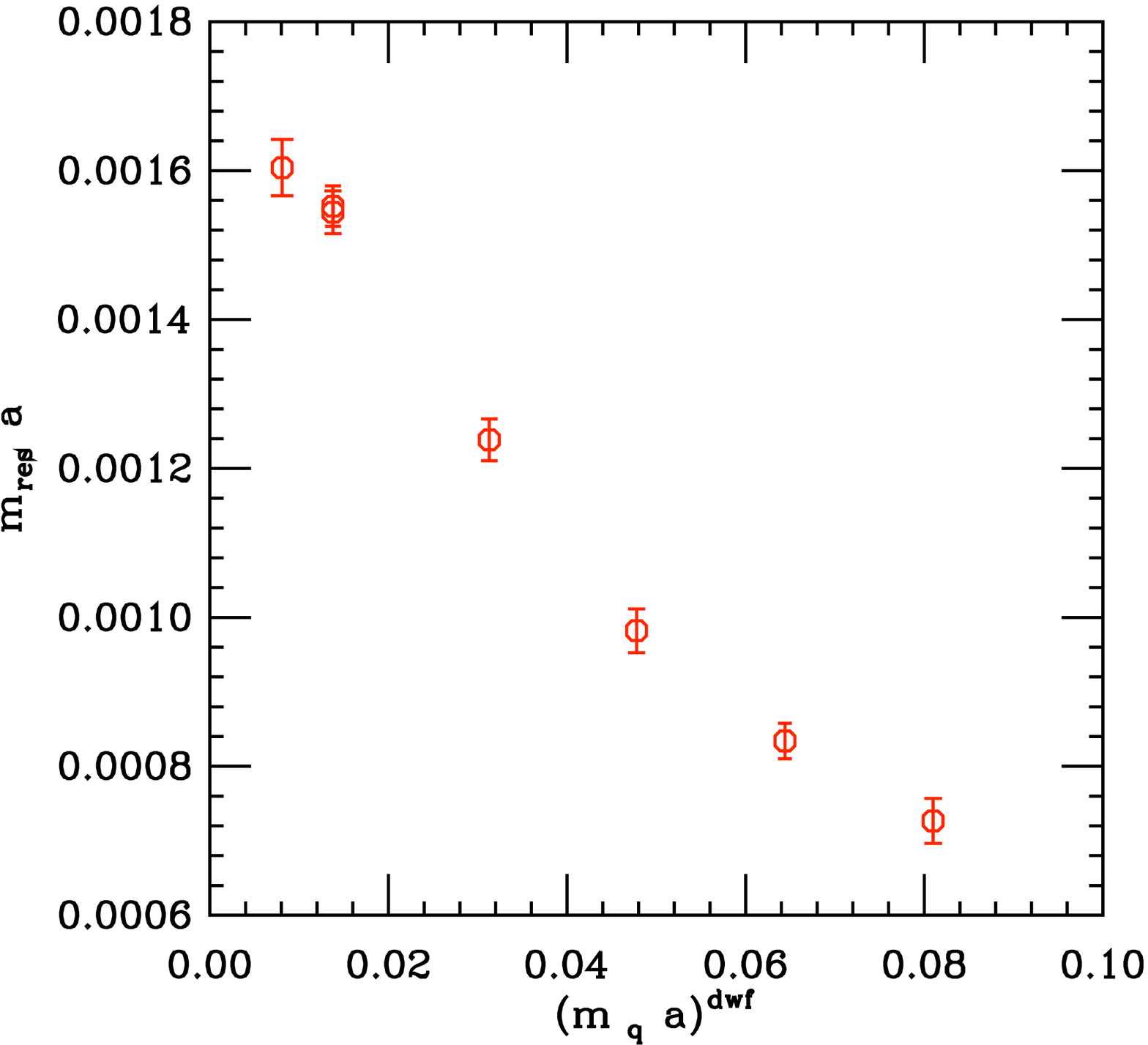}
\\
$(a)$ & $(b)$
\end{tabular}
\caption{The left-hand figure shows the residual mass determined from
the ratio $R(t)$ in Eq.~\eqref{eq:rt}, for the data at $\mud =
0.010$; the quoted value of $m_{\rm res}$ is obtained from a constant
fit to the data.  Note that the effect of the Dirichlet boundary at $t
= 22$ relative to the source is apparent as far out as $t = 14$.  The
right-hand figure shows the quark-mass dependence of the residual
mass. The main reason for the increase in $m_{\rm res}$ with
decreasing $\mud$ is the increased roughening of the gauge field; in
typical quenched calculations, a smaller dependence of the residual
mass on the quark mass is observed.  As a result, chiral symmetry is
satisfied to a lesser degree at fixed $L_s$ as the pion mass decreases.
\label{fig:mres}}
\end{figure}
are four dimensional fields constructed at the midpoint of the 5th
dimension.  The Ward-Takahashi identity of Eq.~\eqref{eq:ward_tak_id}
is the same as the continuum counterpart, except for an additional term
$2 \langle \bar q_{mp}(x)\tau^a \gamma_5 q_{mp}(x) O(y) \rangle$. This
term is there only at finite $L_s$~\footnote{For the flavor singlet
current this term survives the infinite $L_s$ limit and gives rise to
the axial anomaly~\cite{Shamir:1993yf,Kikukawa:1999sy}.} and it is a measure of the explicit chiral symmetry
breaking. At long distances this term is proportional to $2 \langle
\bar q(x)\tau^a \gamma_5 q(x) O(y) \rangle$.  Using the pseudo-scalar
density as a probe operator $O(y)$ the residual mass is defined as
\begin{equation}
{\rm m}_{\rm res} = \frac{1}{t_{max}-t_0}\sum_{t_0}^{t_{max}}\frac{\langle  \bar q_{mp}(t)\tau^a \gamma_5 q_{mp}(t)  \bar q(0)\tau^a \gamma_5 q(0)\rangle}{\langle  \bar q(t)\tau^a \gamma_5 q(t)  \bar q(0)\tau^a \gamma_5 q(0)\rangle}\;,
\end{equation}
where $\left[ t_0, t_{max} \right]$ is the time interval where only the ground state pion contributes
to the two correlators in the ratio.
The ratio
\begin{equation}
R(t) = \frac{\langle  \bar q_{mp}(t)\tau^a \gamma_5 q_{mp}(t)  \bar q(0)\tau^a \gamma_5 q(0)\rangle}{\langle  \bar q(t)\tau^a \gamma_5 q(t)  \bar q(0)\tau^a \gamma_5 q(0)\rangle}\;, \label{eq:rt}
\end{equation}
from which the residual mass is determined, is plotted in the left panel of 
Figure~\ref{fig:mres}  for the $\mud =0.01$ ensemble; $m_{\rm res}$ is obtained from a fit of the data to a constant.  The right-hand panel shows the
quark-mass dependence of $m_\textrm{res}$. The statistical errors on the residual mass are determined using the jackknife method.

\section{Results \label{sec:spectrum}}
\subsection{Fitting methodology\label{sec:fit}}

An additional complication in performing fits to correlation functions constructed using the DWF valence-quark action arises from the oscillating terms in the DWF transfer matrix. As explained in Ref.~\cite{Syritsyn:2007mp}, for the choice of domain wall height $M_5 = 1.7$, lattice artifacts at the cutoff scale produce negative eigenvalues in the transfer matrix that cause temporal oscillations in correlation functions at short times. Since these modes at the  cutoff scale decay rapidly and do not contribute to physical low mass states, one can fit the correlation functions used for our spectroscopic measurements by the usual sum of decaying exponentials plus additional oscillating exponentials $e^{-Mt}(-1)^t$.  Hence, in this work, we have used the following form~\cite{Syritsyn:2007mp,Renner:2007pb} that includes the physical state of interest, the next excited physical state, and one oscillating artifact state:
\begin{equation}
C(t) = A_0 e^{- M_0 (t - t_{\rm src})} + A_1 e^{- M_1 (t - t_{\rm
    src})}  + A_{\rm osc} (-1)^t e^{- M_{\rm osc} (t - t_{\rm
    src})}.\label{eq:osc}
\end{equation}

Results for the spectrum of both mesons and baryons are obtained from
fits to a single correlator smeared both at the source and at the
sink.  Our fitting method is as follows. We perform fits using
the form Eq.~\eqref{eq:osc} introduced above, which enables the use of
data close to the source, but the need for six fitting
parameters. We also perform single-exponential, two-parameter fits to
correlators smeared both at the source and at the sink, with a final
fit range chosen so as to ensure a acceptable quality of fit.  In
general the masses are insensitive to the largest time extent used in
the fit.  We check the consistency between the two fitting methods and except where noted, obtain most final results using the single-exponential method.  

In general, the quoted errors are obtained using single-elimination
jack-knife.  However, for the study of the chiral behavior of the
baryon spectrum, we adopt a different procedure.  We chose a method that aids in accounting for the correlations among the fit parameters for the chiral extrapolations of the baryon masses.  In the case of a calculation in the quenched approximation, the same number of
configurations would be employed at each value of the valence quark
mass, and therefore this could be accomplished using the jack-knife
procedure above.  For our full QCD calculation, we have different
numbers of configurations at each quark mass, and therefore we adopt a
bootstrap procedure in which we use 10,000 bootstrap samplings on
each ensemble, irrespective of the number of configurations.  In this
case, the quoted errors are obtained using the 68\% confidence level
in the bootstrap distribution.

\subsection{Meson spectrum \label{sec:mesons}}

\begin{figure}[t]
\begin{tabular}{cc}
\includegraphics[width=0.42\textwidth]{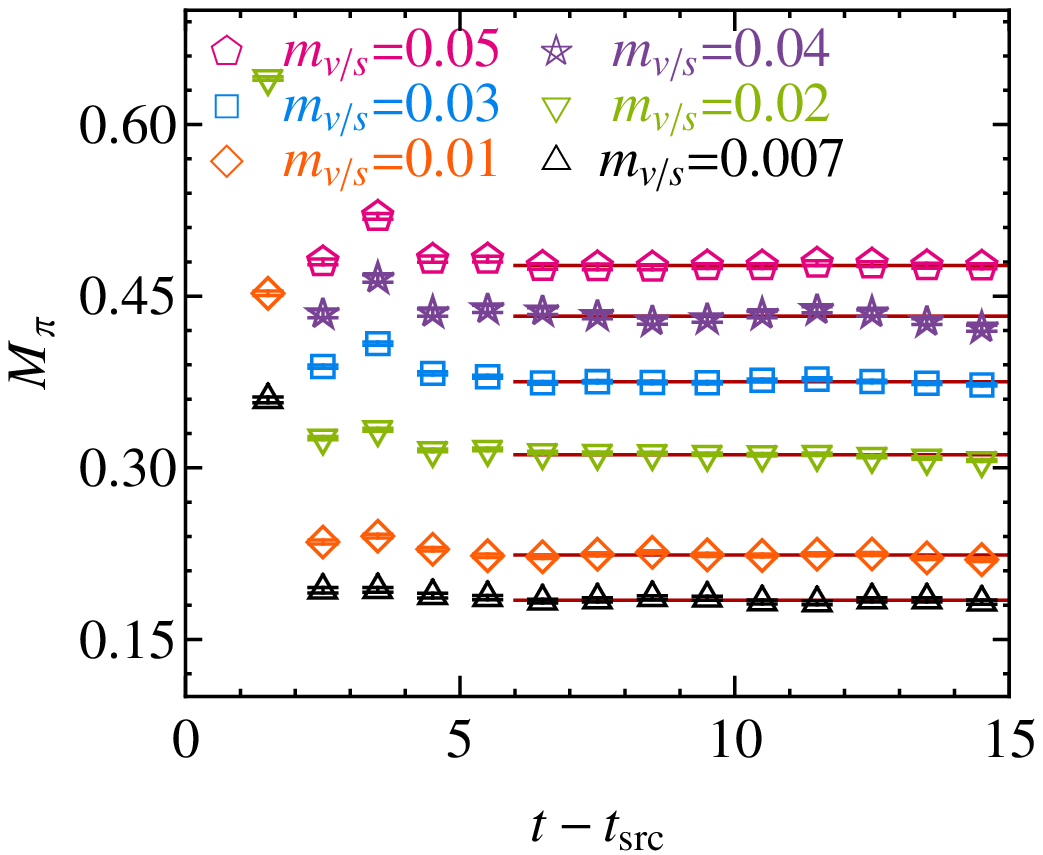}
&\includegraphics[width=0.42\textwidth]{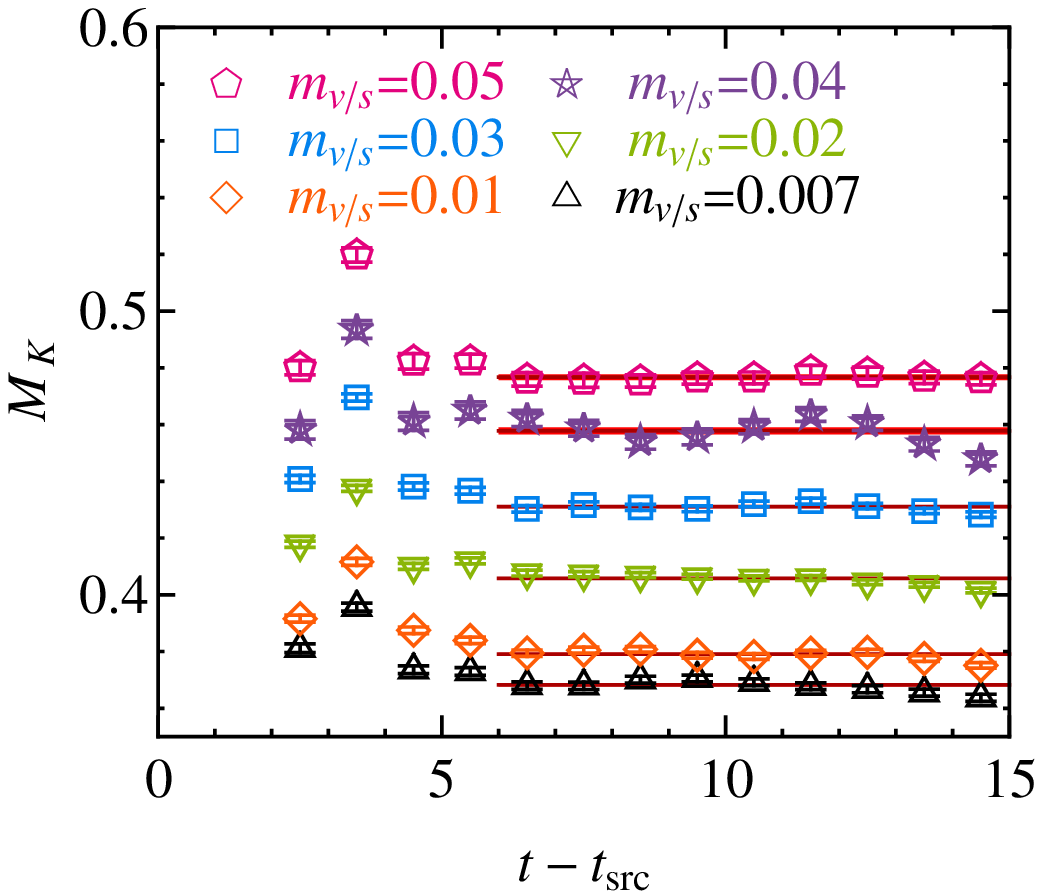}
\\
$(a)$ & $(b)$
\end{tabular}
\caption{\label{fig:PS-meson} The pion (left) and kaon (right)
effective masses at each value of the light-quark mass, together
with the single-exponential fits to the correlators, as described in
the text.  The oscillatory terms in the transfer matrix are evident in the effective mass close to the source.}
\end{figure}

\begin{figure}
\begin{tabular}{cc}
\includegraphics[width=0.45\textwidth]{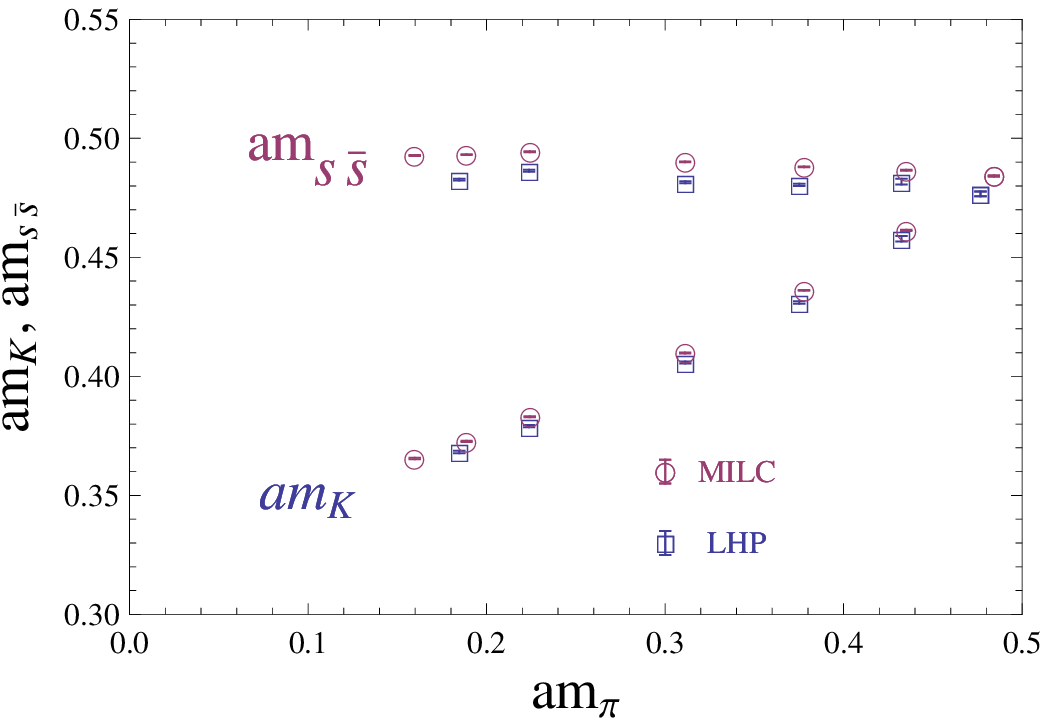}
&\includegraphics[width=0.45\textwidth]{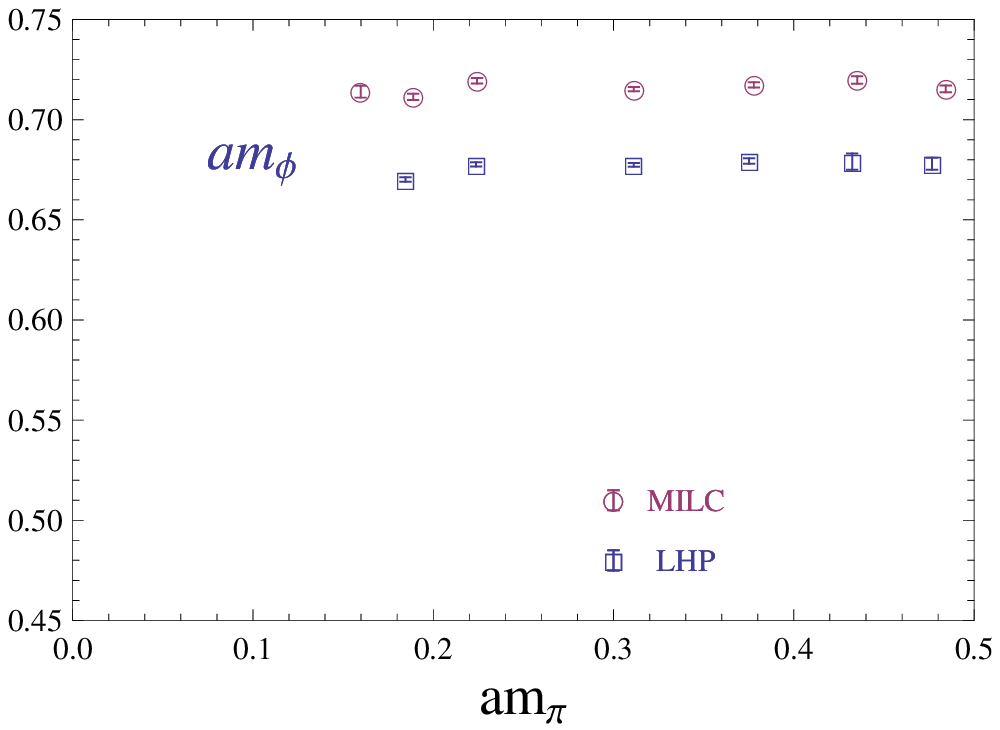}
\\$(a)$ & $(b)$ \\
\includegraphics[width=0.45\textwidth]{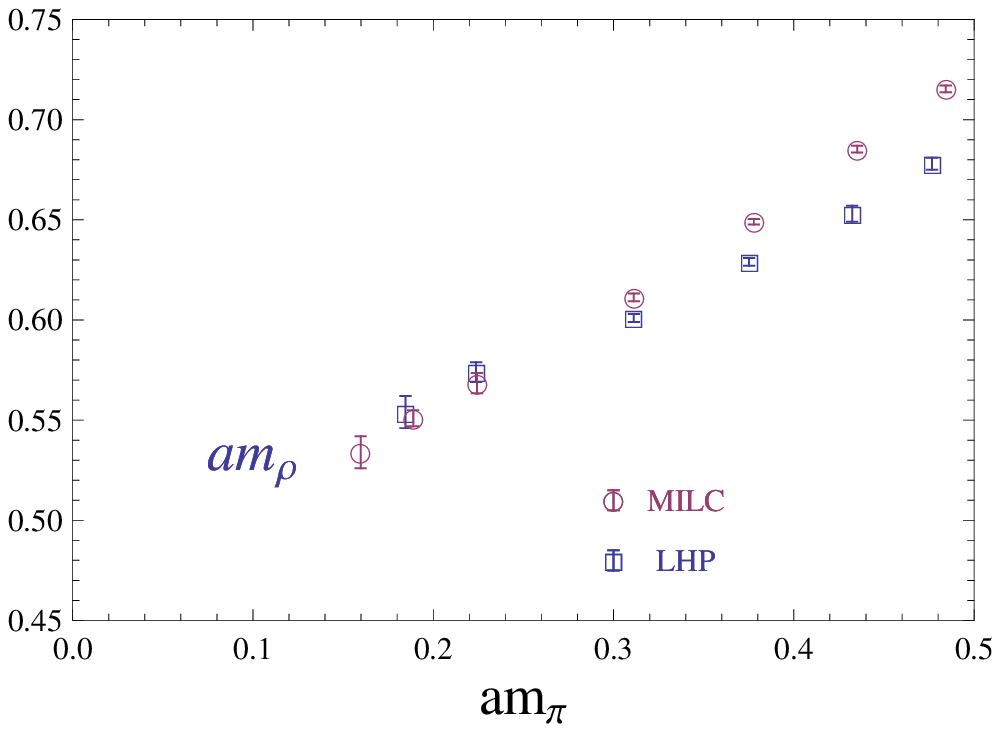}
&\includegraphics[width=0.45\textwidth]{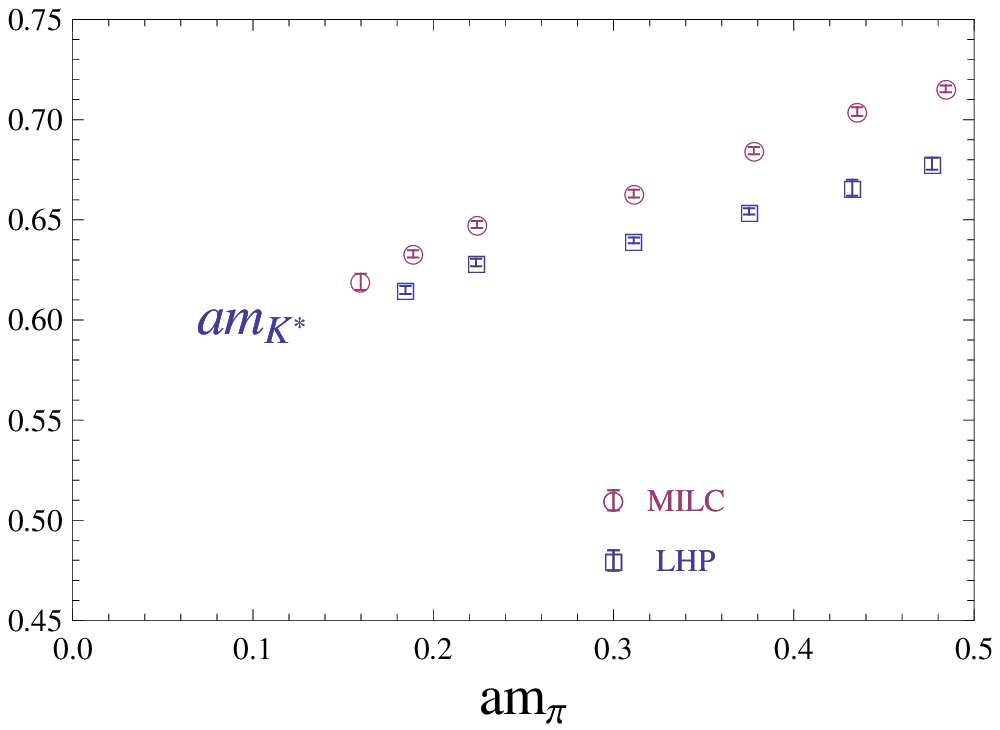}
\\ $(c)$ & $(d)$
\end{tabular}
\caption{\label{fig:VMesons} Comparison between pseudo-scalar and vector masses with domain wall valence quarks (LHP) and staggered valence quarks (MILC) computed on the coarse ($a\approx 0.125$~\texttt{fm}) MILC ensembles.}
\end{figure}

We use meson interpolating fields of the form $\bar{\psi}\Gamma\psi$, which overlap with the physical states listed in Table~\ref{tab:meson_mass}; charge conjugation $C$ applies only to particles with zero net flavor.  Our single-exponential fits using the procedure described above are also summarized in Table~\ref{tab:meson_mass}.  In Figure~\ref{fig:PS-meson}, we show the effective masses of the pion and kaon, together with the corresponding fits.  
In the case of the flavor-neutral states, for example $s\bar{s}$, $\phi$, \textit{etc.}, we only compute the connected contributions to the correlation functions.  In Figure~\ref{fig:VMesons}, we plot our resulting kaon, $s\bar{s}$ and vector meson masses along with those of the coarse ($a\sim0.124$~\texttt{fm}) MILC calculations~\cite{Bernard:2001av,Aubin:2004wf}.  The differences in these correlation functions are the direct result of differences in the valence fermion actions used in the calculations, and thus a measure of discretization effects. 

\begin{table}
\caption{\label{tab:meson_mass}Meson masses in lattice units, calculated from the correlation functions created with the interpolating operators, $\bar{\psi}\G\psi$.  Masses we were unable to extract are denoted ``N/A".} 
\begin{ruledtabular}
\begin{tabular}{cccccccccc}
$Particle$ & $J^{PC}$ & $\G$
& $m007$ & $m010$ & $m020$ & $m030$ & $m040$ & $m050$ \\
\hline
$\pi$ & $0^{-+}$ & $\g_5$ 
& 0.1842(7) &  0.2238(5) &  0.3113(4) &  0.3752(5) &  0.4324(12) &  0.4767(10) \\
$K$ &$0^{-+}$ & $\g_5$ 
&0.3682(5) &  0.3791(5) &  0.4058(4) &  0.4311(5) &  0.4578(12) &  0.4767(10) \\
$s\bar{s}$ &$0^{-+}$ & $\g_5$ 
&0.4827(4) &  0.4846(4) &  0.4816(4) &  0.4805(5) &  0.4818(12) &  0.4767(10) \\
$\rho$ &$1^{--}$ & $\g_i$ 
&0.554(8) &  0.574(5) &  0.601(2) &  0.629(2) &  0.653(4) &  0.678(3) \\
 $K^*$ &$1^{--}$ & $\g_i$ 
&0.615(2) &  0.6287(19) &  0.6397(14) &  0.6542(16) &  0.666(4) &  0.678(3) \\
$\phi$ &$1^{--}$ & $\g_i$ 
&0.6701(11) &  0.6777(10) &  0.6774(10) &  0.6794(14) &  0.679(4) &  0.678(3) \\
$a_0$ &$0^{++}$ & $\mathbf{1}$ 
&N/A &  N/A & N/A &  1.08(6) &  0.93(9) &  0.88(4) \\
$K_0^*$ &$0^{++}$ & $\mathbf{1}$ 
&N/A &  1.05(6) &  0.93(3) &  0.95(3) &  0.91(6) &  0.88(3) \\
$f_0$ &$0^{++}$ & $\mathbf{1}$ 
&0.809(14) &  0.833(13) &  0.879(18) &  0.93(3) &  0.88(7) &  0.86(6) \\
$a_1$ &$1^{++}$ & $\g_i \g_5$ 
&0.78(5) &  0.84(4) &  0.93(2) &  0.920(19) &  0.94(4) &  0.97(3) \\
$K_1$ &$1^{++}$ & $\g_i \g_5$ 
&0.89(2) &  0.915(17) &  0.960(14) &  0.954(15) &  0.96(4) &  0.97(3) \\
$f_1$ &$1^{++}$ & $\g_i \g_5$ 
&0.929(8) &  0.956(9) &  0.983(9) &  0.982(12) &  0.97(3) &  0.97(3) \\
$b_1$ &$1^{+-}$ & $\g_i \g_j$ 
&1.1(4) &  1.04(11) &  0.91(3) &  0.94(3) &  1.03(6) &  1.03(4) \\
$h_1$ &$1^{+-}$ & $\g_i \g_j$ 
&0.92(2) &  0.98(2) &  0.97(3) &  0.99(4) &  1.09(13) &  1.05(10) \\
\end{tabular}
\end{ruledtabular}
\end{table}

In Figure~\ref{fig:VMesons}$(a)$, one observes that the kaon and $s\bar{s}$ masses are the same within a few percent over the entire range of light quark masses used.  This is the direct result of the tuning we have employed, discussed in Sec.~\ref{sec:tuning} and displayed in Table~\ref{tab:mpi}.  The strange quark mass was tuned on the $m005$ ensemble and held fixed both in all the coarse MILC ensembles as well as in all the valence domain-wall propagators calculated for this work.  As the pion, kaon and $s\bar{s}$ mesons are pseudo-scalars, the good chiral properties of the domain-wall propagators and the taste-5 staggered pseudo-scalar mesons protects these masses from additive mass renormalization resulting in the good numerical agreement observed.  All other hadron masses are expected to suffer from additive lattice spacing mass corrections (beginning at $\mc{O}(a^2)$), the size of which depends on the specifics of the discretization method.  Evidence of this is given, for example by calculations of the $\phi$ mass, which we compare to those of MILC in Fig.~\ref{fig:VMesons}$(b)$.
However, while we observe the mass splitting of the $\phi$ is largely independent of $m_\pi$, as expected, this is not the case for the $\rho$ and $K^*$ vector meson masses.  Our calculational results tend to converge with those of the coarse MILC calculations, depicted in Fig.~\ref{fig:VMesons}$(c)$ and $(d)$.

\begin{figure}
\begin{tabular}{cc}
\includegraphics[width=0.45\textwidth]{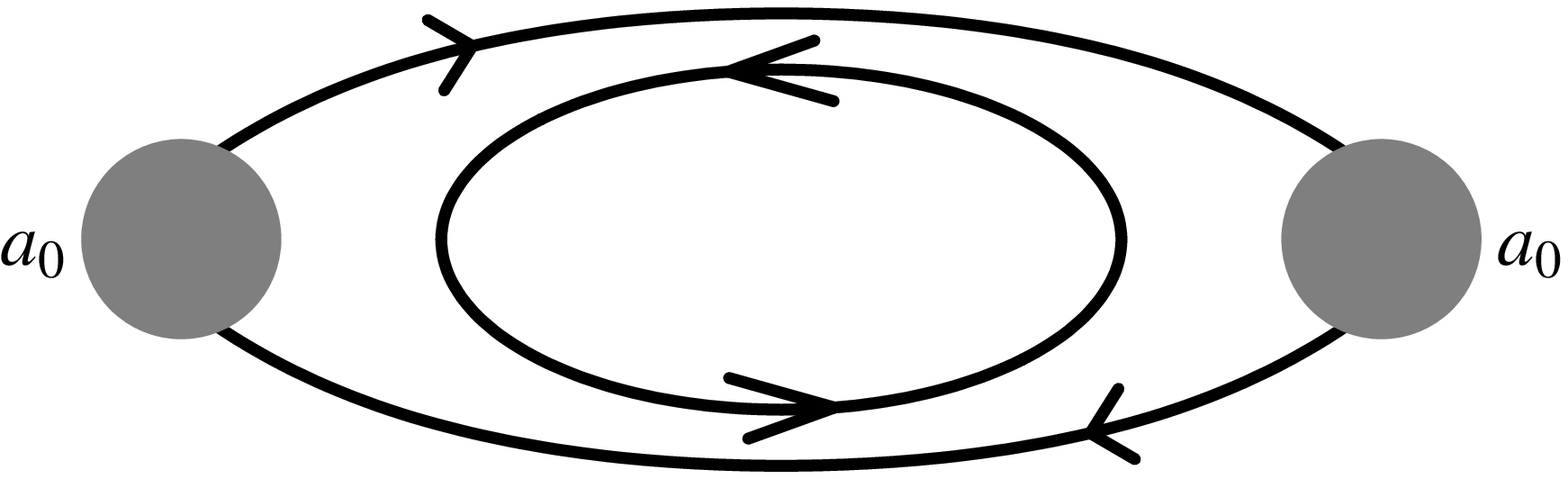}
&\includegraphics[width=0.45\textwidth]{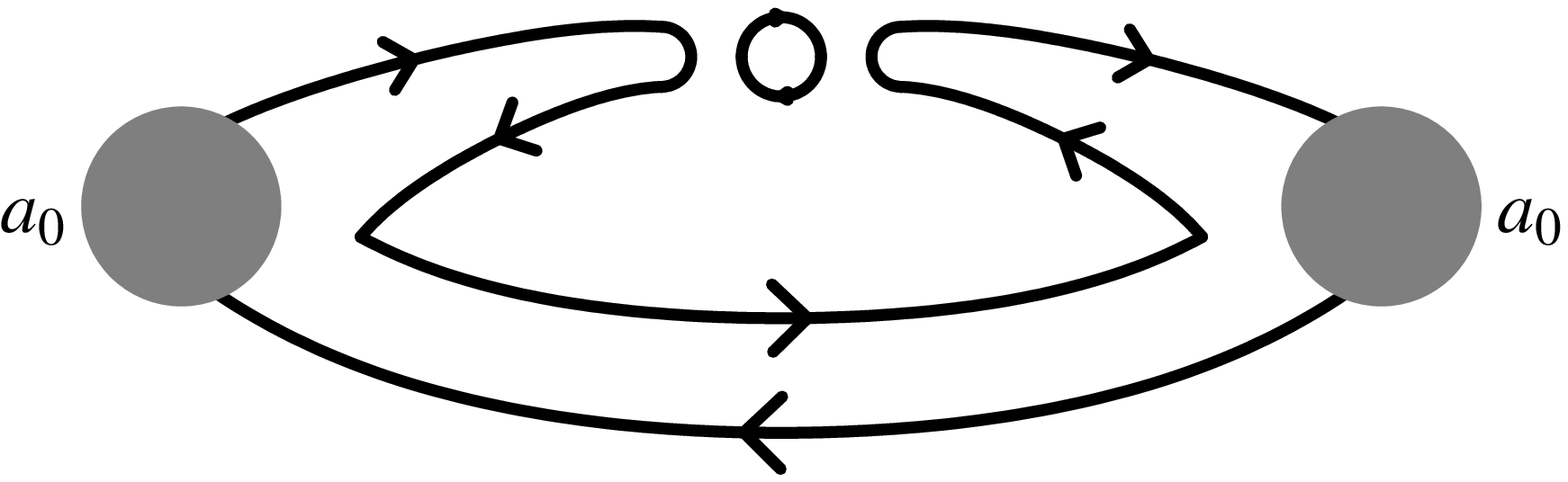}
\\
$(a)$ & $(b)$
\end{tabular}
\caption{\label{fig:a0-bubble} Two types of ``bubble'' contributions to the scalar meson: the left one is $B_1$ in Eq.~\eqref{eq:a0-bubble} while the right panel is $B_2$ in Eq.~\eqref{eq:a0-bubble}.
}
\end{figure}

The comparison of these meson mass calculations is complicated by the fact that they are unstable particles, at least at the lightest two mass points. 
\footnote{Strictly speaking, in a Euclidean finite volume, there are not unstable particles, but rather a mixing of single and multiple particle states with forbidden energy level crossings~\cite{Luscher:1991cf}.  Then, a particle which in an infinite Minkowski volume is unstable, in Euclidean finite volume may experience power law corrections to its energy levels instead of the standard corrections which scale exponentially as $\textrm{exp}\left(-m_\pi L\right)$.  In this work we will adopt the perhaps misleading, but intuitively guided language of describing these states with Minkowski-space terminology.  Because of the constrained kinematics in the finite volumes we work with, these would-be \textit{unstable} particles are kinematically forbidden from \textit{decaying}.  Or, in more appropriate terminology, the mixing of the eigenstates of the hamiltonian experiencing power law volume dependence with the states experiencing exponentially small volume dependence is suppressed for the typical hadron correlation functions calculated on the lattices we use in this work.} 
%
While we can not explain in detail the observed diminishing of the mass splitting of the $\rho$ and $K^*$ masses at lighter pion masses, we do believe we understand the origin of the noted different behavior.  There are two important issues related to the comparison of our calculation with that of MILC, which arise from the lack of unitarity present in both our mixed action calculation and that of MILC.  These are in fact the same issues which must be addressed for the calculation of the $a_0$ correlator~\cite{Prelovsek:2005rf,Bernard:2007qf} (which we address in more detail shortly).  For both discretization methods, there are hairpin interactions in the decay channel, similar to Fig.~\ref{fig:a0-bubble}$(b)$ for $a_0$.  However, in the case of staggered valence propagators, there are two additional hairpin interactions than for the mixed action calculation.  Calculated in the appropriate EFT, these hairpin interactions are known provide a negative contribution to the correlation functions and become particularly relevant when the intermediate two-particle states are close to going ``on-shell."  In the continuum limit, the hairpin interactions in the two lattice calculations would be the same, and would vanish on the QCD line of degenerate valence and sea quarks.  The second issue is related to the density of two-particle states near the decay threshold in the two different lattice calculations, those states which contribute to the $\rho$ and $K^*$ correlators similar to those for the $a_0$ depicted in Fig.~\ref{fig:a0-bubble}$(a)$.  In the case of our mixed-ation calculation, the only two-particle states which couple are those of QCD, $\pi\pi$ for the $\rho$ and $K\pi$ for the $K^*$, however their masses are heavier as they are mixed valence-sea mesons~\cite{Orginos:2007tw}.  In the case of the MILC calculation, there are 16 intermediate two-particle states for each one that exists in QCD, due to the extra staggered taste degrees of freedom, with masses given by the known taste-splittings on the MILC ensembles~\cite{Aubin:2004fs}.  It is plausible that these two issues give rise to the strong mass dependence observed in these vector meson mass splittings.  One would expect similar issues with the $\phi$, however due to the strange quark being too heavy on these configurations, the dominant decay mode of the $\phi$, the $K\bar{K}$ state, is significantly above threshold.  It would have been nice to resolve this issue with a comparison of the fine ($a\approx0.09$~\texttt{fm}) and super-fine ($a\approx0.06$~\texttt{fm}) MILC calculations.  However, the super-fine results are not available yet and the fine have a reduced strange quark mass as well as smaller lattice spacing.


\begin{figure}
\begin{tabular}{cc}
\includegraphics[width=0.45\textwidth]{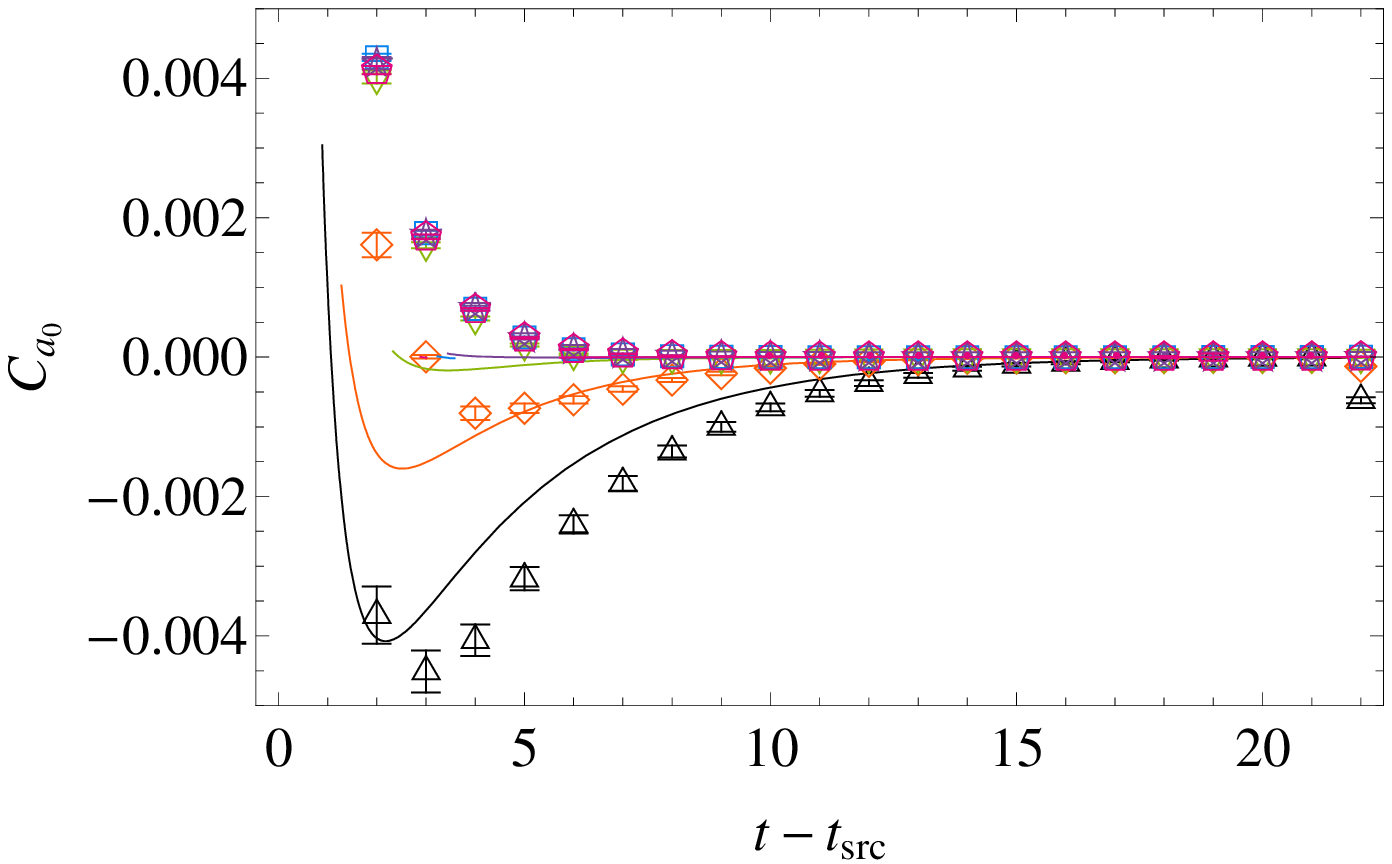}
&\includegraphics[width=0.45\textwidth]{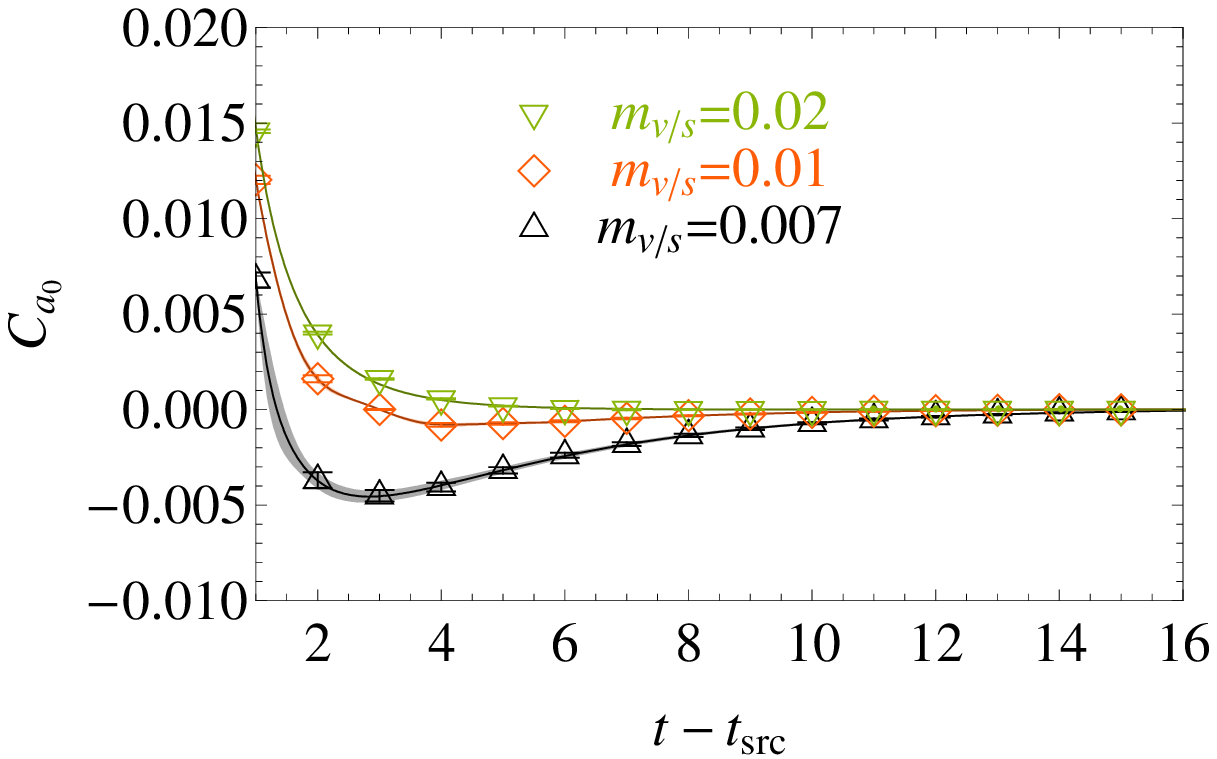}
\\
$(a)$ & $(b)$
\end{tabular}
\caption{\label{fig:a0-corr} The left panel of this figure shows the ``effective'' point-point scalar meson correlator plot (as defined in Eq.~\eqref{eq:C_PP-eff}), along with the mixed-action bubble contribution from Eq.~\eqref{eq:a0-bubble}. Note that the symbols are the same as in Fig.~\ref{fig:PS-meson}.
The right hand panel shows the fits to data from the lightest three ensembles where the largest ``bubble'' contributions dominate.
}
\end{figure}

Returning to the mass of the scalar meson, we note that it has been notoriously difficult to determine, not only on the lattice but also experimentally. In the case of our mixed action, as mentioned above, the scalar meson is sensitive to taste-breaking terms in the staggered sea sector. Such an effect can be described by effective field theory and subtracted from the measured correlators to extract the scalar meson mass correctly in the point-source and point-sink case using the formulation by Prelovsek~\cite{Prelovsek:2005rf}.  This has been performed in detail in Ref.~\cite{Aubin:2008wk} for the case of domain wall valence fermions and the staggered sea fermions.  There are two major ``bubble'' types of contributions (see Fig.~\ref{fig:a0-bubble}) that one should include in order to extract the scalar mass (in the notation of Refs.~\cite{Prelovsek:2005rf,Aubin:2008wk}):
\begin{equation}\label{eq:a0-bubble}
B(t) = \frac{\mu_{dw}^3}{L^3} (B_1(t)+B_2(t))\, ,
\end{equation}
where
\begin{align}\label{eq:a0-bubble1a}
B_1(t) = \sum_{\vec{k}=-L}^L \Bigg\{&
	-\frac{e^{-2 \w_{\text{vv}} t} (t\, \w_{\text{vv}}+1 )
		[(m_{\text{UI}}^2 -m_{\text{vv}}^2 ) (m_{\text{SI}}^2 -m_{\text{vv}}^2 )]}
		{2 \w_{\text{vv}}^4 (m_{\eta_\text{I}}^2 -m_{\text{vv}}^2 )}
\nonumber\\
	&-\frac{e^{-2 \w_{\text{vv}} t}
		\left(2 m_{\text{SI}}^4+m_{\text{UI}}^4+3 m_{\text{vv}}^2
		\left(m_{\text{vv}}^2-2 m_{\text{$\eta
		$I}}^2\right)\right)}{m_{\text{vv}}^2 \left(3 \left(m_{\text{$\eta
		$I}}^2-m_{\text{vv}}^2\right)^2\right)}
	+\frac{2 e^{-\left(\w_{\text{vv}}+\w_{\text{$\eta $I}}\right) t}
		\left(m_{\text{SI}}^2-m_{\text{UI}}^2\right)^2}{9 \left(\w_{\text{vv}} \w_{\text{$\eta $I}}\right) 
		\left(\w_{\text{vv}}^2-\w_{\text{$\eta $I}}^2\right)^2}
	\Bigg\}
\end{align}
%
with $\vec{k}$ running over the allowed triplet integers of lattice momenta and
\begin{eqnarray}\label{eq:a0-bubble1b}
B_2(t) &=& \sum_{\vec{k}=-L}^L \left(\frac{3 e^{-2 \w_{\text{vs}} t}}{4\w_{\text{vs}}^2}+\frac{3 e^{-2 \w_{\text{vu}} t}}{2
\w_{\text{vu}}^2}\right)\, .
\end{eqnarray}
In these equations, $m_{\rm vv}$ is the pion mass calculated from the DWF valence sector,
\begin{align}
&m_{UI}=\sqrt{\Delta_I+2\mu_{\rm stag} m_l}\, ,&
&m_{\rm vu}=\sqrt{\Delta_{\rm mix}+m_{\rm vv}^2+2\mu_{\rm stag} m_l}\, ,&
\nonumber\\
&\w_x=\sqrt{m_x^2+\left(\frac{2 \pi}{L} \right)^2 \vec{k}^2}\, ,&
&m_{\eta\textrm{I}}=\sqrt{\frac{m_{UI}^2+m_{SI}^2}{3}}\, .&
\end{align}
Similarly, $m_{SI}$ and $m_{\rm vs}$ are obtained by replacing the light quark mass, $m_l$, with the strange one, $m_s$, in the staggered sector.  The quantities $\mu_{\rm stag}$, $\mu_{\rm dw}$, $\Delta_{\rm mix}$ and $\Delta_I$ can be obtained from Refs.~\cite{Aubin:2004wf,Aubin:2004fs,Orginos:2007tw}.  
However, in our work, we only compute the Gaussian smeared propagator with Gaussian and point sink. A naive combination of these two correlators provides an ``effective" point-point correlator,
\begin{eqnarray}\label{eq:C_PP-eff}
C_{PP}^{\rm eff} = C_{GP}(t)^2/C_{GG}(t).
\end{eqnarray}
Unlike the close match between the scalar correlator and the bubble term in Ref.~\cite{Aubin:2008wk}, when compared to our ``effective'' point-point correlators, the bubble-term contribution from Eq.~\eqref{eq:a0-bubble} does not fully describe the unphysical contributions, see Fig.~\ref{fig:a0-corr}. This is not completely surprising, since our correlators use Gaussian smearing.
Even though we project onto zero momentum in our calculation, there may be non-trivial contributions to these bubble terms. Therefore, we consider an alternative fit form including $B(t)=d_1 B_1(t) + d_2 B_2(t)$ in Eq.~\eqref{eq:osc}. Unfortunately, even though we are able to fit all the correlators with this new formula, the large number of parameters in the fit results in fitted masses with large error bars.

Similarly, abnormal behavior is observed in the strange partner  $K_0^*$ on the lightest sea ensemble. The $K_0^*$ decays primarily to $K\pi$ and only on the heaviest ensemble is it above this threshold. On the other hand, $f_0$ is the most stable of all the scalar mesons; there is no negative dipping behavior, and decay into two $\rho$ mesons cannot happen here.
As with the p-wave mesons, we found that the $b_1$ masses are within the statistical error bar of the $a_1$ meson on the heaviest 4 ensembles but are heavier on the lightest two ensembles. This is in contradiction with experimental expectations. Their strange partners, $h_1$ and  $f_1$, are almost consistent within their statistical error bars.  However, it is difficult for us to extract similarly accurate masses from these correlators.

\subsection{Decay constants}

Recently, the utility of performing chiral extrapolations in terms of $m_\pi/f_\pi$ has been demonstrated in a number of papers~\cite{Beane:2005rj,Beane:2006kx,Beane:2006gj,Beane:2007xs,Beane:2007uh,Edwards:2006qx,Hagler:2007xi}.  In addition to alleviating scale setting issues, for meson physics, the benefit of this type of extrapolation is understood by using the mixed action EFT in terms of renormalization and the use of ``lattice-physical" parameters, which eliminate the leading and much of the sub-leading lattice spacing corrections~\cite{Chen:2006wf}.  Less restrictive but similar arguments also apply to other hadronic quantities including baryon quantities~\cite{Chen:2007ug}.  To utilize this method, we determine $f_\pi$ with our set of propagators following NPLQCD~\cite{Beane:2006kx}.  In addition, we use the values of $f_K/f_\pi$ in Ref.~\cite{Beane:2006kx} to determine $f_K$ on the $m007$--$m030$ lattices so that we can explore baryon mass extrapolations in terms of $m_\pi/f_K$, providing insight into $SU(3)$ breaking effects.  We also calculate $f_K$ on the $m040$ lattice ensemble since this was not done in Ref.~\cite{Beane:2006kx}.  The resulting values of the decay constants are collected in Table~\ref{tab:decayConst}.
\begin{table}
\caption{\label{tab:decayConst} Pseudoscalar meson decay constants.  An asterisk denotes a value of $f_K$ determined from $f_\pi$ and the values of $f_K/f_\pi$ calculated in Ref.~\cite{Beane:2006kx}.}
\begin{ruledtabular}
\begin{tabular}{c|ccccccc}
$ID$ & $m007$ & $m010$ & $m020$ & $m030$ & $m040$ & $m050$ \\
\hline
 $m_\pi/f_\pi$ &  1.983(12) &  2.325(9) &  3.035(8) &  3.489(9) &  3.82(2) &  4.101(20) \\
 $m_K/f_\pi$ &  3.964(18) &  3.938(13) &  3.957(9) &  4.008(10) &  4.05(2) &  4.101(20) \\
 $f_\pi$ &  0.0929(4) &  0.0963(3) &  0.1026(2) &  0.1076(3) &  0.1131(7) &  0.1162(6) \\
 $f_K$ & $0.1079(4)^*$ & $0.1087(3)^*$ & $0.1103(2)^*$ & $0.1122(3)^*$ &  0.1155(7) &  0.1162(6) \\
\end{tabular}
\end{ruledtabular}
\end{table}

\subsection{Chiral extrapolation of $m_\pi$ and $f_\pi$}
One of the ultimate goals of the combined efforts of lattice QCD calculations and effective field theory is to determine the numerical values of the LECs of the chiral Lagrangians.  This would allow us to test the range of applicability of  chiral perturbation theory and to make predictions of observables not readily computable with lattice QCD.  Perhaps two of the simplest quantities one could imagine performing this chiral extrapolation analysis for are the pion mass and decay constant.  Including the physical value of the pion mass, further allows for a determination of the light quark mass, $ \hat{m}$.  The relevant two-flavor extrapolation formulae for our mixed action calculation are known to NLO~\cite{Bar:2005tu} and given by, 
\begin{align}
m_\pi^2 &=
	2B \hat{m}\left\{
	1+\frac{2B \hat{m}}{(4\pi f)^2}\ln \left( \frac{2B \hat{m}}{\mu^2} \right)
	+4 l_3^r(\mu) \frac{2B\hat{m}}{f^2}
	-\frac{\tilde{\D}_{PQ}^2}{(4\pi f)^2} \left[ 1+\ln \left( \frac{2B\hat{m}}{\mu^2} \right) \right]
	+l_a^m(\mu) \frac{a^2}{f^2}
	\right\}\, ,
\\
f_\pi &=
	f \left\{ 1
	- \frac{2 \tilde{m}_{ju}^2}{(4\pi f)^2} \ln \left( \frac{\tilde{m}_{ju}^2}{\mu^2} \right)
	+2 l_4^r(\mu) \frac{m_\pi^2}{f^2}
	+l_a^f(\mu) \frac{a^2}{f^2}
	\right\}\, .
\end{align}
In these expressions, $l_3^r$ and $l_4^r$ are the renormalized LECs~\cite{Gasser:1983yg}.  The unphysical coefficients $l_a^m$ and $l_a^f$ arise from discretization effects and cannot be determined with one lattice spacing.  However, at this order they act simply to renormalize the LO values of the chiral condensate and decay constant, $B$ and $f$.  The hairpin partial quenching parameter is simply given by (on our specific mixed action calculation) 
\begin{equation}
	\tilde{\D}_{PQ}^2 = a^2 \D_\mathrm{I}\, ,
\end{equation}
where $a^2 \D_\mathrm{I}$ is the known coarse staggered taste splitting~\cite{Aubin:2004fs}.  The mixed valence-sea meson masses are given by 
\begin{equation}
	\tilde{m}_{ju}^2 = m_\pi^2 + a^2 \D_\mathrm{Mix}\, ,
\end{equation}
where the mixed mass splitting has been calculated for the domain-wall valence fermions on the coarse MILC ensembles~\cite{Orginos:2007tw}.  Unfortunately, our numerical calculation is insufficient to reliably extract the values of these LECs.  For example, using the lightest three mass points, corrected for NLO finite volume shifts, one finds a value $\bar{l}_3 = 4.00(5)$ where $\bar{l}_3$ is standardly defined~\cite{Gasser:1983yg}, 
\begin{equation}
	\bar{l}_3 = -4(4\pi)^2 l_3^r(\mu) - \ln \left( \frac{m_{\pi,phys}^2}{\mu^2} \right)\, .
\end{equation}
We have not made any attempt at determining a systematic error bar to emphasize our lack of confidence in this number.  While this number is consistent with other lattice determinations~\cite{Leutwyler:2008ma}, in order to have confidence in the determination of $\bar{l}_3$ and $\bar{l}_4$ from this mixed action method, one will need calculations on at least one additional lattice spacing and likely lighter pion masses. 

\subsection{Baryon spectrum \label{sec:baryons}}

\begin{table}[b]
\centering
\caption{The number of linearly independent interpolating operators
  that can be formed from $u/d$ and $s$ quark quasi-local fields in
  each irreducible representation of the cubic group. The number in
  brackets denotes the dimension of the representation.}
\label{tab:operators}
\begin{tabular}{cccc}
Flavor & $G_{1g/u} (2)$ & $H_{g/u}(4)$ & $G_{2g/u}(2)$\\ \hline
$N$ & 3 & 1 &  - \\
$\Delta$ & 1 & 2 & - \\
$\Lambda$ & 4 & 1 & - \\
$\Sigma$ & 4 & 3 & - \\
$\Xi$ & 4 & 3 & - \\
$\Omega$ & 1 & 2 & -
\end{tabular}
\end{table}
The extraction of the baryon spectrum from a lattice calculation is
obscured by the small number of double-valued irreducible
representations admitted by the cubic symmetry of the lattice, namely
$G_{1 g/u} (2)$, $H_{g/u}(4)$ and $G_{2 g/u} (2)$, where the $g$ and
$u$ denote positive and negative parity, respectively, and the number
in brackets denotes the dimension of the representation.  We follow
the technique introduced in Ref.~\cite{Basak:2005aq} to construct all
the possible baryon interpolating operators that can be formed from
local or quasi-local $u/d-$ and $s$-quark fields; the number of such
interpolating fields is listed in Table~\ref{tab:operators}; the
$G_{2g/u}$ irreducible representations are not accessible with only local or quasilocal
fields.  To yield the largest possible statistical ensemble, the
correlators are averaged over the rows of the irreducible representation.  In this paper,
we present the calculations of the masses of the lowest-lying
positive-parity states in both the octet and decuplet sectors; for the
case of three flavors of degenerate quarks, $SU(3)$ symmetry reduces
the spectrum to single spin-$1/2$ and spin-$3/2$ ground states.


\subsubsection{Comparison of fitting forms}
\begin{figure}[t]
\begin{tabular}{cc}
\includegraphics[width=0.45\textwidth]{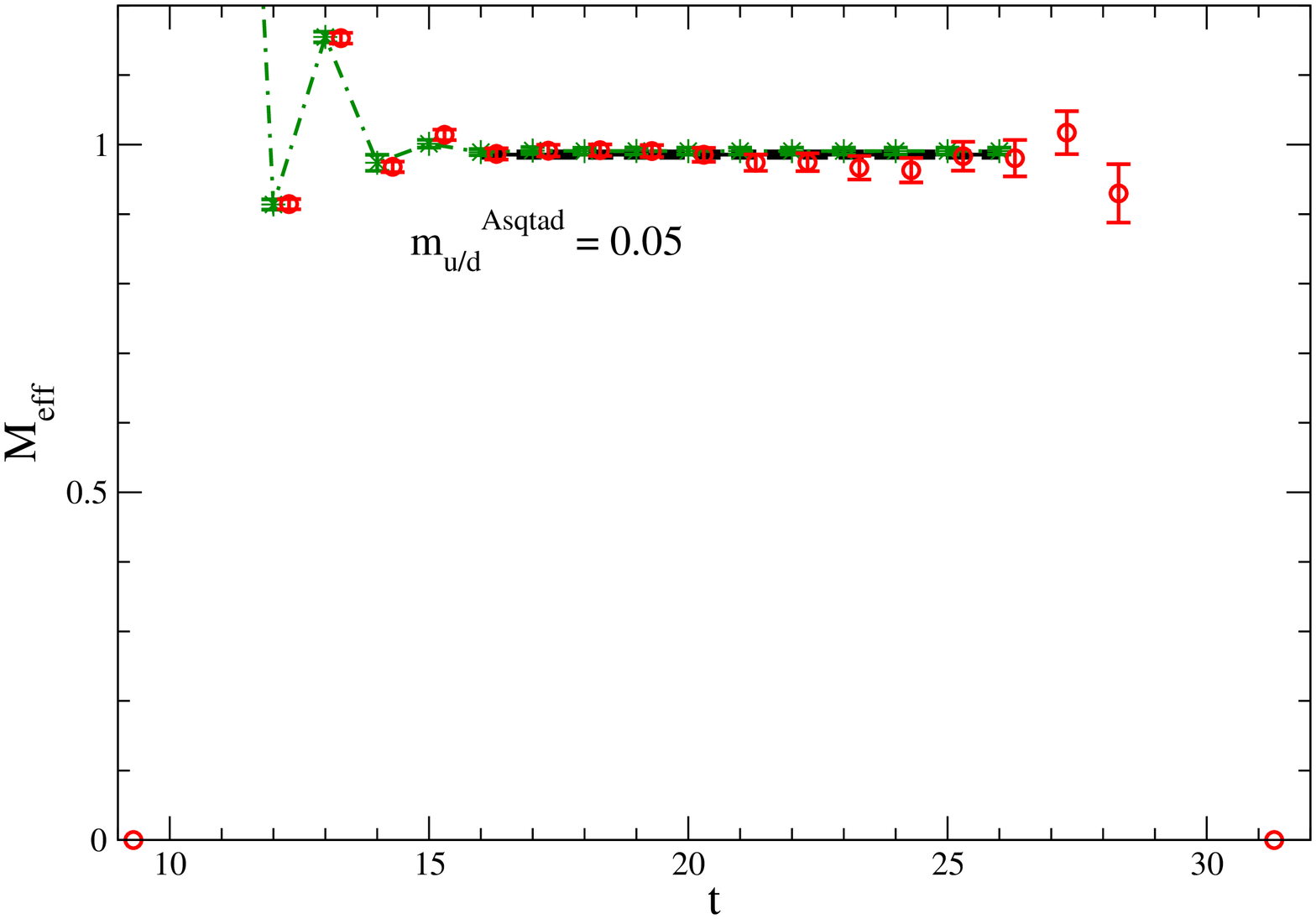}\hspace{1cm}
&\includegraphics[width=0.45\textwidth]{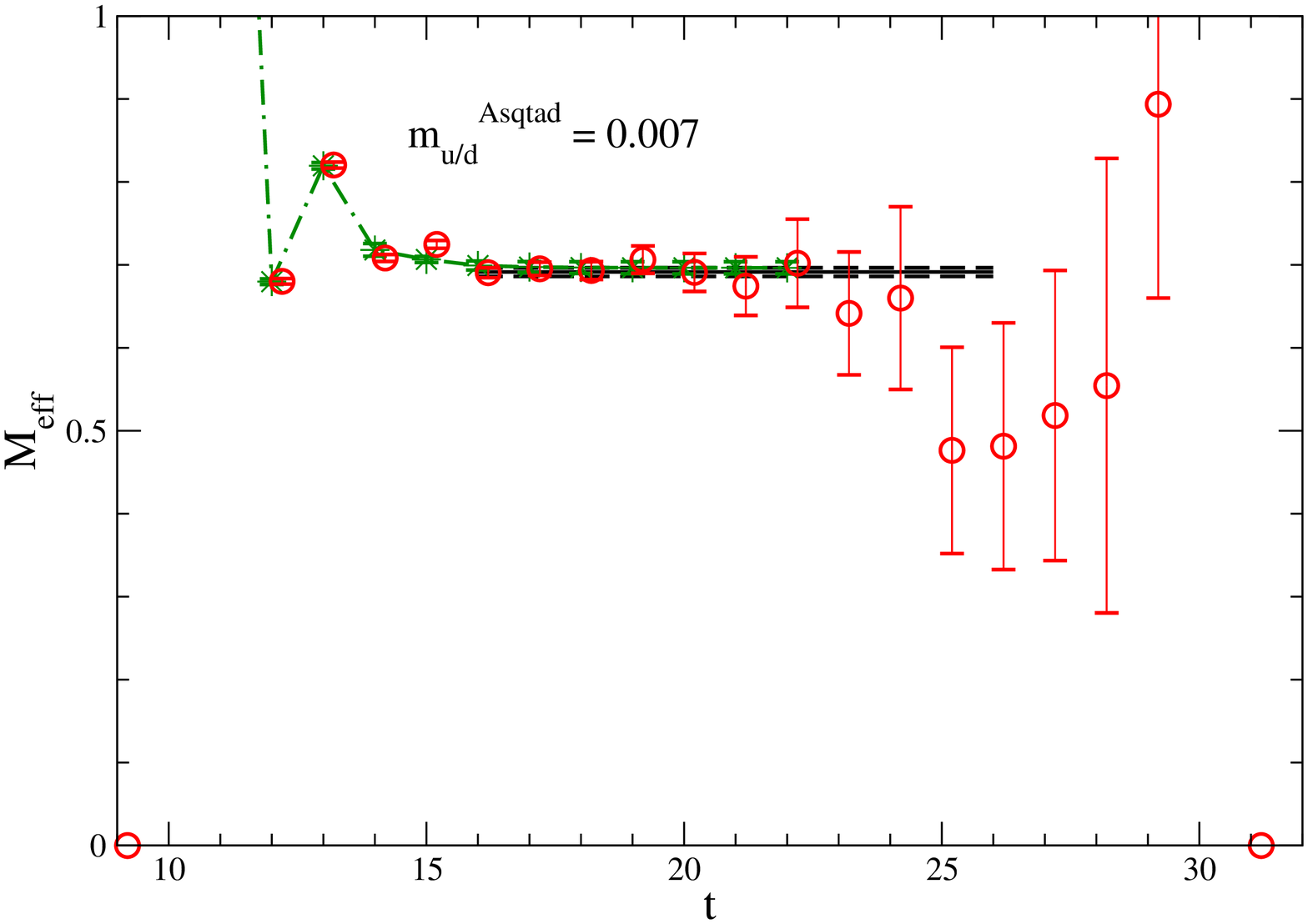}
\\
$(a)$ & $(b)$
\end{tabular}
\caption{The left- and right-hand figures show the nucleon effective
  mass at $\mud = 0.05$ and $\mud =0.007$, respectively, from the smeared-smeared correlators.  The circles
  denote the lattice data for the effective mass, and the solid line and
  dashed-dot line denote the single-exponential, and oscillating fit,
  respectively; note that lattice data is offset for clarity.
  \label{fig:Nosc}}
\end{figure}

\begin{figure}
\vspace{1.1cm}
\begin{tabular}{cc}
\includegraphics[width=0.45\textwidth]{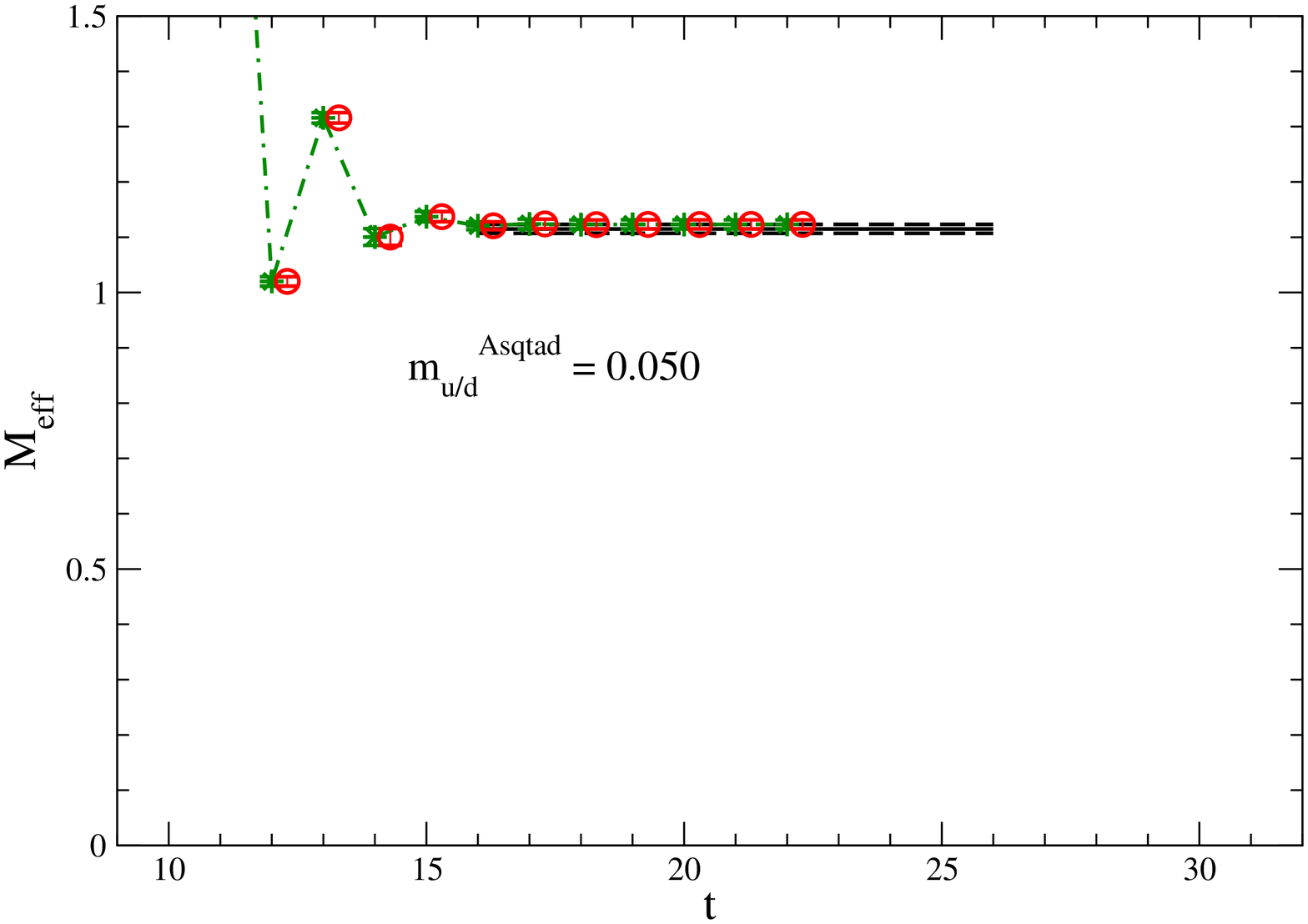}\hspace{1cm}
&\includegraphics[width=0.45\textwidth]{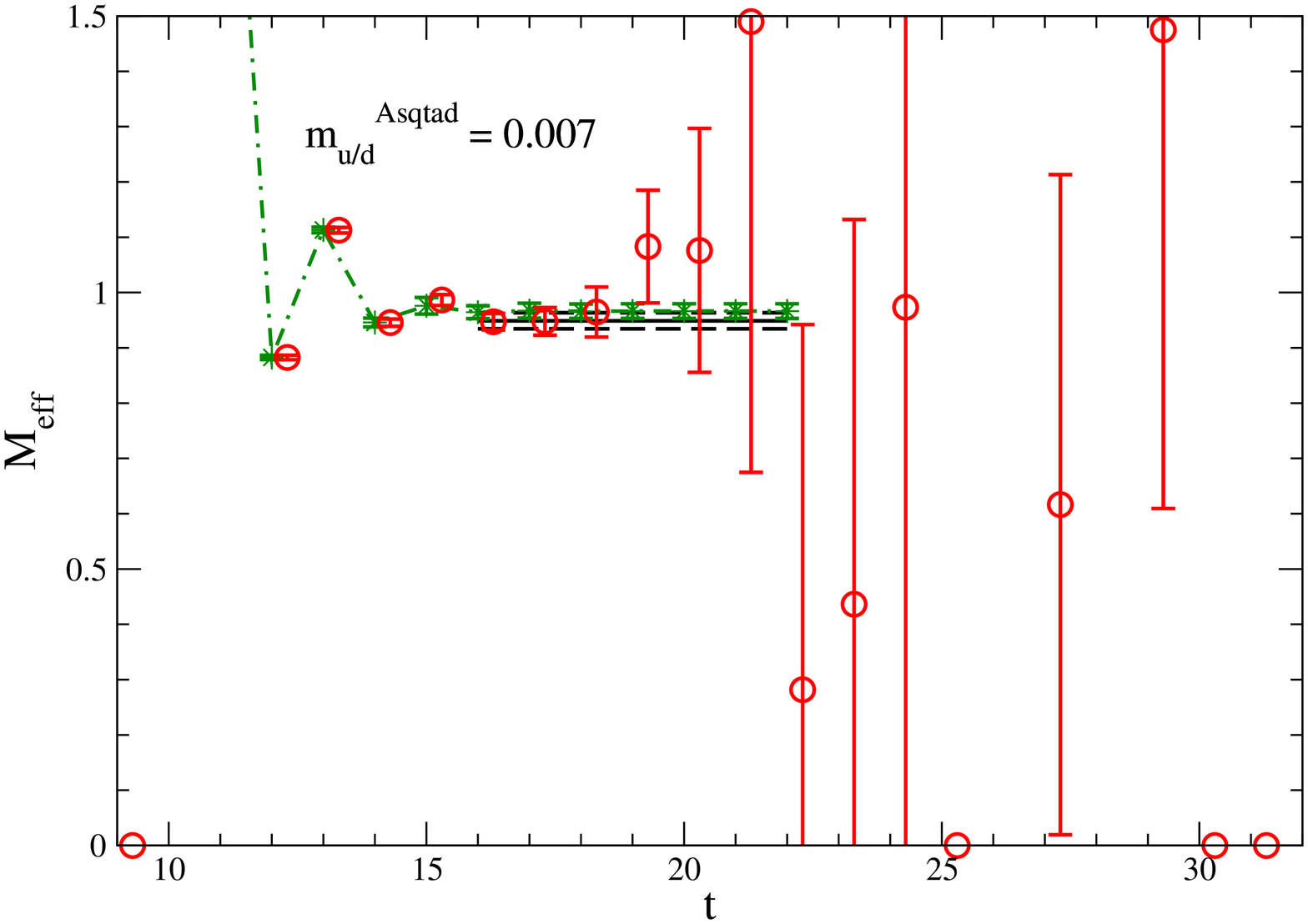}
\\
$(a)$ & $(b)$
\end{tabular}
\caption{The left- and right-hand figures show the delta effective
  mass at $\mud = 0.05$ and $\mud =0.007$, respectively, from the smeared-smeared correlators.  The circles
  denote the lattice data for the effective mass, and the solid line and
  dashed-dot line denote the single-exponential, and oscillating fit,
  respectively; note that lattice data (the circles) are slightly offset for clarity.
  \label{fig:Dosc}}
\end{figure}

We begin by a comparison of the two fitting procedures described in
Section~\ref{sec:fit}: ``oscillating fits'' of the form of
Eq.~\eqref{eq:osc}, and single-exponential fits with errors
derived from the bootstrap distribution.  In Figures~\ref{fig:Nosc} and
\ref{fig:Dosc} we show the effective masses for the nucleon and delta,
respectively, at the SU(3)-symmetric quark masses, and at $\mud =
0.007$.  Also shown are the effective masses obtained from the fits
using our two chosen procedures.  There is a notable consistency between
the two procedures.  Whereas we will employ the single-exponential
procedure in most of the remainder of the discussion, we show the masses
obtained using fit form Eq.~\eqref{eq:osc} and use the oscillating fit in analyzing the volume dependence of the delta mass in Fig.~\ref{fig:MDelta_20vs28}.
\begin{table}[!t]
\begin{center}
\caption{The upper and lower tables show ground-state masses in the
  octet and decuplet sectors, respectively, obtained from fits of form
  Eq.~\eqref{eq:osc}, as described in the text.\label{tab:osc}}
\begin{tabular}{c|cccccc}
& \multicolumn{6}{c}{$\mud$} \\
Channel & 0.007 & 0.010 & 0.020 & 0.030 & 0.040 & 0.050 \\
\hline
 $N$ &  0.696(7) &  0.726(5) &  0.810(5) &  0.878(5) &  0.941(6) &  0.991(5) \\
 $\Sigma$ &  0.837(4) &  0.850(4) &  0.886(4) &  0.925(5) &  0.963(6) &  0.991(5) \\
 $\Xi$ &  0.891(2) &  0.902(2) &  0.924(3) &  0.951(4) &  0.976(6) &  0.991(5) \\
 $\Lambda$ &  0.784(4) &  0.804(3) &  0.861(4) &  0.911(4) &  0.957(6) &  0.991(5) \\
\hline
\end{tabular}\\[1.0ex]

\begin{tabular}{c|cccccc}
& \multicolumn{6}{c}{$\mud$} \\
Channel & 0.007 & 0.010 & 0.020 & 0.030 & 0.040 & 0.050 \\
\hline
 $\Delta$ &  0.966(13) &  0.974(9) &  1.005(12) &  1.056(11) &  1.076(14) &  1.123(8) \\
 $\Sigma$ &  0.998(9) &  1.026(6) &  1.043(9) &  1.080(10) &  1.090(13) &  1.123(8) \\
 $\Xi$ &  1.053(5) &  1.073(4) &  1.081(7) &  1.104(8) &  1.104(12) &  1.123(8) \\
$\Omega$ &  1.103(3) &  1.117(3) &  1.117(5) &  1.127(8) \
&  1.115(11) &  1.121(9) \\
\hline
\end{tabular}
\end{center}
\end{table}

\subsubsection{Single-exponential fits}

\begin{figure}
\begin{tabular}{cc}
\vspace{1cm}
\includegraphics[width=0.45\textwidth]{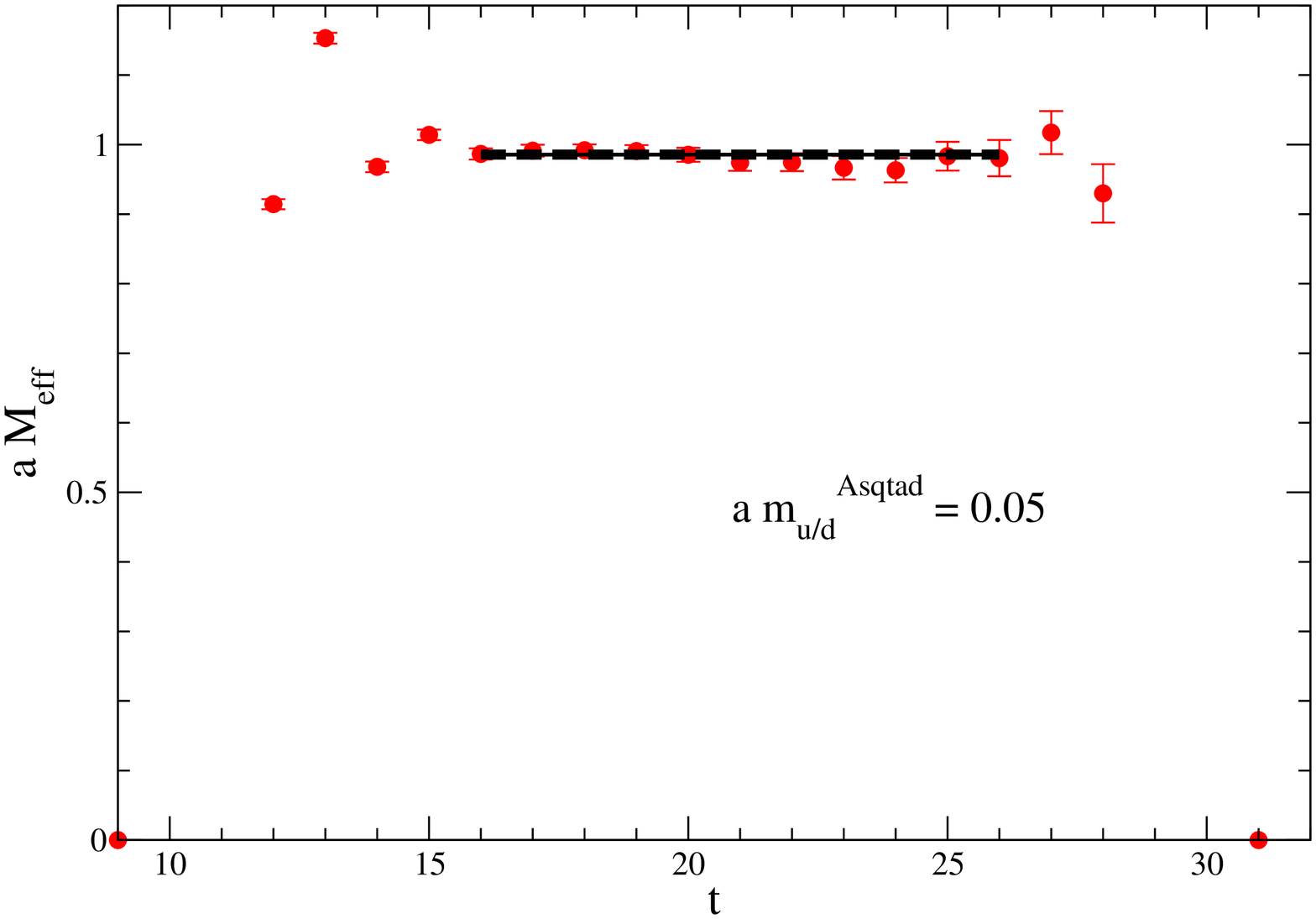}\hspace{1cm}
&\includegraphics[width=0.45\textwidth]{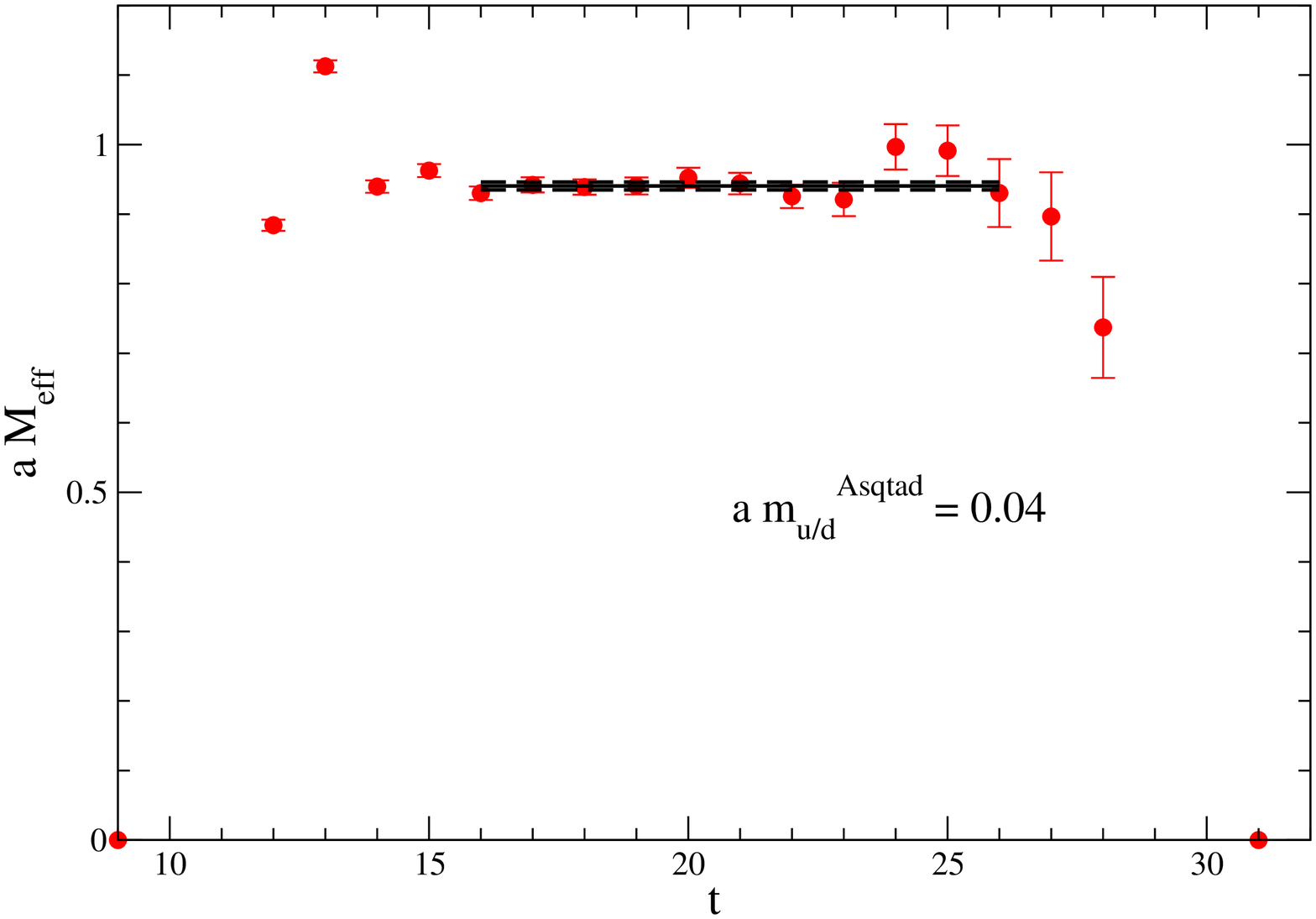}
\\ \vspace{1cm}
\includegraphics[width=0.45\textwidth]{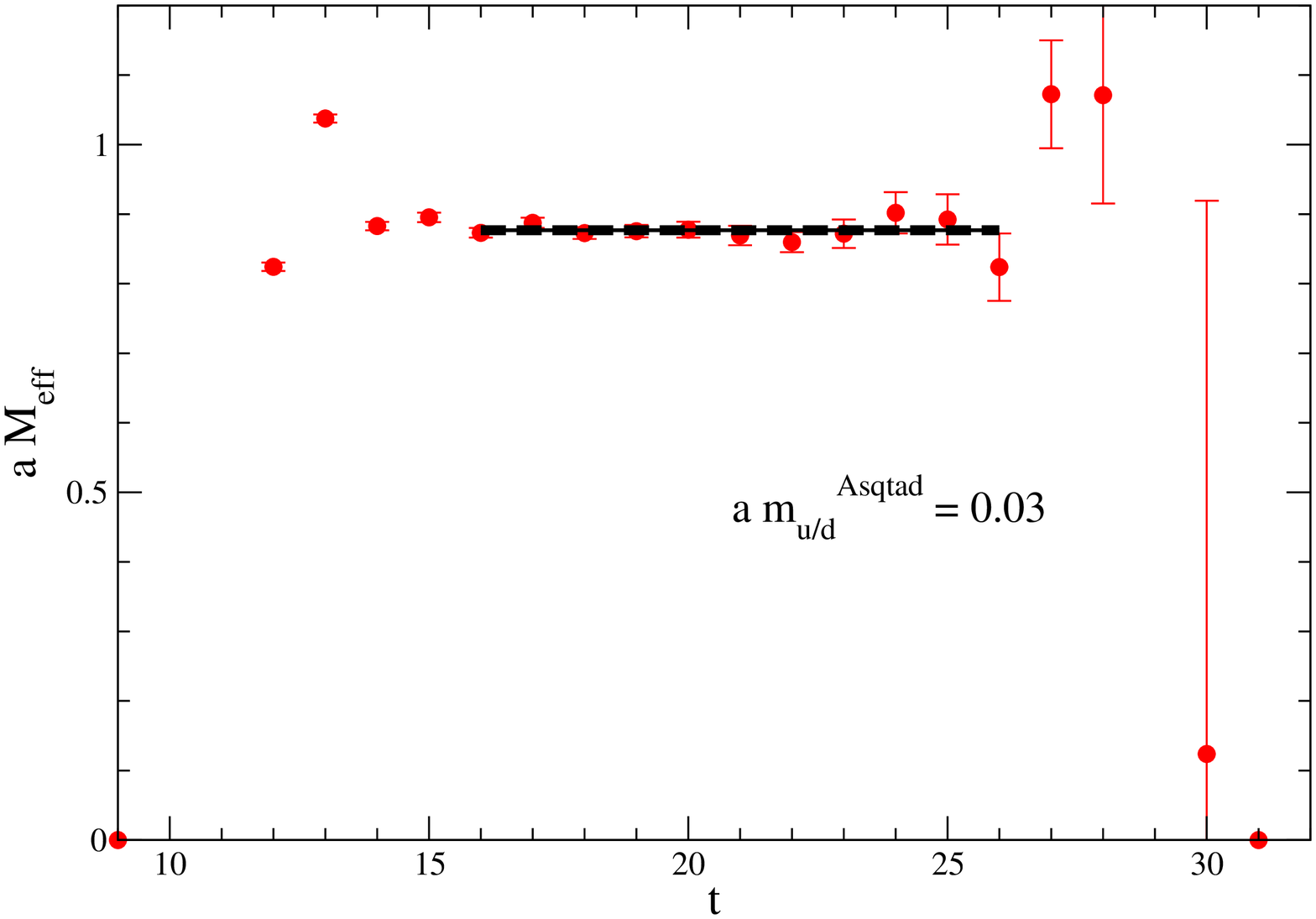}\hspace{1cm}
&\includegraphics[width=0.45\textwidth]{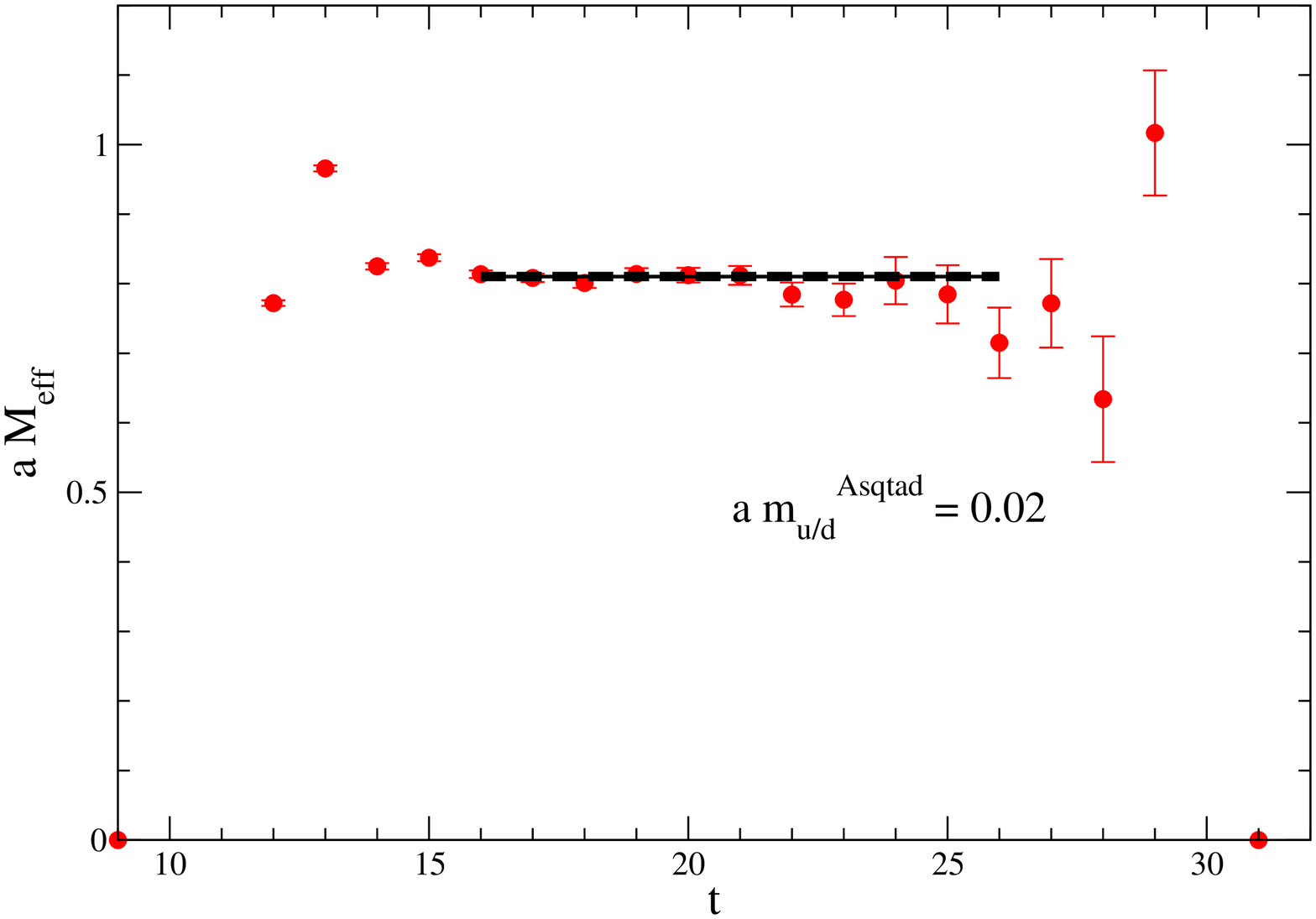}
\\ \vspace{1cm}
\includegraphics[width=0.45\textwidth]{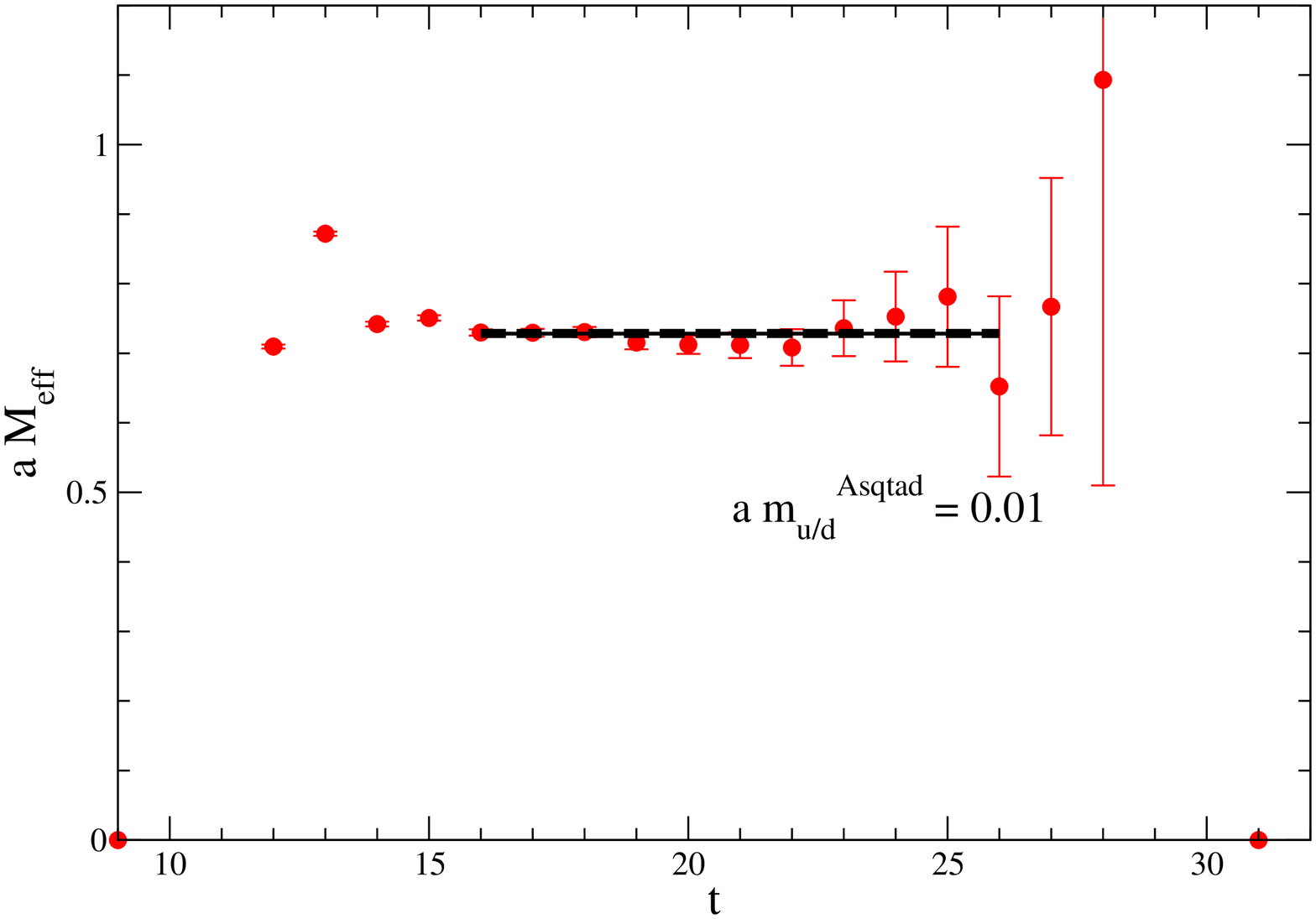}\hspace{1cm}
&\includegraphics[width=0.45\textwidth]{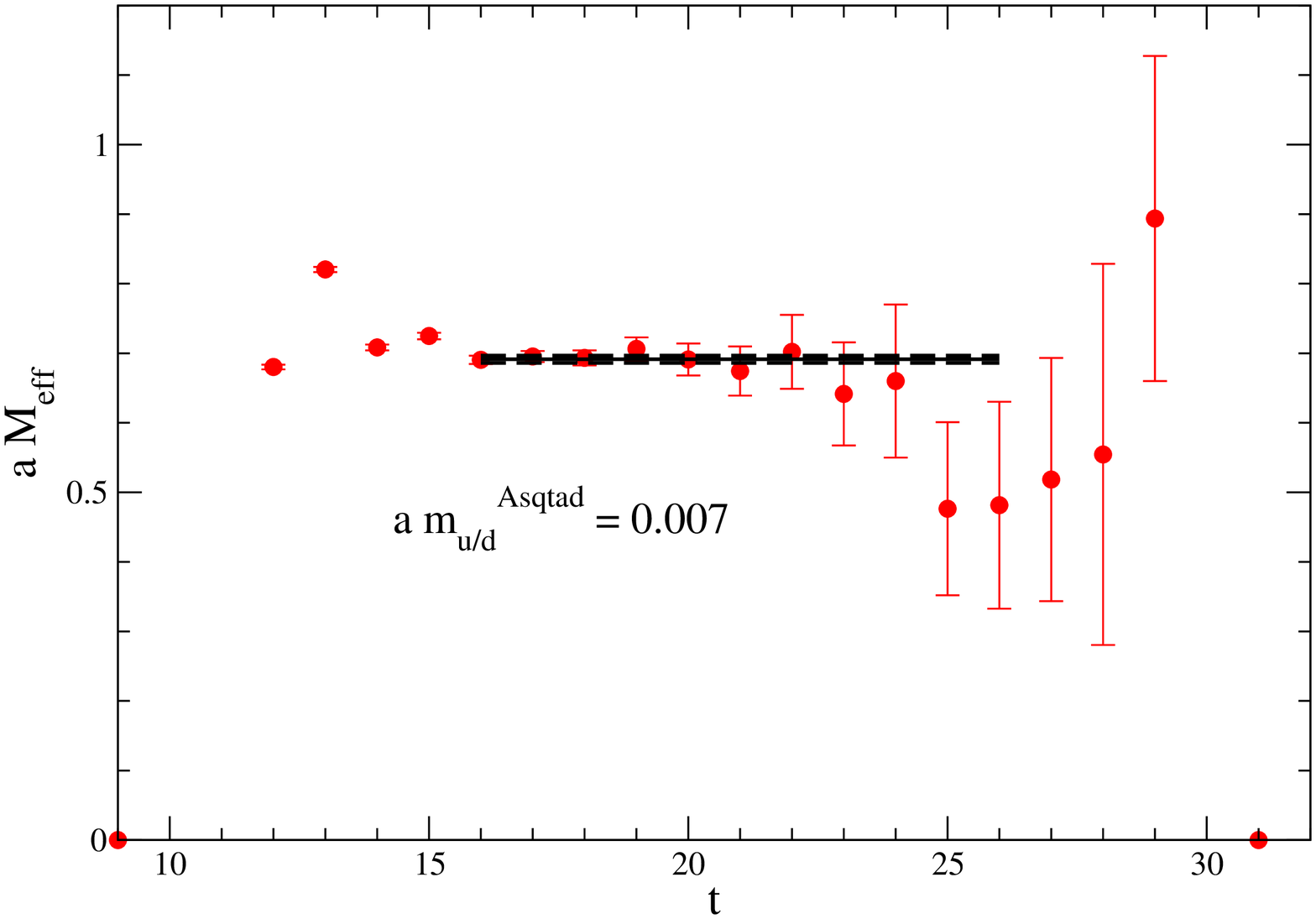}
\end{tabular}
\caption{The figure shows the nucleon effective mass obtained from a
  single correlator smeared at both source and sink at each of the
  values of the light-quark masses used in the calculation; the lines
  show single-exponential fits to the correlator, with the dashed
  lines bootstrap errors corresponding to the 68\% confidence levels
  on the bootstrap distributions.\label{fig:Neff}}
\end{figure}

\begin{figure}
\begin{tabular}{cc}
\vspace{1cm}
\includegraphics[width=0.45\textwidth]{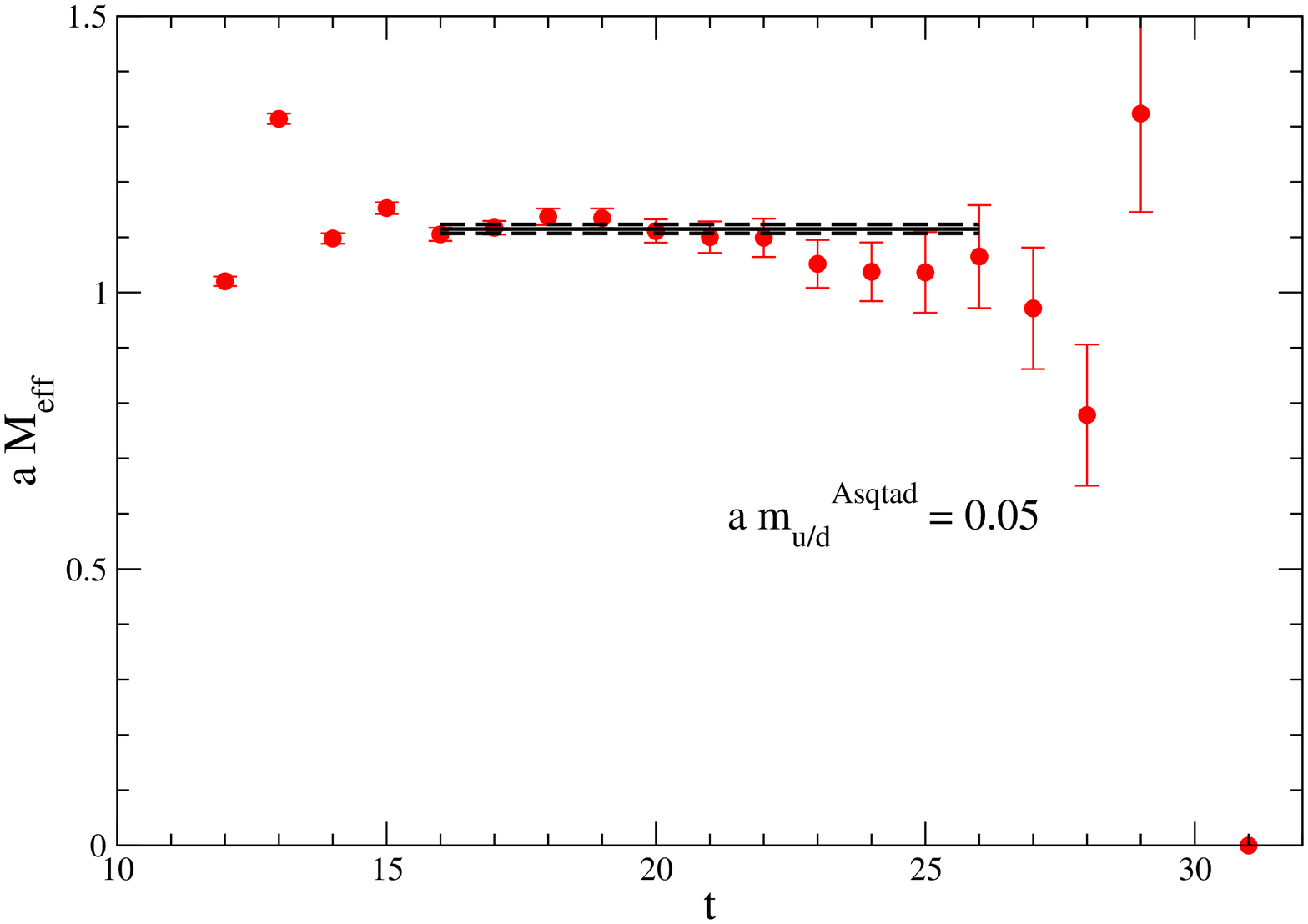}\hspace{1cm}
&\includegraphics[width=0.45\textwidth]{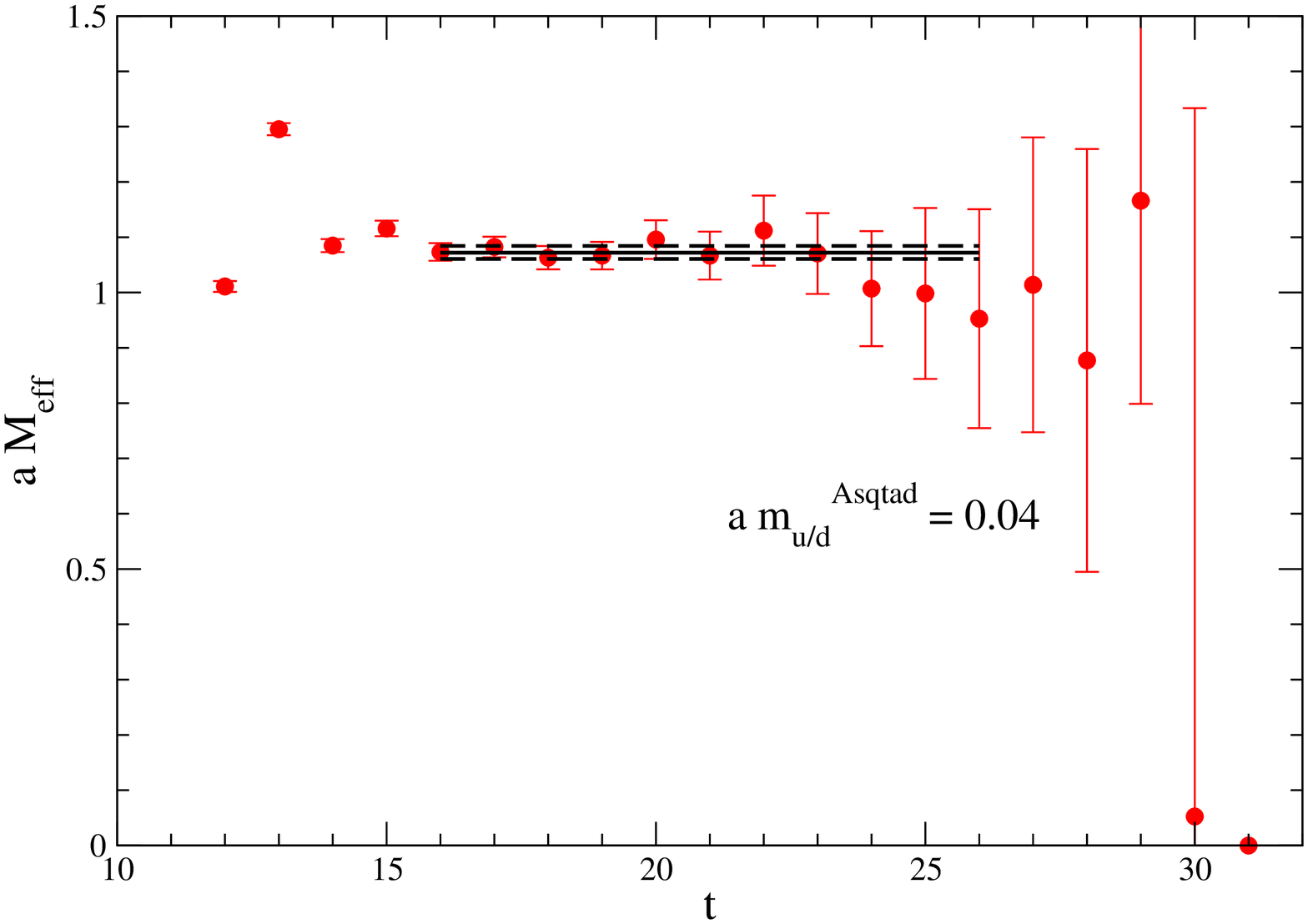}\\
\\ \vspace{1cm}
\includegraphics[width=0.45\textwidth]{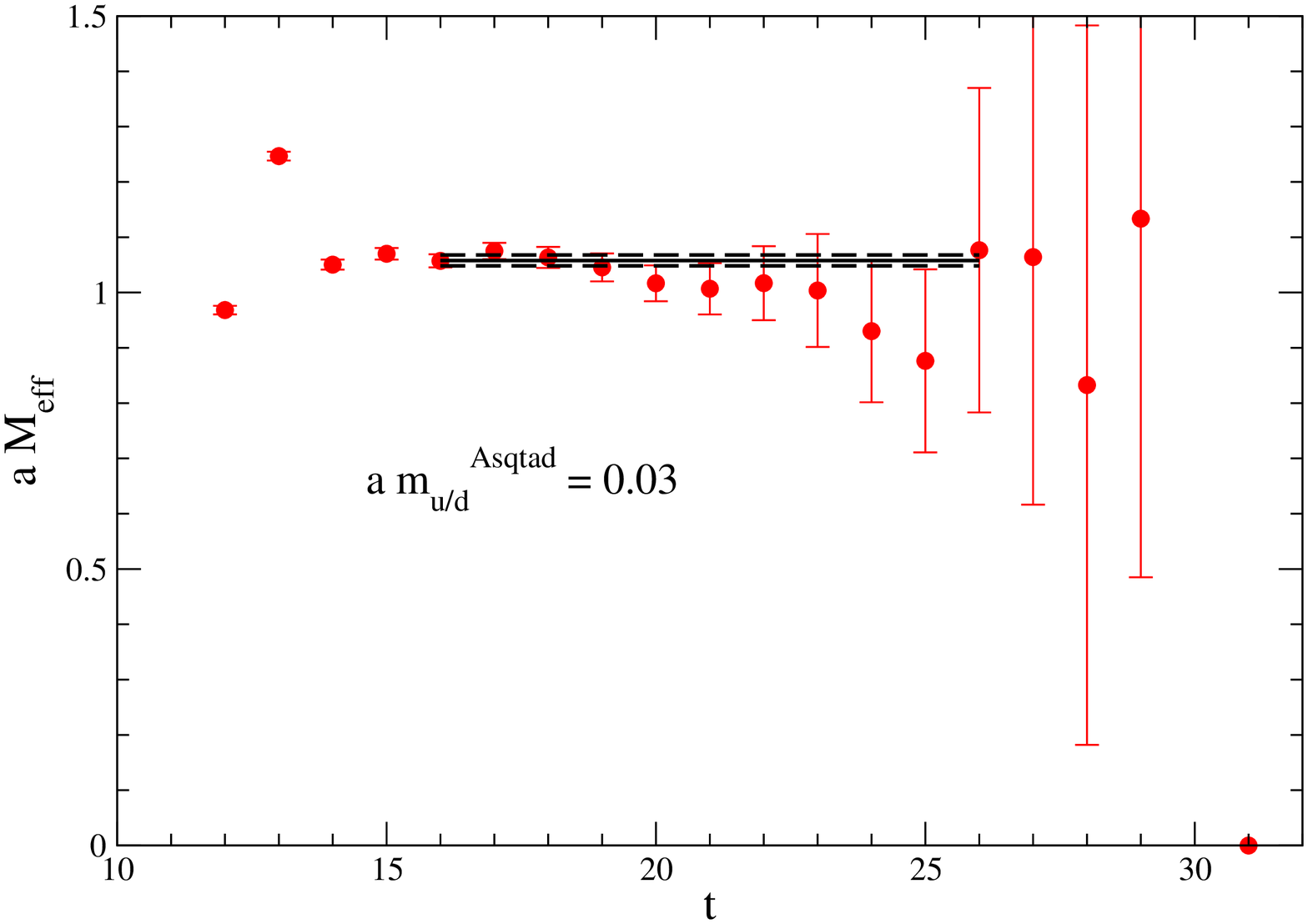}\hspace{1cm}
&\includegraphics[width=0.45\textwidth]{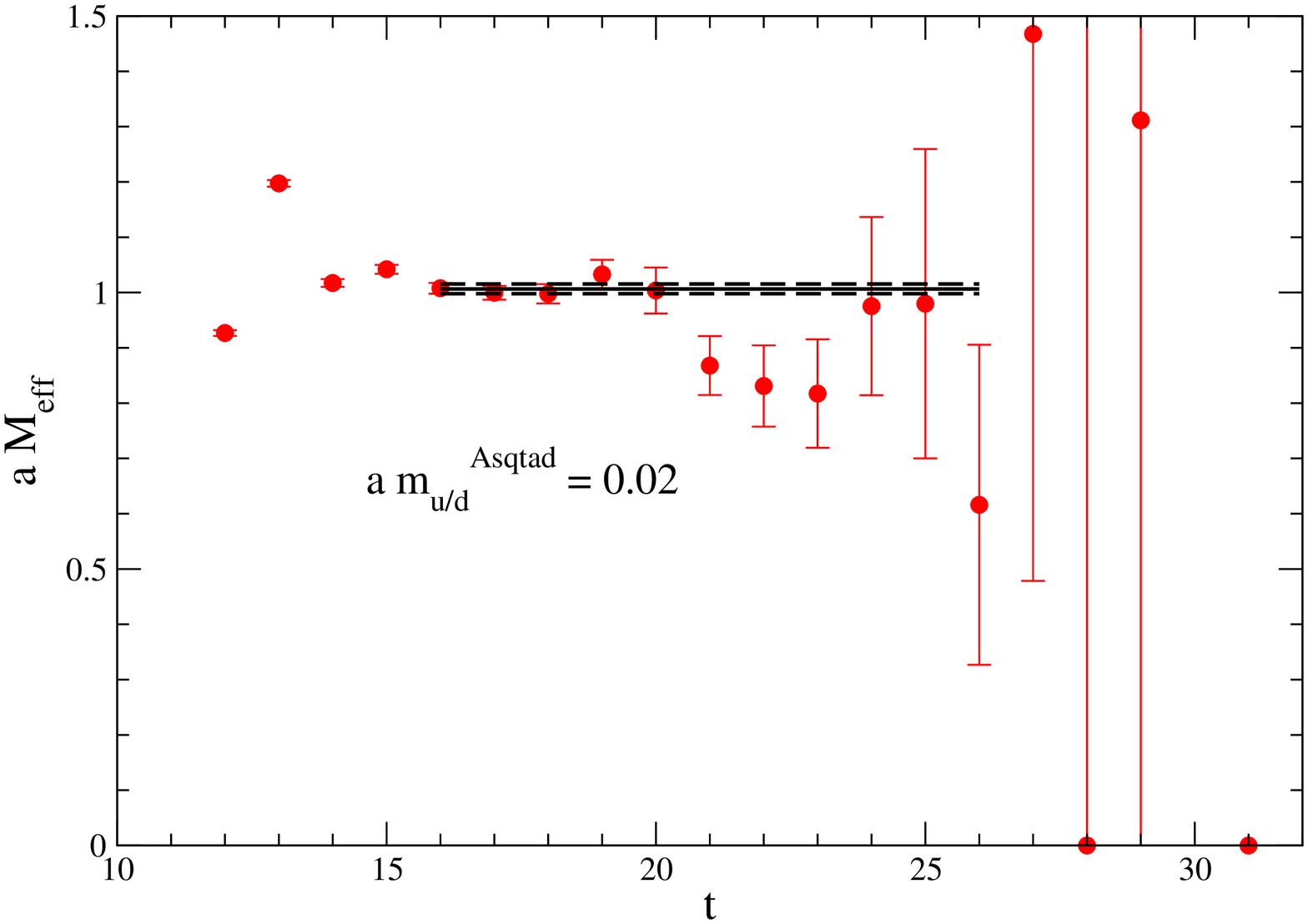}\\
\\ \vspace{1cm}
\includegraphics[width=0.45\textwidth]{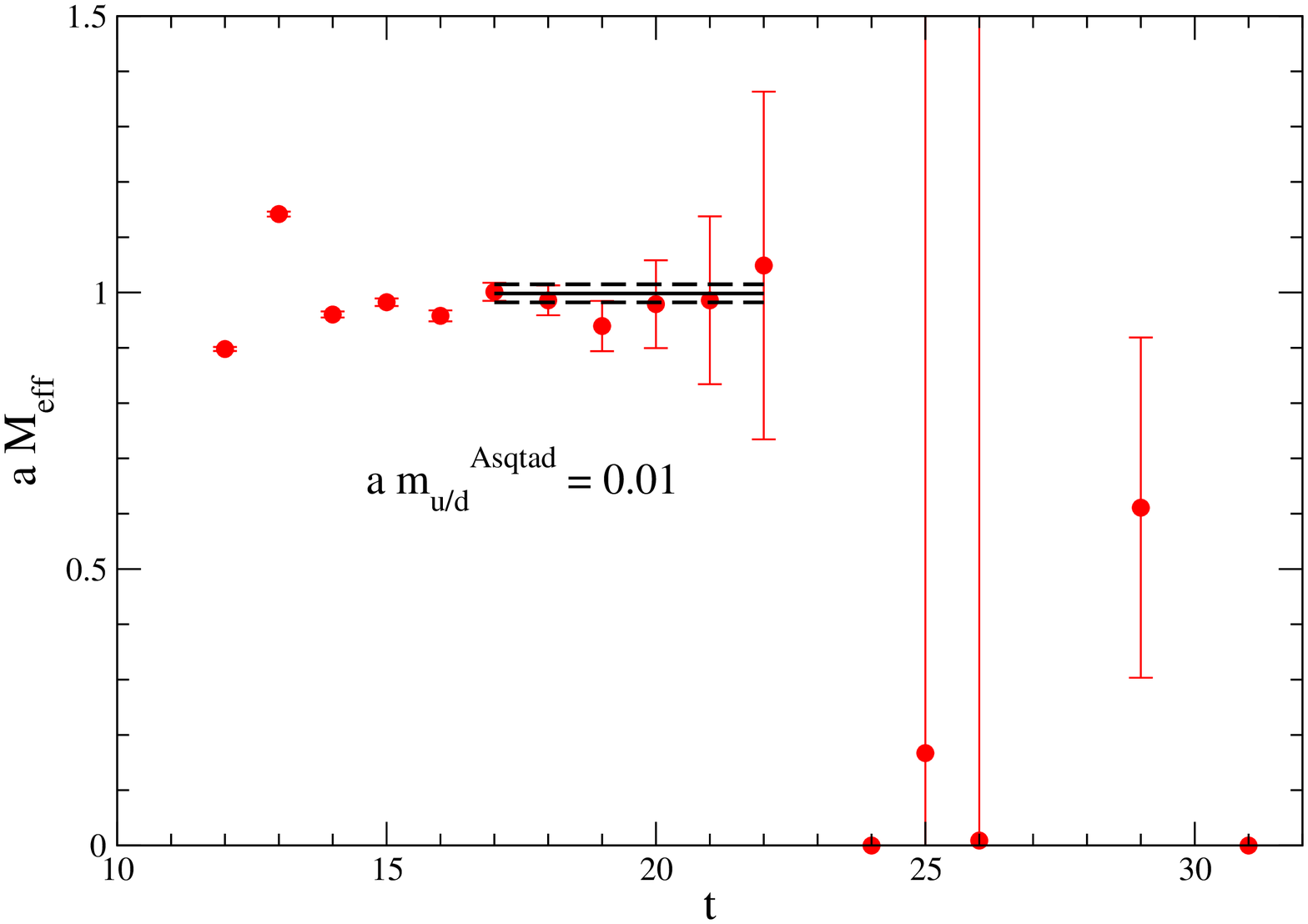}\hspace{1cm}
&\includegraphics[width=0.45\textwidth]{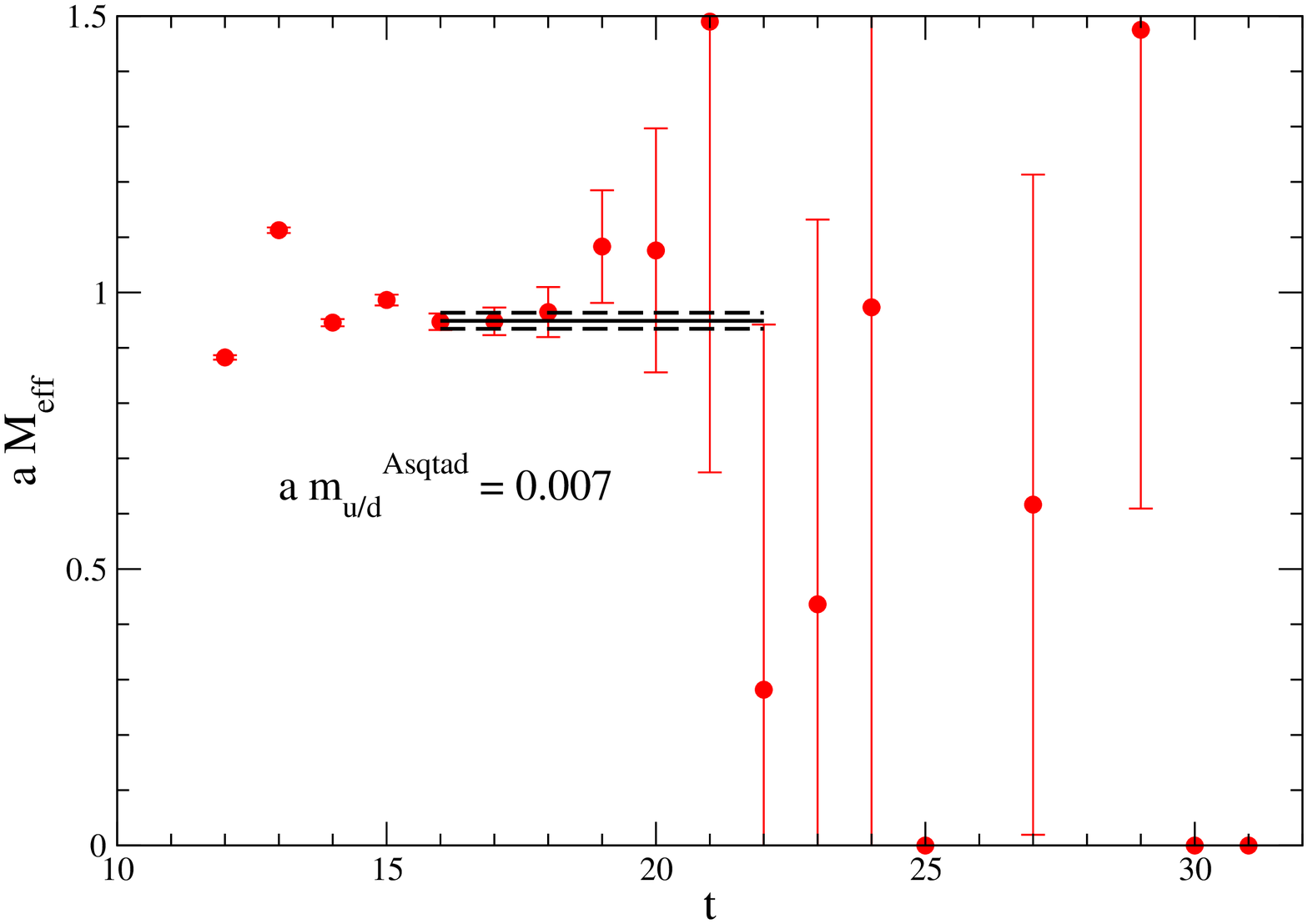}\\
\end{tabular}
\caption{The figure shows the $\Delta$ effective mass obtained from a
  single correlator smeared at both source and sink at each of the
  values of the light-quark masses used in the calculation.  The lines
  show single-exponential fits to the correlator, with the dashed
  lines bootstrap errors corresponding to the 68\% confidence levels
  on the bootstrap distributions.\label{fig:Deff}}
\end{figure}

Our final computations of the masses are obtained from the
single-exponential fits to correlators smeared at both the source and
at the sink, using for each flavor sector the operator that yields the
smallest statistical uncertainty in the correlator.  In order to
illustrate the quality of the data on each ensemble, we show in
Figures~\ref{fig:Neff} and \ref{fig:Deff} the effective mass plots in the
nucleon and delta channels for each of our $u/d$
masses.  The quality of the data degrades appreciably with decreasing
light-quark mass, despite the increasing ensemble size, most notably
in the case of the delta mass.
One minor exception to the consistency between ``oscillating fits" and ``single exponential fits" occurs for the case of the delta effective mass at $\mud = 0.01$, where, as seen in Fig.~\ref{fig:Deff},  the ``single exponential fit'' is visibly higher than the ``oscillating fit''.  However, as seen in Tables~\ref{tab:osc} and \ref{tab:decupletfit}, the masses agree within errors, and the case of the delta mass is addressed in more detail in connection with Fig.\ref{fig:MDelta_20vs28}.

Our results for the ground-state
masses for the octet and decuplet baryons are listed in
Tables~\ref{tab:octetfit} and \ref{tab:decupletfit}, respectively.
The fitting ranges
are chosen so as to provide an acceptable $\chi^2/{\rm dof}$ and quality of
fit.  The central
values for the quoted masses are obtained from a fit to the complete
ensemble.  The error bars are obtained from the 68\% confidence level
in the bootstrap distributions of the masses, using 10,000 bootstrap
sub-ensembles.
\begin{table}[!tbh]
\begin{center}
\caption{The table lists the masses of the lowest-lying,
  positive-parity spin-$1/2$ octet baryons at each value of the light
  $(u/d)$ quark mass, together with the time range used in the fit.  The quoted
  errors are obtained from the 68\% confidence level using 10,000
  bootstrap sub-samples.\label{tab:octetfit}}
\begin{tabular}{r|rl|rl|rl|rl|rl|rl}
& \multicolumn{12}{c}{$\mud$} \\
Channel & \multicolumn{2}{c}{0.007} &
\multicolumn{2}{c}{ 0.010} & \multicolumn{2}{c}{0.020} & 
\multicolumn{2}{c}{ 0.030} &
\multicolumn{2}{c}{0.040} & \multicolumn{2}{c}{0.050}\\ \hline 
$N$ & 16-26 & $0.691\err{5}{5}$ & 16-26 &
0.728\err{4}{4} & 16-26 & $0.810\err{4}{4}$ & 16-26 &$0.877\err{4}{4}$
& 17-26 &$0.945\err{7}{7}$ & 16-26 & $0.986\err{5}{4}$ \\ $\Lambda$ &
16-26&$0.780 \err{3}{3}$ & 16-26 & 0.805\err{3}{3} & 16-25 &
$0.861\err{3}{3}$ & 18-26 & $0.906\err{5}{5}$ & 17-26 &
$0.961\err{6}{6}$ & & \\ $\Sigma$ & 16-26 & $0.831\err{4}{4}$ & 16-26
& 0.849\err{3}{3} & 16-26 & $0.886\err{4}{4}$ &18-26 &
$0.920\err{6}{6}$ & 17-26 & $0.966\err{6}{6}$ & & \\ $\Xi$ & 16-26 &
$0.888\err{2}{2}$ & 18-26 & 0.901\err{2}{2} & 16-25 &
$0.924\err{3}{3}$ & 18-26 & $0.946\err{5}{5}$& 17-26 &
$0.979\err{6}{6}$ & & \\ \hline
\end{tabular}
\end{center}
\end{table}
\begin{table}[!th]
\begin{center}
\caption{The table lists the masses of the lowest-lying,
  positive-parity spin-$3/2$ decuplet baryons at each value
  of the light $(u/d)$ quark mass, together with the time range used
  in the fit.  The quoted errors are obtained from the 68\% confidence
  level using 10,000 bootstrap sub-samples.\label{tab:decupletfit}}
\begin{tabular}{r|rl|rl|rl|rl|rl|rl}
& \multicolumn{12}{c}{$\mud$} \\
Channel & \multicolumn{2}{c}{0.007} &
\multicolumn{2}{c}{ 0.010} & \multicolumn{2}{c}{0.020} & 
\multicolumn{2}{c}{ 0.030} &
\multicolumn{2}{c}{0.040} & \multicolumn{2}{c}{0.050}\\ \hline 
$\Delta$ & 16-22 & $0.948\err{15}{14}$
 & 17-22 & 0.998\err{16}{16} & 16-26 & $1.007\err{9}{9}$& 16-26 &
$1.058\err{10}{10}$& 16-26 & $1.072\err{12}{11}$ & 16-26 & $1.115\err{8}{8}$\\
$\Sigma^*$ & 16-26 & $0.991\err{7}{7}$ & 17-26 & 1.035\err{8}{8} & 16-26 
& $1.044\err{7}{7} $&16-26 & $1.081\err{9}{9}$ & 
16-26 & $1.085\err{11}{11}$ & & \\
$\Xi^*$ & 16-26 & $1.047\err{4}{4}$& 17-26 & 1.069\err{6}{6} & 16-26 &
$1.082\err{6}{6}$ & 16-26 & $1.104\err{8}{8}$ & 16-26 & 
$1.099\err{11}{10}$ & & \\
$\Omega$ &16-26 & $1.101\err{3}{3}$ & 18-26 & $1.113\err{4}{4}^*$ & 
16-26 & 
$1.119\err{5}{5}$ & 18-26 & $1.118\err{10}{10}$& 16-26 & 
$1.112\err{10}{9}$ & & \\
\end{tabular}
\end{center}
\end{table}

%
%
\section{Baryon Mass Chiral Extrapolations \label{sec:ChExtrap}}

In this section, we perform chiral extrapolations of the ground-state octet and decuplet baryon masses. %
%
We begin with a brief summary of our main results before presenting the details of our chiral extrapolation analysis in Secs.~\ref{sec:SU2extrap} and \ref{sec:SU3extrap}.  
We use a generalization of heavy baryon chiral perturbation theory (HB$\chi$PT)~\cite{Jenkins:1990jv,Jenkins:1991ne,Jenkins:1992ts} that includes the relevant lattice spacing effects for this mixed action lattice calculation~\cite{Tiburzi:2005is}.  We also perform a two-flavor extrapolation of the nucleon and delta masses, using continuum HB$\chi$PT.  At $\mc{O}(m_\pi^4)$, even the continuum two-flavor formulae have too many unknown low energy constants (LECs) to be determined from our lattice results alone.  Therefore we must input various values to the formulae, including the nucleon, delta and nucleon-delta axial couplings, denoted in this work as $g_A$, $g_{\D\D}$ and $g_{\D N}$ respectively.  At this order, the two-flavor mixed action formula for the nucleon mass has an additional three unknown parameters, not including those from lattice spacing effects~\cite{Tiburzi:2005na,Chen:2007ug}, making the fit impossible with our results, even with the input of the LECs as in the continuum fit.  Additionally, we only have numerical results at one lattice spacing, and therefore we can not control the continuum extrapolation.  Despite these limitations, as explained below, we find at a pion mass of $m_\pi=137$~\texttt{MeV} (and $m_K = 497.6$~\texttt{MeV}),
\begin{align}
	M_N &= 954 \pm 42 \pm 20 \textrm{ MeV} \qquad \textrm{from NNLO $SU(2)$ HB$\chi$PT}
\nonumber\\
	&= 960 \pm 24 \pm 8 \textrm{ MeV} \qquad \textrm{from NLO $SU(6|3)$ mixed action HB$\chi$PT}\, .
\end{align}
where the first uncertainty is statistical and the second uncertainty is an estimate of the systematic error in the extrapolation.  While  our extrapolated nucleon mass agrees with the physical value, using either the two-flavor continuum formula or the mixed action three-flavor formula, we must add a  cautionary note.  In the case of the three-flavor extrapolations, the values of the octet axial couplings, $D$ and $F$, as well as the value of the octet-decuplet axial coupling, $C$, which we determine in the mass fits, are significantly different from the known phenomenological values fit with $SU(3)$ HB$\chi$PT~\cite{Jenkins:1991es,Butler:1992pn,Savage:1996zd,FloresMendieta:1998ii,Cabibbo:2003cu,Ratcliffe:2004jt} and they are also inconsistent with the more recent lattice calculation of the octet axial charges~\cite{Lin:2007ap}.  The discrepancy with respect to the phenomenologically determined values of $D$, $F$ and $C$ is independent of whether one uses the continuum $SU(3)$ HB$\chi$PT formulae, or some generalization thereof,%
\footnote{These include partially quenched~\cite{Chen:2001yi} or mixed action~\cite{Tiburzi:2005is} HB$\chi$PT, whether or not the decuplet degrees of freedom are explicitly included in the theory, and the so called ``covariant" baryon $\chi$PT~\cite{Becher:1999he} as well as a different regularization scheme such as finite range regularization~\cite{Donoghue:1998rp,Leinweber:1999ig,Young:2002ib,Leinweber:2003dg}.} 
raising questions about the convergence of the three flavor heavy baryon theory.  

A salient result of our calculations is the observation that, without any known theoretical explanation, the octet baryon masses are surprisingly linear in $m_\pi$ over the range of pion masses studied in this work. 
For example, using the fit ansatz,%
\footnote{Plotting our resulting pion masses (squared) versus the input domain-wall quark masses, one can verify the expected scaling, $m_\pi^2 \sim m_q$.} 
\begin{align}\label{eq:MNstraightmpi}
	M_N &= \a_0^N + \a_1^N\, m_\pi\, ,
	\nonumber\\&
	\sim \a_0^N +\tilde{\a}_1^N \sqrt{m_q}\, ,
\end{align}
and fitting to the lightest five of our nucleon mass results, we find
\begin{equation}
	M_N = 938\pm 9 \textrm{ MeV}\, .
\end{equation}
Furthermore, examining the nucleon mass calculations of several other lattice groups, we find that this is not unique to our results, and in fact this trend of the nucleon mass data is common to most other lattice calculations, employing a variety of different lattice actions, with different lattice spacings, some with three light dynamical quarks and some with two, and all with relatively light pion masses.  

Before presenting the details of our analysis, we first summarize the mixed action extension of HB$\chi$PT we use to perform the chiral extrapolations as well as discuss our general fitting procedure.  Fitting is an issue with the three-flavor extrapolations because our analysis involves fits with up to 7 unknown parameters that are highly correlated.

\bigskip
In heavy baryon $\chi$PT~\cite{Jenkins:1990jv,Jenkins:1991ne}, the octet baryon masses have a quark mass expansion given by
\begin{equation}\label{eq:MBexpansion}
	M_B = M_0 + \d M_B^{(1)} + \d M_B^{(3/2)} + \d M_B^{(2)} + \dots
\end{equation}
where $M_0$ is the quark mass independent contribution to the baryon masses in the chiral limit.  The corrections, $\d M_B^{(n)}$, denote corrections to the mass of the baryon $B$, which scale as $m_q^n$.
The inclusion of the spin-$3/2$ states (resonances) in the theory somewhat complicates the expansion, since their inclusion introduces a new scale into the theory, the decuplet-octet (delta-nucleon) mass splitting in the chiral limit, which is a chiral singlet and leads to a renormalization of all parameters in the Lagrangian~\cite{WalkerLoud:2004hf,Tiburzi:2004rh,Tiburzi:2005na}.  With these states included, one typically adopts a power counting~\cite{Jenkins:1990jv,Jenkins:1991ne,Hemmert:1997ye} that includes the mass splitting
\begin{equation}
\Delta \equiv M_\Delta - M_N \sim m_\pi\, ,
\end{equation}
which is certainly relevant for our lattice calculations as well as the physical point.  The decuplet baryon masses then have a similar expansion given by
\begin{equation}\label{eq:MTexpansion}
	M_T = M_0 +\Delta_0 + \d M_T^{(1)} + \d M_T^{(3/2)} + \d M_T^{(2)} + \dots
\end{equation}
with $\Delta_0 = \Delta |_{m_q=0}$.  At finite lattice spacing, these extrapolation formulae are modified in a known way.  The two-flavor mixed action extrapolation formula for the nucleon has been determined to next-to-leading order (NLO) in Ref.~\cite{Tiburzi:2005is}.  To determine the formulae for the delta masses as well as the rest of the members of the octet and decuplet baryons, one can either extend the work of Ref.~\cite{Tiburzi:2005is} or use the prescription in Ref.~\cite{Chen:2007ug} to convert the known partially quenched formulae~\cite{Chen:2001yi,Beane:2002vq,WalkerLoud:2004hf,Tiburzi:2004rh,Tiburzi:2005na} to the relevant mixed action formulae.  At NLO, one needs knowledge of the staggered taste-identity meson mass splitting~\cite{Aubin:2004fs} as well as the mixed meson mass splitting, which has been determined in Ref.~\cite{Orginos:2007tw}.  Furthermore, supplemented by a treatment of the one loop flavor disconnected diagrams to the baryon masses~\cite{WalkerLoud:2006sa}, one can determine the mixed action extrapolation formula to next-to-next-to-leading order (NNLO), $\mc{O}(m_\pi^4)\sim \mc{O}(a^2 m_\pi^2)$, in the mixed action effective theory.  Here and throughout, we use a shorthand for the dimensionless expansion parameter, $\varepsilon^2 \sim a^2 \L_\chi^2 \sim m_\pi^2 / \L_\chi^2$, in which we drop the relevant $\L_\chi$'s.

%
%
\subsection{Fitting method for global chiral extrapolations}

When performing the chiral extrapolations, in addition to a standard $\chi^2$ minimization, we perform, for lack of a better name, a \textit{bootstrap chiral extrapolation}.  As discussed above, for a given baryon mass, there are too many LECs in the NLO chiral extrapolation formulae to be determined from our lattice results alone.  However, by considering a global $\chi^2$ minimization of all four non-degenerate octet baryon masses (or the four non-degenerate decuplet baryon masses), we can perform a fit to all seven (five) LECs that appear in the three flavor extrapolation formulae for the octet (decuplet) masses at NLO.  In principle, we can even perform the NNLO fits that have 18 (14) LECs for the octet (decuplet) masses, although we find these fits generally do not converge and are not stable to changes in the fit ranges and the values of possibly fixed parameters.  When performing the global $\chi^2$ minimization, there are two correlations that are important to include, one of which makes these fits quite challenging.  The first correlation is among the lattice results.  For a given quark mass ensemble, $m007$, $m010$, \textit{etc.}, the various baryon masses are all correlated. Therefore one either should construct an error correlation matrix that accounts for this, or use the \textit{jackknife} or \textit{bootstrap} analysis techniques to handle these correlations.  The second correlation is among the LECs being determined in the minimization, which is in fact the most challenging part of the analysis.  In terms of the re-scaled dimensionless error correlation matrix, many of the off-diagonal elements are close to unity.  Here we describe our bootstrap chiral extrapolation method that takes into account both of these correlations, and as far as we are aware, has not been implemented before.

We begin by constructing bootstrap lists for all the baryon masses, with an equal number of bootstrap samples on each of the different quark mass ensembles, $m007, m010, m020, m030, m040$ and $m050$.  For this work, we generated $N_{bs} = 10,000$ bootstrap samples at each quark mass.  With these bootstrap lists, we then construct a bootstrap list of the global $\chi^2$,
\begin{equation}\label{eq:bs_chi2}
\chi^2[bs] = \sum_{B} \sum_{m_q} \left( 
	\frac{m_B[bs, m_q] - g(m_B: f_\pi[m_q], m_\pi[m_q], m_K[m_q], \dots, \lambda_i )}{\s_{m_B}[m_q]} \right)^2\, ,
\end{equation}
where the sums run over $B = \{N,\L,\S,\Xi\}$ (and similarly for the decuplet), $m_q = \{0.007,0.010,0.020,0.030,\dots\}$, up to the heaviest mass set used in a given minimization and the $g(m_B: f_\pi[m_q], m_\pi[m_q], m_K[m_q], \dots, \lambda_i)$ are the various baryon mass, $m_B$, chiral extrapolation functions that depend upon the masses and decay constants as well as the LECs, $\l_i$.  In constructing the bootstrap lists of these global $\chi^2[bs]$, we fix the meson masses and decay constants,  $f_\pi[m_q]$, $m_\pi[m_q]$ and $m_K[m_q]$ to their central values, since their bootstrap fluctuations are an order of magnitude smaller than those of the baryon masses.  We also do ignore the bootstrap fluctuations of the error of a given baryon mass, $\s_{m_B}[m_q]$, taking their central values from tables \ref{tab:meson_mass}, \ref{tab:decayConst}, \ref{tab:octetfit} and \ref{tab:decupletfit} respectively.%
\footnote{The inclusion of these fluctuations amounts to an error on the error, which is beyond our consideration here.} 
To construct this list of global $\chi^2$ functions, as can be seen with Eq.~\eqref{eq:bs_chi2}, it is essential to have an equal number of bootstrap samples on the different quark mass ensembles, precluding the use of the jackknife method, at least with our sets of ensembles.  

The advantages of this method then follow naturally.  Because the baryon masses on the different quark mass ensembles are statistically independent, this amounts to adding independent noise, weighted by the statistical error $\s_{m_B}[m_q]$, to the central value of a given baryon mass on each of the different quark mass ensembles, mimicking a well known method of handling fits with highly correlated parameters.  We note that for our calculations, the bootstrap samples appear Gaussian distributed about the mean.  Furthermore,  the correlations among the different baryon masses on a given quark mass ensemble are automatically taken into account by use of the bootstrap distributions.  Therefore, by minimizing each of the $N_{bs}$ entries in the global $\chi^2[bs]$ list, we generate a bootstrap list of the fit parameters, or the determined LECs, $\{ \l_i[bs] \}$.%
\footnote{We could also generate the bootstrap list of the error correlation matrix, but this would also amount to an error on the error, so we only retain the central values of this matrix.} 
Using the bootstrap error analysis, we can then make predictions for the resulting LECs, as well as the extrapolated baryon masses, which accounts both for the correlations among the ensembles at a given quark mass as well the correlations among the LECs from the minimization procedure,
\begin{equation}
\l_i = \l_{i_0} \pm \s_{\l_i}, \qquad
	\l_{i_0} = \frac{1}{N_{bs}} \sum_{bs}^{N_{bs}} \l_i[bs], \qquad
	\s_{\l_i} = \sqrt{\frac{1}{N_{bs}-1}\sum_{bs}^{N_{bs}}\left( \l_i[bs] - \l_{i_0} \right)^2},
\end{equation}
and
\begin{align}
&m_B = m_{B_0} \pm \s_{m_B},& 
&m_{B_0} = 	g(m_B: f_\pi^{phys}, m_\pi^{phys}, m_K^{phys}, \dots, \l_{i_0}),&
\nonumber\\
&&&\s_{m_B} = \sqrt{\frac{1}{N_{bs}-1} \sum_{bs}^{N_{bs}} 
	\left( m_{B_0} - g(m_B: f_\pi^{phys}, m_\pi^{phys}, m_K^{phys}, \dots, \l_{i}[bs]) \right)^2}\, .&
\end{align}
To perform the minimization, we use both \texttt{Mathematica} and  \texttt{MINUIT}.

%
%
\subsection{Two flavor chiral extrapolations \label{sec:SU2extrap}}

In this section, we perform two-flavor chiral extrapolations of our nucleon and delta mass results.  
From the point of view of testing predictions from HB$\chi$PT, \textit{i.e.} looking for non-analytic chiral behavior, one would ultimately like to determine values of $g_A$, $g_{\D\D}$ and $g_{\D N}$ directly from the nucleon and delta mass extrapolations.  These LECs represent the leading order (LO) axial charges of the nucleon, delta and nucleon-delta transitions.  The leading virtual pion cloud contributions to the nucleon and delta masses are proportional to these couplings, contributing at NLO.  For example, the nucleon mass takes the following form at NLO
\begin{equation}\label{eq:MN_NLO}
M_N =\ M_0 -2\a_M(\mu)\, m_\pi^2 
	-\frac{3 \pi g_A^2}{(4\pi f_\pi)^2}m_\pi^3
	-\frac{8 g_{\D N}^2}{3(4\pi f_\pi)^2} \mc{F}(m_\pi,\D,\mu)\, ,
\end{equation}
with
\begin{equation}\label{eq:F}
\mc{F} (m,\Delta,\mu) = ( \Delta^2-m^2 +i\epsilon )^{3/2} 
	\ln \left( \frac{\Delta + \sqrt{\Delta^2 - m^2 + i \epsilon}}{\Delta - \sqrt{\Delta^2 - m^2 + i \epsilon}} \right)
	- \frac{3}{2}\Delta\, m^2 \ln \left( \frac{m^2}{\mu^2} \right)
	- \Delta^3 \ln \left( \frac{4 \Delta^2}{m^2} \right)\, .
\end{equation}
However, fits to our lattice results for the nucleon mass with this NLO formula return values of $g_A$ and $g_{\D N}$ that are inconsistent with our knowledge of these LECs from either phenomenology or from lattice QCD.%
\footnote{The use of the mixed action expression for the nucleon mass at this order~\cite{Tiburzi:2005is}, supplemented by the known valence-sea meson mass splitting~\cite{Orginos:2007tw} does not qualitatively change this conclusion.  See Table~\ref{tab:MN_SU2_LECs} for details.} 
In large part, this can be understood from the observation that our lattice data of the nucleon mass are well approximated by Eq.~\eqref{eq:MNstraightmpi}, $M_N = \a_0^N + \a_1^N m_\pi$ (see Table~\ref{tab:MN_SU2_LECs} and Figure~\ref{fig:MNExtrap} for more details).  Therefore, in order for the $SU(2)$ HB$\chi$PT nucleon mass expression to fit our lattice results, the different orders in the heavy baryon expansion of $M_N$, which is a polynomial series in $m_\pi$ beginning at $\mc{O}(m_\pi^2)$ and supplemented by chiral logarithms, must conspire to form this straight line.  Viewed in this light, it is not surprising that an NLO analysis fails to return values of $g_A$ and $g_{\D N}$ that are consistent with their known values.  Therefore, in accord with the expectations from Refs.~\cite{Beane:2004ks,McGovern:2006fm}, we must use the extrapolation formula to at least NNLO and ideally next-to-next-to-next-to-leading order (NNNLO) to test if the values have stabilized.  At this order, unfortunately, there are too many LECs in the formula to be determined from a fit to the lattice data points alone, a problem that is only exacerbated with the use of the mixed action formula.  Hence, we must resort to fixing the values of some of these parameters using results either from phenomenology or other lattice calculations, ideally determined with the same lattice action.

\subsubsection{The nucleon mass}

To study the chiral extrapolation of the nucleon mass, we perform fits to our lattice data with the following formulae;
\begin{itemize}
\item
LO: $SU(2)$ HB$\chi$PT, 
\begin{equation}\label{eq:MN_LO}
M_N = M_0 -2\a_M m_\pi^2\, ,
\end{equation}

\item
NLO: $SU(2)$ HB$\chi$PT, Eq.~\eqref{eq:MN_NLO}~\cite{Jenkins:1990jv,Jenkins:1991ne,Jenkins:1992ts}.

\item
NLO: $SU(4|2)$ mixed action HB$\chi$PT~\cite{Tiburzi:2005is} with
\begin{align}\label{eq:MN_NLO_MA}
M_N =&\ M_0 -2 \a_M(\mu)\, m_\pi^2
	-\frac{(g_A^2 - 4 g_A g_1 -5g_1^2)\pi}{3(4\pi f_\pi)^2}\, m_\pi^3
	-\frac{(8g_A^2 +4g_A g_1 +5g_1^2)\pi}{3(4\pi f_\pi)^2}\, \tilde{m}_{ju}^3
\nonumber\\&
	-\frac{4 g_{\D N}^2}{3(4\pi f_\pi)^2}\, \mc{F}(m_\pi,\D,\mu)
	-\frac{4 g_{\D N}^2}{3(4\pi f_\pi)^2}\, \mc{F}(\tilde{m}_{ju},\D,\mu)
	-\frac{3\pi(g_A+g_1)^2}{2(4\pi f_\pi)^2}\, m_\pi\, \tilde{\D}_{PQ}^2\, ,
\end{align}
where $g_1$ is an additional axial coupling appearing in the mixed action/partially quenched Lagrangian~\cite{Beane:2002vq}, $\tilde{\D}_{PQ}^2 = a^2 \D_\mathbf{I}$ is the taste-singlet staggered meson splitting, and $\tilde{m}_{ju}^2 = m_\pi^2 + a^2\D_\mathrm{Mix}$, with the mixed valence-sea meson mass splitting given in Ref.~\cite{Orginos:2007tw}.

\item
NNLO: $SU(2)$ HB$\chi$PT with explicit delta degrees of freedom (explicit deltas)~\cite{Tiburzi:2005na} (and a slightly modified renormalization prescription from Ref.~\cite{Tiburzi:2005na})%
\footnote{The coefficients $b_A(\mu)$ and $b_M(\mu)$ are not the renormalized coefficients as defined in Ref.~\cite{Tiburzi:2005na}.  There are additional operators which contribute to the nucleon mass at NNLO, notably the operators which are responsible for the LO contribution to the delta mass.  However, the contribution to the nucleon mass from these other operators is parametrically (in $m_\pi$) the same as those proportional to $b_A$ and $b_M$.  Since we are not doing a combined fit of the nucleon and delta mass, we have absorbed these effects with a re-definition of these coefficients.} 
\begin{align}\label{eq:MN_NNLO}
M_N =&\ M_0 -2\a_M(\mu) m_\pi^2 
	-\frac{3 \pi g_A^2}{(4\pi f_\pi)^2}m_\pi^3
	-\frac{8 g_{\D N}^2}{3(4\pi f_\pi)^2} \mc{F}(m_\pi,\D,\mu) 
	\nonumber\\&
	+m_\pi^4 
		\left[ b_M(\mu) +\frac{8 g_{\D N}^2 \a_M(\mu)}{(4\pi f_\pi)^2}- \frac{9g_{\D N}^2}{4M_0(4\pi f_\pi)^2}
		-\frac{45 g_A^2}{32M_0(4\pi f_\pi)^2} \right]
	+\frac{8 g_{\D N}^2 \a_M}{(4\pi f_\pi)^2}\, m_\pi^2 \mc{J}(m_\pi,\D,\mu)
	\nonumber\\&
	+\frac{m_\pi^4}{(4\pi f_\pi)^2}\, \ln \left( \frac{m_\pi^2}{\mu^2} \right)
		\left[ 6\a_M(\mu) - \frac{3 b_A(\mu)}{4\pi f_\pi} -\frac{27 g_A^2}{16 M_0} -\frac{5 g_{\D N}^2}{2M_0} \right]\, ,
\end{align}
with $\mc{F}$ given in Eq.~\eqref{eq:F}, and
\begin{equation}
\mc{J}(m,\Delta,\mu) = m^2 \ln \left( \frac{m^2}{\mu^2} \right)
	-2\D \sqrt{\D^2 -m^2 +i\e}\, \ln \left( \frac{\D + \sqrt{\D^2 -m^2 +i\e}}{\D -\sqrt{\D^2 -m^2 +i\e}} \right)
	+2\D^2 \ln \left( \frac{4\D^2}{m^2} \right)\, .
\end{equation}

\item
NNLO: $SU(2)$ HB$\chi$PT without explicit deltas, Eq.~\eqref{eq:MN_NNLO} with $g_{\D N}=0$.

\item
NNLO: $SU(2)$ covariant baryon $\chi$PT formula without explicit deltas, expanded to $\mc{O}(m_\pi^5)$~\cite{Procura:2003ig}
\begin{align}\label{eq:MN_NNNLO}
M_N =&\  M_0 - 4c_1 m_\pi^2 -\frac{3\pi g_A^2}{(4\pi f_\pi)^2}m_\pi^3
	+\frac{3\pi g_{A}^2}{8M_0^2 (4\pi f_\pi)^2}m_\pi^5
	\nonumber\\&
	+m_\pi^4 \left[ e_1(\mu) - \frac{3}{2(4\pi f_\pi)^2}\left( \frac{g_{A}^2}{M_0} - \frac{c_2}{2}\right)
	-\frac{3}{2(4\pi f_\pi)^2} \left(\frac{g_{A}^2}{M_0} - 8c_1 +c_2 +4c_3\right)
		\ln \left( \frac{m_\pi^2}{\mu^2} \right) \right]\, .
\end{align}

\item
$M_N = \a_0^N + \a_1^N m_\pi$: 
an empirical form motivated by the observed nucleon mass results, not motivated by any understanding of low energy QCD we currently have.

\end{itemize}

%
%
\begin{table}[t]
\caption{\label{tab:MN_SU2_LECs} Various two-flavor nucleon mass chiral extrapolation results.  In the left column, we provide the particular fit function as well as the values of LECs fixed in the minimization.  In the far right column, we provide the resulting nucleon mass at $m_\pi=137$~\texttt{MeV}.  In the fit parameters as well as the predicted nucleon masses, the first uncertainty is statistical.  In the NNLO fits, the second uncertainty is systematic determined by varying the fixed LECs over their given ranges.  The last fit function is motivated purely by the observed lattice results for the nucleon mass, and not by any understanding of low-energy QCD we currently have.  We have set the renormalization scale to $\mu=1$~GeV.}
\begin{ruledtabular}
\begin{tabular}{ccccccccc}
FIT: LO & range & $M_0 [\textrm{GeV}]$ & $\a_M [\textrm{GeV}^{-1}]$ & & & $\chi^2$ & d.o.f. & $M_N [\textrm{MeV}]$ \\
\hline
$M_N = M_0 - 2 \a_M m_\pi^2$ & 007-020 & 1.00(1) & -0.57(3) &&& 1.4 & 1 & $1028 \pm9$ \\
& 007-030 & 1.02(1) & -0.53(2) &&& 4.6 & 2 & $1037 \pm8$ \\
& 007--040 & 1.02(1) & -0.51(1) &&& 6.8 & 3 & $1043\pm7$ \\
& 007--050 & 1.04(1) & -0.47(1) &&& 21 & 4 & $1056 \pm6$ \\
\\
\hline
NLO $SU(2)$, Eq.~\eqref{eq:MN_NLO} & range &$M_0 [\textrm{GeV}]$ & $\a_M [\textrm{GeV}^{-1}]$ & $g_A$ & $g_{\D N}$ & $\chi^2$ & d.o.f. & $M_N [\textrm{MeV}]$ \\
\hline
$f_0 = 121.9(8.8)$~\texttt{MeV} & 007--040 & 
	0.98(2) & -0.80(12) & 0.43(9) & 0.00(1.86) & 1.39 & 1 & $1013\pm 15$ \\
&007--050 & 
	0.98(1) & -0.84(8) & 0.47(6) & 0.00(2.48) & 1.60 & 2 & $1009\pm 12$ \\
\\
\hline
NLO $SU(4|2)$, Eq.~\eqref{eq:MN_NLO_MA} & range &$M_0 [\textrm{GeV}]$ & $\a_M [\textrm{GeV}^{-1}]$ & $(g_A,g_1)$ & $g_{\D N}$ & $\chi^2$ & d.o.f. & $M_N [\textrm{MeV}]$ \\
\hline
$f_0 = 121.9(8.8)$~\texttt{MeV} & 007--050 & 
	1.01(4) & -0.95(13) & (0.6(2) , -0.5(1.4)) & 0.03(3.60) & 1.66 & 1 & $1046\pm 38$ \\
\\
\hline
NNLO, Eq.~\eqref{eq:MN_NNLO} & range &$M_0 [\textrm{GeV}]$ & $\a_M [\textrm{GeV}^{-1}]$ & $b_M [\textrm{GeV}^{-3}]$ & $b_A$ & $\chi^2$ & d.o.f. & $M_N [\textrm{MeV}]$ \\
\hline
$f_0 = 121.9(8.8)$~\texttt{MeV} & 007--040 & 
	0.87(6)(3) & -3.1(7)(8) & 62(11)(30) & -29(8)(16) & 0.06 & 1 & $941\pm 42 \pm 17$ \\
$g_A = 1.2(1)$, $g_{\D N} = 1.5(3)$ &007--050 & 
	0.90(4)(5) & -2.7(4)(9) & 55(7)(24) & -24(5)(17) & 0.75 & 2 & $966\pm 43 \pm20$  \\
\\
\hline
NNLO, Eq.~\eqref{eq:MN_NNLO} & range &$M_0 [\textrm{GeV}]$ & $\a_M [\textrm{GeV}^{-1}]$ & $b_M [\textrm{GeV}^{-3}]$ & $b_A$ & $\chi^2$ & d.o.f. & $M_N [\textrm{MeV}]$ \\
\hline
$f_0 = 121.9(8.8)$~\texttt{MeV} & 007--040 & 
	0.91(5)(0) & -1.8(5)(1) & 4.6(0.4)(1.0) & -7.6(4.2)(0.9) & 0.00 & 1 & $964\pm 41 \pm 20$ \\
$g_A = 1.2(1)$, $g_{\D N} = 0$ &007--050 & 
	0.96(4)(5) & -1.4(3)(6) & 4.8(0.3)(1.1) & -3.6(2.3)(4.1) & 1.36 & 2 & $996\pm 30 \pm30$  \\
\\
\hline
NNLO, Eq.~\eqref{eq:MN_NNNLO} & range & $M_0 [\textrm{GeV}]$ & $c_1 [\textrm{GeV}^{-1}]$ & $e_1 [\textrm{GeV}^{-3}]$ && $\chi^2$ & d.o.f. & $M_N [\textrm{MeV}]$ \\
\hline
$f_0 = 121.9(8.8)$~\texttt{MeV}  & 007--030 & 
	0.90(2)(1) & -0.97(4)(8) & 2.8(5)(9) && 0.02 & 1 & $958\pm 15 \pm 9$\\
$g_A = 1.2(1)$ & 007--040 & 
	0.90(1)(2) & -0.97(2)(8) & 2.7(2)(8) && 0.07 & 2 & $956\pm 12 \pm 11$ \\
$[c_2,c_3] = [3.2(4),-3.4(4)] \textrm{ GeV}^{-1}$ & 007--050 & 
	0.88(1)(2) & -1.01(1)(10) & 2.2(1)(8) && 6.5 & 3 & $940\pm 9 \pm 14$\\
\\
\hline
Eq.~\eqref{eq:MNstraightmpi}, $M_N = \a_0^N + \a_1^N m_\pi$ & range & $\a_0^N [\textrm{GeV}]$ & $\a_1^N$ & & & $\chi^2$ & d.o.f. & $M_N [\textrm{MeV}]$ \\
\hline
& 007--020 & 0.83(2) & 0.93(5) & & & 0.00 & 1 & $953\pm 13$ \\
& 007--030 & 0.82(2) & 0.94(4) & & & 0.18 & 2 & $950\pm 11$\\
& 007--040 & 0.80(1) & 0.99(3) & & & 4.39 & 3 & $938\pm 9$ \\
& 007--050 & 0.80(1) & 1.01(2) & & & 5.40 & 4 & $933\pm 8$
\end{tabular}
\end{ruledtabular}
\end{table}

%
%
\begin{figure}
\begin{tabular}{cc}
\includegraphics[width=0.48\textwidth]{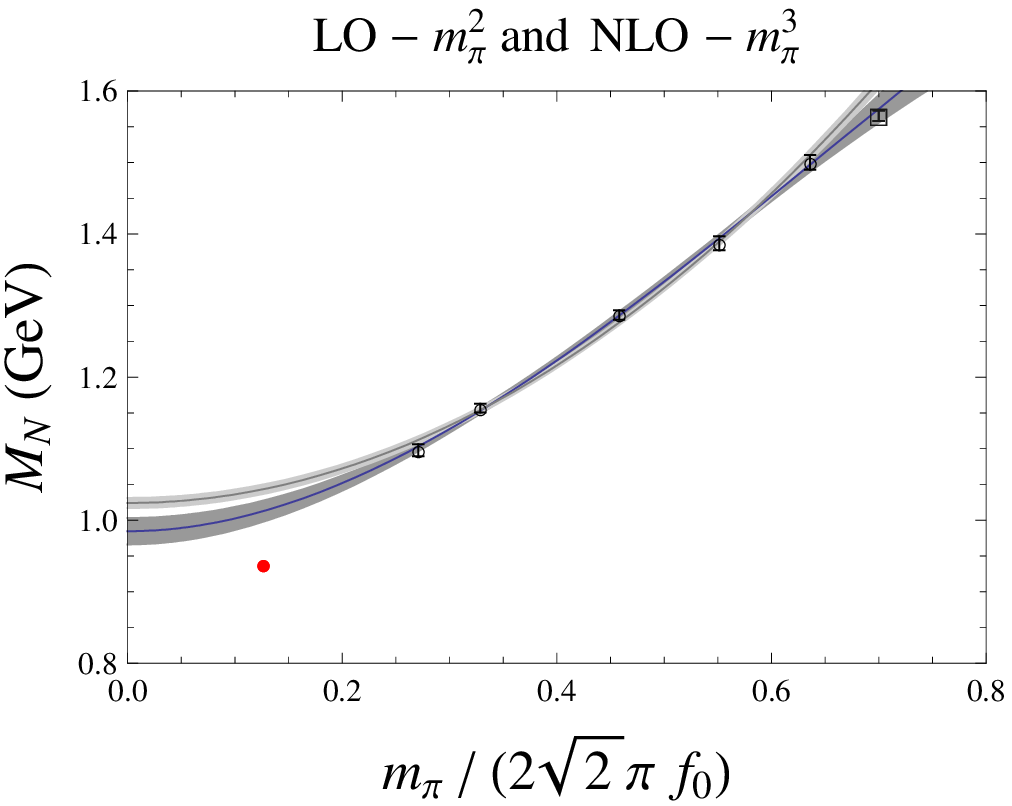}
&
\includegraphics[width=0.48\textwidth]{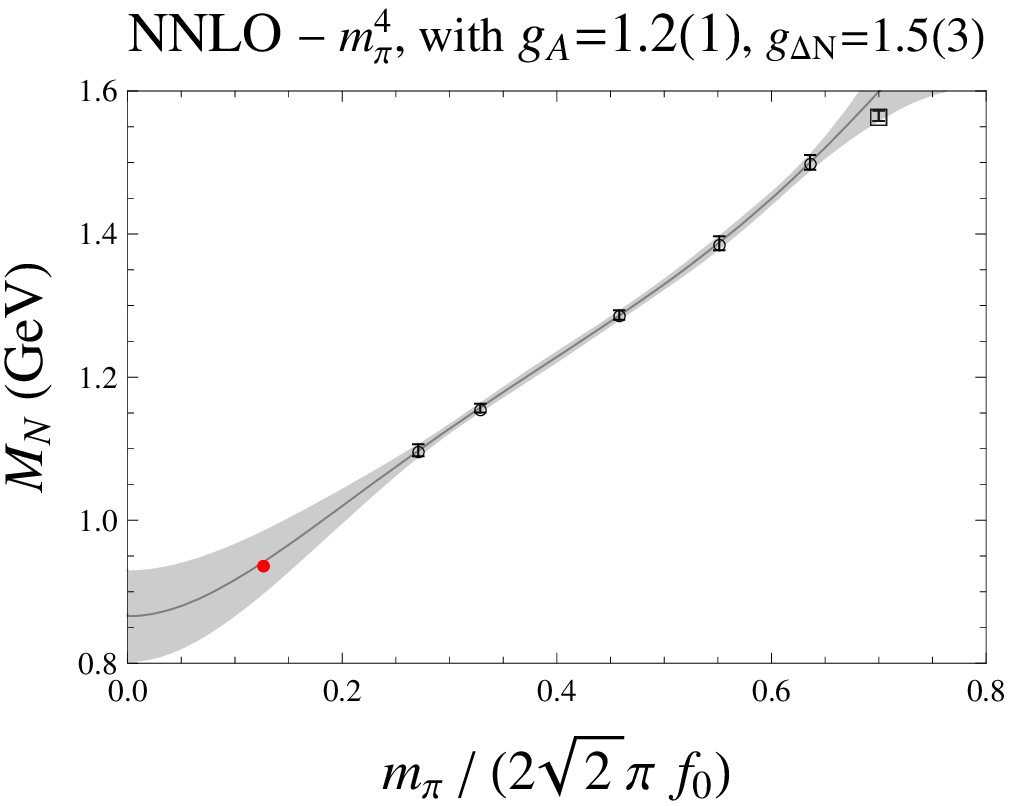}
\\ $(a)$ & $(b)$ \\
\includegraphics[width=0.48\textwidth]{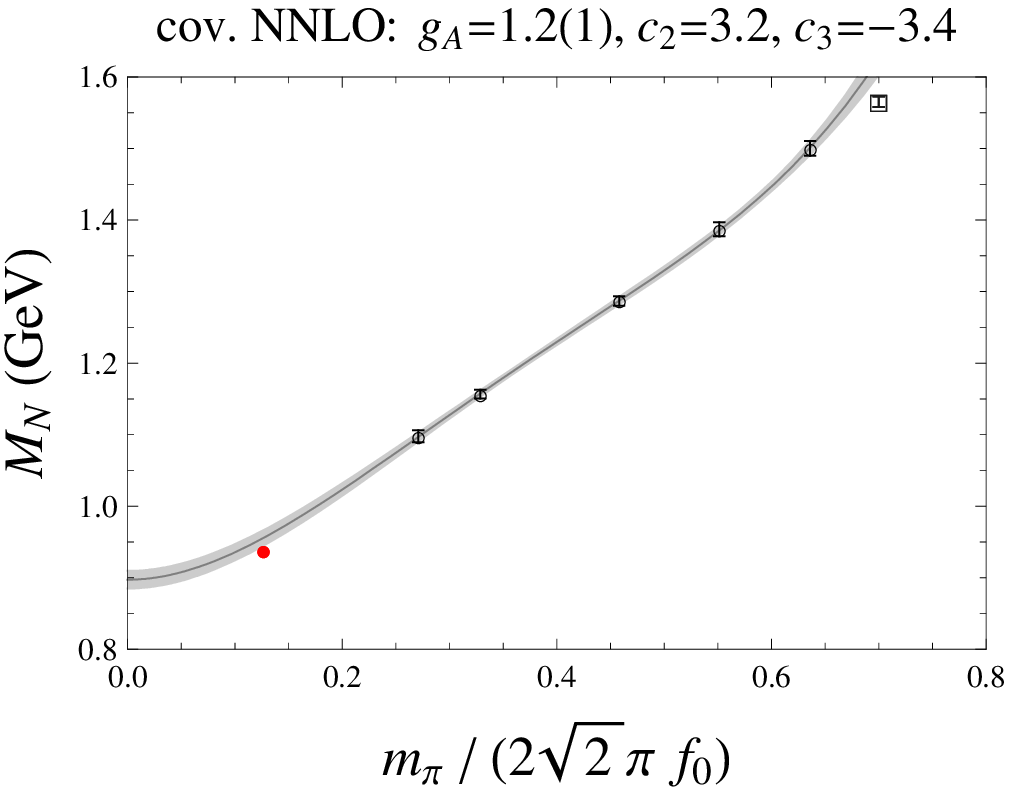}
&
\includegraphics[width=0.48\textwidth]{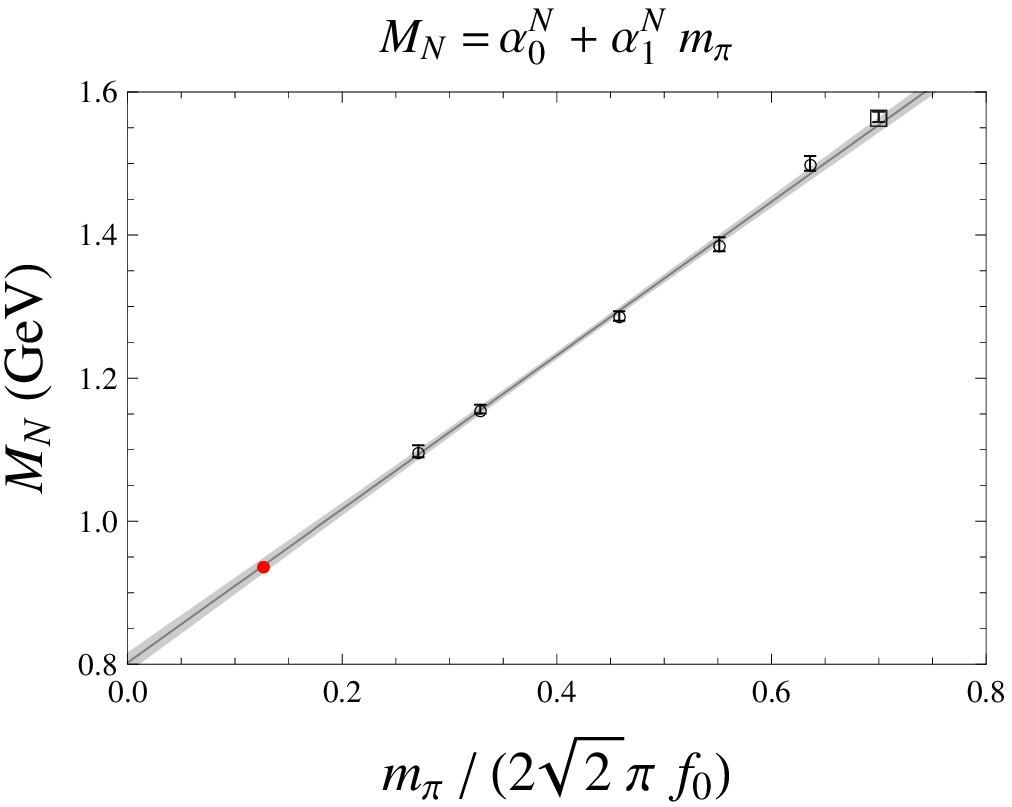}
\\
$(c)$ & $(d)$
\end{tabular}
\caption{\label{fig:MNExtrap} Chiral extrapolations of the nucleon mass, plotted vs the approximate expansion parameter $m_\pi / 2\sqrt{2}\pi f_0$, with $f_0 = 121.9$~\texttt{MeV}.  For comparison purposes, in all figures, we display the results of fits to the $m007$--$m040$ mass points denoted by the small black circles with error bars.  The black square is the $m050$ mass point not included in any of these fits.  The filled (red) circle is the physical nucleon mass, taken to be $939.6$~\texttt{MeV}, at a pion mass $m_\pi=137$~\texttt{MeV}, which is never included in the minimization.  The gray bands represent the 68\% confidence interval, and are only determined from the statistical error bar in the lattice results.  In Fig.~\ref{fig:MNExtrap}$(a)$, we plot both the LO and NLO $SU(2)$ HB$\chi$PT results.  The light shaded band is from LO and the darker shaded band is NLO.  In Fig.~\ref{fig:MNExtrap}$(b)$, we plot the results of the NNLO $SU(2)$ fit including explicit deltas.  In Fig.~\ref{fig:MNExtrap}$(c)$, we plot the NNLO $SU(2)$ covariant fit without deltas.  In comparing Figs.~\ref{fig:MNExtrap}$(b)$ and $(c)$ one needs to note the size of the error band is dictated by the number of free parameters and not by the use of infrared-regularization (covariant expression).  In Fig.~\ref{fig:MNExtrap}$(d)$, we plot the straight line fit, Eq.~\eqref{eq:MNstraightmpi}.  All of these fits, except the LO fit of $(a)$, are statistically consistent with our lattice results, as can be seen in Table~\ref{tab:MN_SU2_LECs}.}
\end{figure}

\noindent
Performing the LO through NNLO fits in principle allows us to study not only the chiral convergence of the nucleon mass, but also to monitor the resulting values of the LECs as higher order terms are added to the expansion.  If things are working as desired, we would find not only that the expression for the nucleon mass is converging order by order, but that the values of the LECs determined in the analysis would shift by only small amounts as we add higher order terms.  Unfortunately, since we must fix certain LECs in the NNLO fits, we are not able to honestly make this comparison.  In the NLO and NNLO expressions, the LECs $f_\pi$, $g_A$ and $g_{\D N}$ can either take their LO values, or one can replace them with their lattice-physical values,%
\footnote{We denote lattice-physical quantities as those that are determined directly from correlation functions, and that have not been extrapolated to the continuum, infinite volume or physical/chiral point.} 
the difference appearing at NNNLO.  To maintain as much consistency between the various fits as possible, including analyses performed by other groups, we always take $f_\pi = 121.9(8.8)$~\texttt{MeV}, consistent with the two loop determination of $f_\pi$ in the chiral limit~\cite{Colangelo:2003hf}.  We consider values of $g_A$ that are consistent with the physical value, the phenomenological value in the chiral limit, and  the lattice value of the nucleon axial charge calculated with this mixed action approach~\cite{Edwards:2005ym}, $g_A = 1.2(1)$, see also Ref.~\cite{Hemmert:2003cb}.  To fix $g_{\D N}$, we use the known width of the delta, combined with the LO expression for the width from HB$\chi$PT~\cite{Jenkins:1991ne,Tiburzi:2005na}
\begin{equation}
	\G_\D = -2 \textrm{Im}[M_\D] = \frac{g_{\D N}^2}{6 \pi f_\pi^2}(\D^2 - m_\pi^2)^{3/2}\, ,
\end{equation}
from which we obtain $g_{\D N} = 1.5(3)$.%
\footnote{There are various sign conventions for the axial couplings used in the literature~\cite{Beane:2004rf,Khan:2006de}.  The nucleon and delta masses are proportional to the square of these couplings, and we take them to be positive.} 
The error we assign also encompasses the NLO determination of $g_{\D N}$~\cite{Butler:1992pn}.  We also  set $\D = 271$~\texttt{MeV} as well as insert the lattice value for $\D = M_\D[m_q] - M_N[m_q]$ into our extrapolation analysis for each quark mass.  When using the NNLO formula, Eq.~\eqref{eq:MN_NNNLO}, we follow Ref.~\cite{Procura:2006bj} and set $c_2 = 3.2(4) \textrm{ GeV}^{-1}$ and $c_3 = -3.4(4) \textrm{ GeV}^{-1}$ (we have slightly inflated the errors from those in Ref.~\cite{Procura:2006bj}). 
%
%
\begin{figure}
\begin{tabular}{cc}
\includegraphics[width=0.48\textwidth]{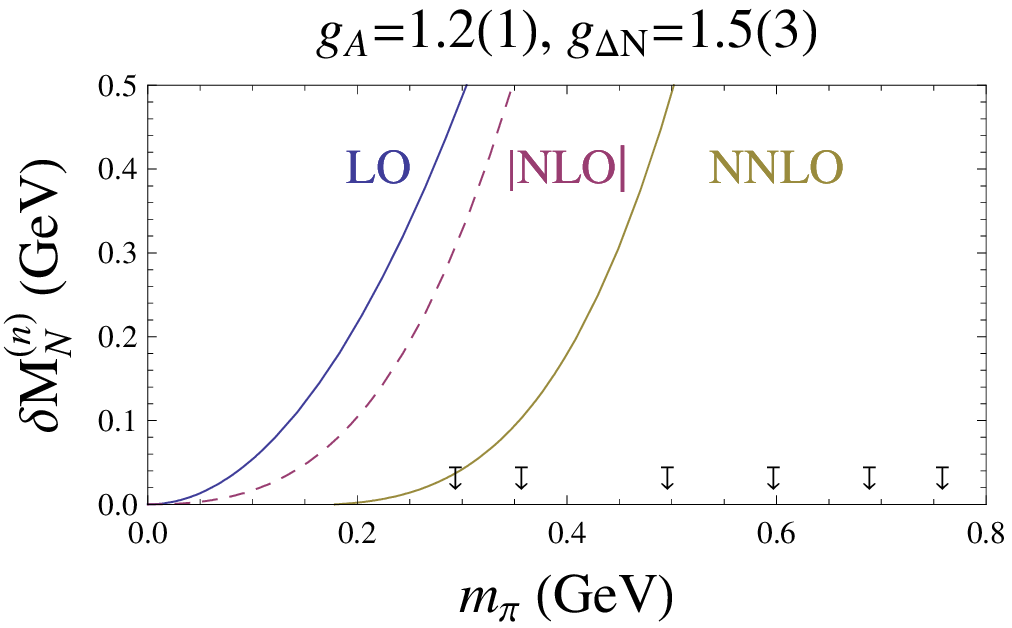}
&
\includegraphics[width=0.48\textwidth]{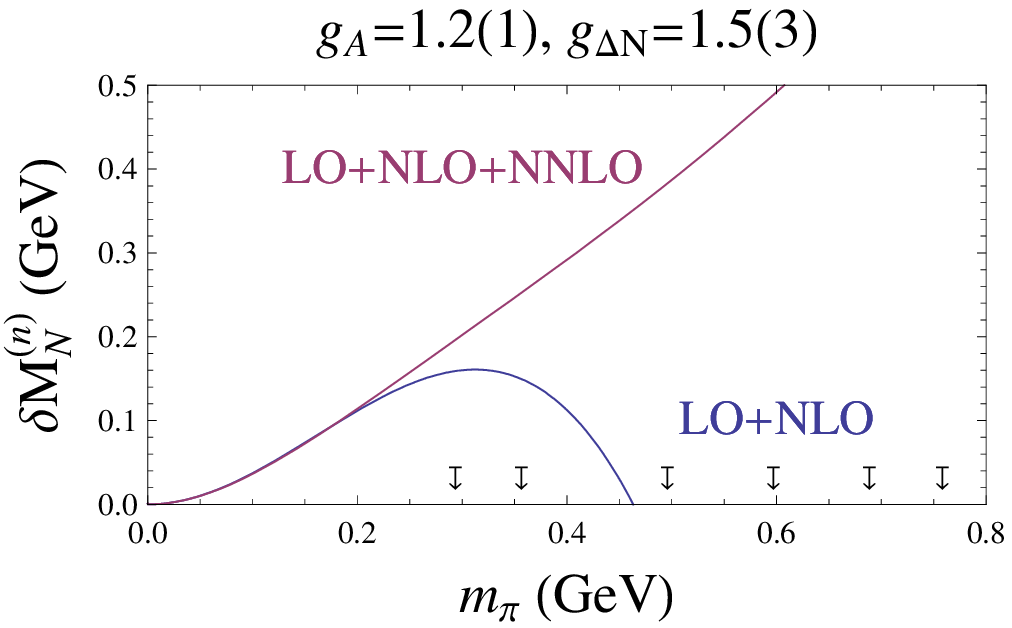}
\\ $(a)$ & $(b)$ 
\end{tabular}
\caption{\label{fig:MN_LONLONNLO} Plot of the contributions to the nucleon mass order-by-order, $\d M_N^{(n)}$ resulting from the NNLO heavy baryon $\chi$PT fit including the delta.  In $(a)$ we plot each order separately, plotting the magnitude of the (negative) NLO results denoted by a dashed line.  In $(b)$ we plot the combined LO+NLO and LO+NLO+NNLO results.  Plotting any of the NNLO fits of the nucleon mass in this work results in similar values for the $\d M_N^{(n)}$.  The arrows denote the values of the pion mass used in this work.  From $(a)$, one notes that the convergence of the resulting NNLO fit may already be breaking down at the second lightest mass point, although the summed contributions in $(b)$ display better convergence.}
\end{figure}

 We collect the results of these various extrapolations in Table~\ref{tab:MN_SU2_LECs}.  
The first uncertainty in the fit parameters and the predicted nucleon mass is statistical.  In the NNLO fits, the second error is a systematic error obtained by independently varying the fixed LECs over their given ranges.  In Figure~\ref{fig:MNExtrap}, we display some of the resulting fits along with their statistical 68\% confidence bands.  For comparison purposes, we have chosen to display fits that include the $m007$--$m040$ mass points.%
\footnote{If we instead perform these fits in $r_1$ units, as commonly employed by the MILC Collaboration, the qualitative features do not change.  For example, converting our nucleon and pion mass results to $r_1$ units using Table 1 of Ref.~\cite{Aubin:2004wf}, using the $m007$--$m040$ ensembles, we find $\a_1^N = 1.03(3)$ as compared to the value in Table~\ref{tab:MN_SU2_LECs}.  The NNLO $SU(2)$ HB$\chi$PT analysis with $g_A=1.2(1)$ and $g_{\D N}=1.5(3)$ also does not change qualitatively, returning for example $M_0 = 0.93(6)$~GeV and $\a_M = -2.5(3) \textrm{ GeV}^{-1}$.} 

From the analysis presented in Table~\ref{tab:MN_SU2_LECs}, we draw the following conclusions.  The  LO nucleon mass formula, Eq.~\eqref{eq:MN_LO}, does not provide a good description of our lattice data, except for the lightest three mass points.  At NLO, we see that both the continuum and the mixed action formulae provide reasonable descriptions of our lattice data, such that with the available degrees of freedom, we can not strongly conclude one fit is better than the other.  However, they return values of $g_A$ and $g_{\D N}$ that are inconsistent with both phenomenological and lattice determinations.  At NNLO, with fixed axial couplings, we see that the explicit inclusion  ($g_{\D N}=1.5(3)$) or exclusion ($g_{\D N}=0$) of the delta does not have a significant impact upon the quality of the fit, or the chirally extrapolated nucleon mass.  However, as expected, the values of the LECs $\a_M$ and more notably $b_M$ and $b_A$ are very sensitive to this change.  The NNLO fit based on covariant baryon $\chi$PT with infrared regularization, without explicit deltas, in fact has only three unknown fit parameters, as compared to the other NNLO  fits, and this is the reason the statistical error bars are much smaller.  The reason we chose to fix more parameters in the covariant NNLO fit is to have a more direct comparison with the chiral extrapolation analysis of other groups, who also chose to fix the values of $c_2$ and $c_3$ from phenomenological determinations~\cite{Procura:2006bj,Gockeler:2007rx,Alexandrou:2008tn}.  It is worth noting that the values of the fit parameters using the NNLO covariant fit agree within errors with those determined in Refs.~\cite{Procura:2003ig,Procura:2006bj,Gockeler:2007rx} which performed the same fit to two-flavor lattice calculations.

When we let either of these LECs float as a free parameter in the extrapolation, then the fit is comparable to the NNLO HB$\chi$PT analysis we have performed.  In Figure~\ref{fig:MN_LONLONNLO}$(a)$, we plot the resulting contributions to the nucleon mass, order by order for the NNLO fit with explicit deltas and in Fig.~\ref{fig:MN_LONLONNLO}$(b)$ we plot the combined LO+NLO and LO+NLO+NNLO mass contributions.  A similar plot with any of the HB$\chi$PT NNLO fits returns a similar expansion, see also Ref.~\cite{Bernard:2003rp}.  For visual aid, the arrows  denote the values of the pion masses used in our lattice calculation.  As can be seen in Table~\ref{tab:MN_SU2_LECs} and Figs.~\ref{fig:MNExtrap} and \ref{fig:MN_LONLONNLO}, even though the resulting fits to the nucleon mass are in good statistical agreement with our lattice data, already at the lightest value of the pion mass, Fig.~\ref{fig:MN_LONLONNLO}$(a)$ suggests the order-by-order convergence is becoming questionable.  In contrast, Fig.~\ref{fig:MN_LONLONNLO}$(b)$ displays much better convergence, which is understood from the oscillatory nature of the expansion.

The last fit we perform is a linear fit in $m_\pi~\sim \sqrt{m_q}$, Eq.~\eqref{eq:MNstraightmpi}.  This  ansatz is not motivated by any theoretical understanding of low energy QCD, but rather by the empirical observation that the results are nearly linear in $m_\pi$.  It is then natural to ask how good this fit is statistically.   It does not describe our lattice data as well as the NNLO analysis, however purely from the $\chi^2$, this fit can not be ruled out.  Here we would like to stress that we are not advocating Eq.~\eqref{eq:MNstraightmpi} as a means of performing the chiral extrapolation.  This fit ansatz is conceptually incorrect near the chiral limit where chiral symmetry dictates the nucleon mass scales as $M_N = M_0 + \alpha^\prime m_q$.  Rather we are highlighting this fit to bring attention to this unexpected phenomenon, and to ask the question, for which values of $m_\pi$ will this fit ansatz fail to describe the nucleon mass?

%
%
\begin{figure}
\begin{tabular}{cc}
\includegraphics[width=0.45\textwidth]{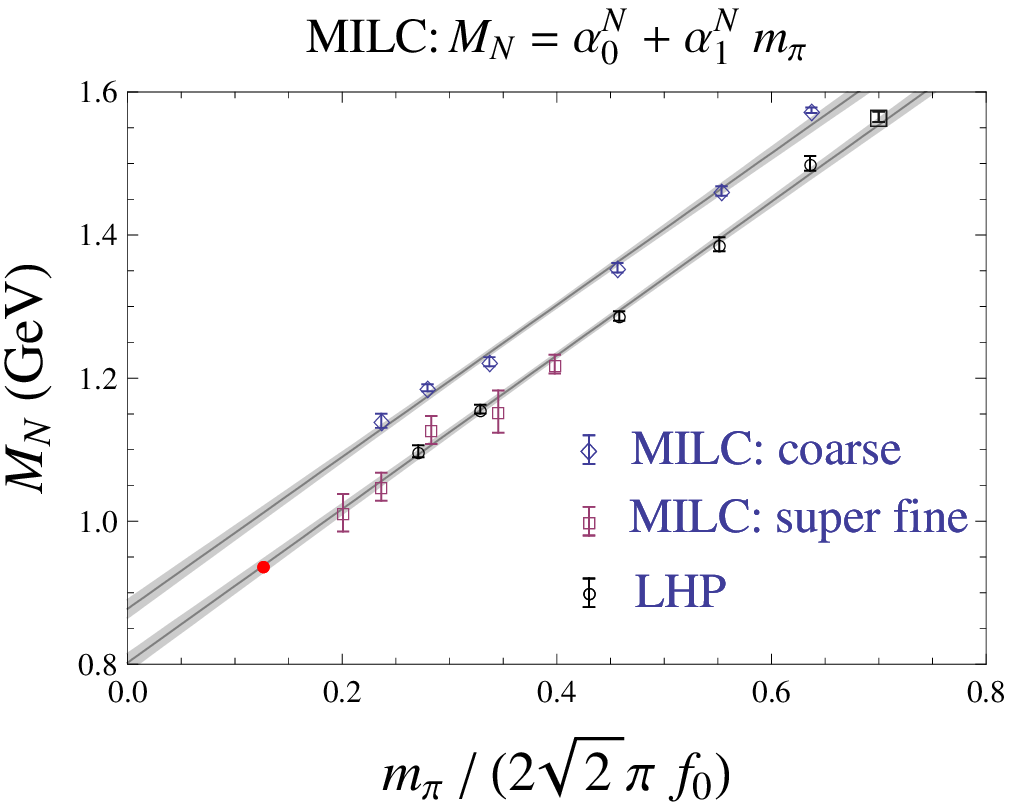}
&
\includegraphics[width=0.45\textwidth]{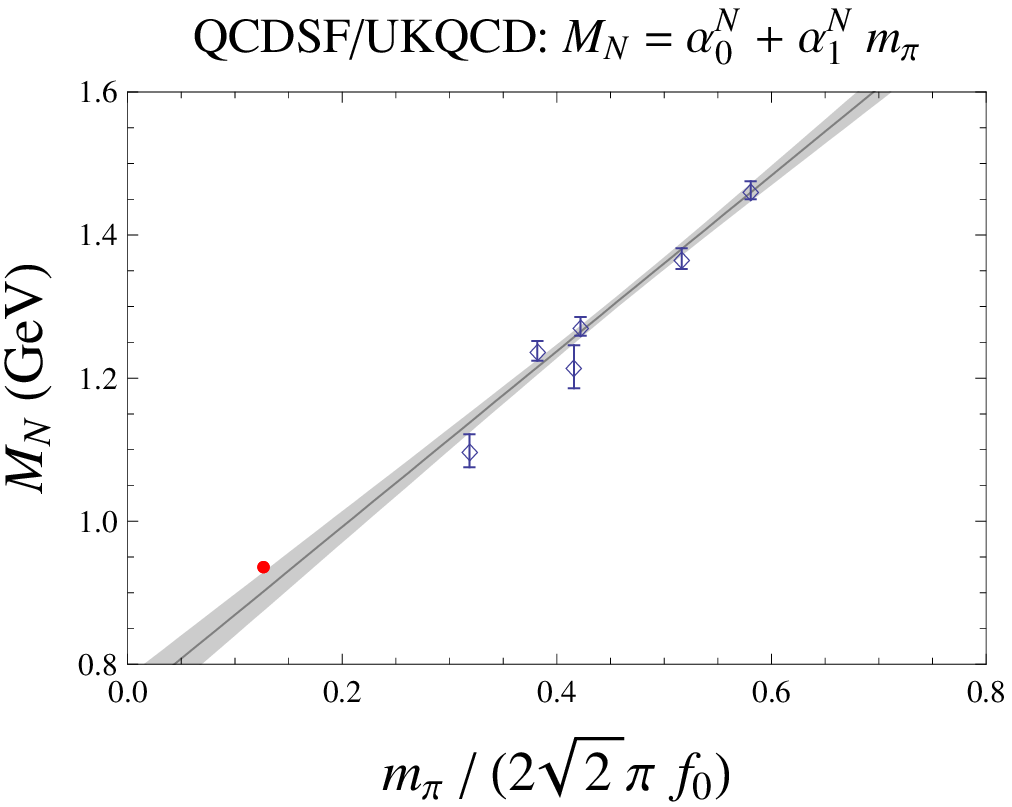}
\\
$(a)$ & $(b)$
\\ \\
\includegraphics[width=0.45\textwidth]{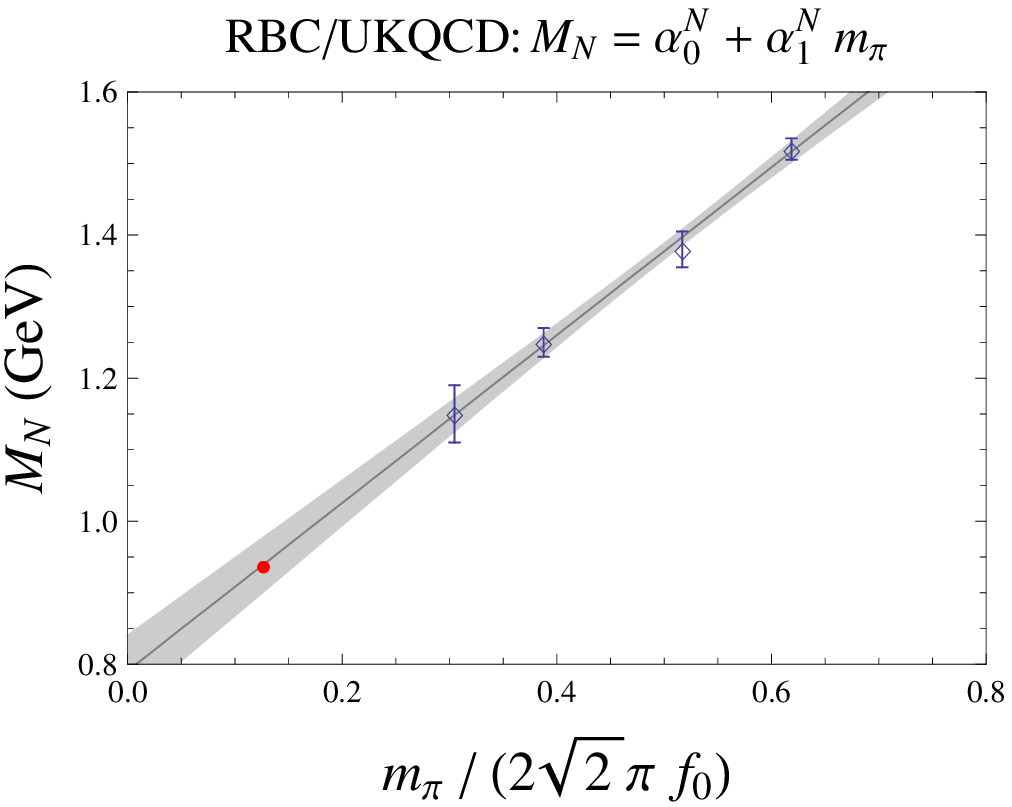}
&
\includegraphics[width=0.45\textwidth]{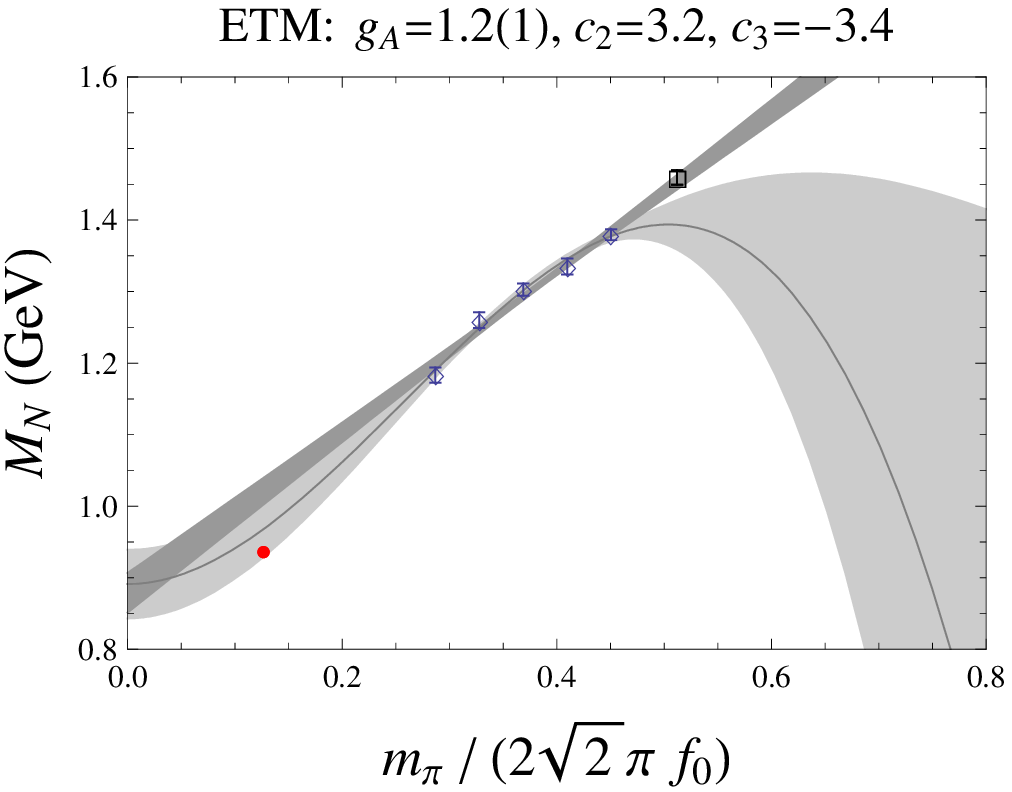}
\\
$(c)$ & $(d)$
\end{tabular}
\caption{\label{fig:MNcompare} Nucleon mass results from other lattice groups, the MILC~\cite{Bernard:2007ux}, QCDSF/UKQCD~\cite{Gockeler:2007rx}, RBC/UKQCD~\cite{Yamazaki:2007mk} and ETM~\cite{Alexandrou:2008tn} Collaborations.  We apply our linear fit in $m_\pi$  to the lattice results from all groups and plot the corresponding best fit and 68\% confidence band.  In all cases, the open diamonds correspond to the points included in the analysis.  In Fig.~\ref{fig:MNcompare}$(a)$, in addition to the coarse MILC ($a\sim 0.124$~\texttt{fm}) nucleon mass results, we also display the same analysis from our lattice results for comparison, as well as the MILC superfine ($a\sim 0.06$~\texttt{fm}) results, which were not used in the analysis.  In Fig.~\ref{fig:MNcompare}$(d)$, we also display a fit to the ETM lattice data using Eq.~\eqref{eq:MN_NNNLO}, which provides a better $\chi^2$ description of their nucleon mass results.}
\end{figure}

It is important to point out again that our lattice calculations have been performed with only one lattice spacing and only one lattice volume.  The volume effects are expected to be about 1\% or less, for our lightest mass as determined at NLO in heavy baryon $\chi$PT~\cite{Beane:2004tw}, and therefore do not impact this analysis significantly.   However, we have no control over the lattice spacing effects, and since we can not perform the NNLO mixed action analysis to compare to the continuum NNLO analysis, we can not provide more than an order of magnitude estimate of their size.  For these reasons, it is important for us to compare our nucleon mass results with other groups, which use a variety of different lattice actions, at different lattice spacings, with slightly different lattice volumes and with various scale setting procedures.

Recently, four other groups have reported  their computational results for the nucleon mass, and in most cases the resulting chiral extrapolations: the MILC~\cite{Bernard:2007ux}, QCDSF/UKQCD~\cite{Gockeler:2007rx}, RBC/UKQCD~\cite{Yamazaki:2007mk} and most recently the ETM~\cite{Alexandrou:2008tn} collaborations, all of which have employed relatively light pion masses in their lattice calculations.  The MILC collaboration uses $N_f=2+1$ flavor staggered valence fermions on rooted, staggered sea fermions, for which the nucleon mass is expected to have complicated lattice spacing dependence~\cite{Bailey:2007iq}.  The MILC Collaboration has results for three lattice spacings, and here we compare to the coarse ($a\sim 0.124$~\texttt{fm}) and superfine ($a\sim 0.06$~\texttt{fm}) MILC results from Ref.~\cite{Bernard:2007ux}.  The QCDSF/UKQCD group results are from $N_f=2$, $\mc{O}(a)$ improved Wilson fermions, with several different lattice spacings, ranging from $a\sim 0.085-0.067$~\texttt{fm}, for which the nucleon mass is expected to have very simple lattice spacing corrections~\cite{Beane:2003xv}.  The RBC/UKQCD group employed $N_f = 2+1$ domain-wall on domain-wall fermions with $a\sim 0.114$~\texttt{fm}, and are expected to have very simple and mild lattice spacing corrections.  The ETM collaboration has used $N_f=2$, $\mc{O}(a)$ improved twisted mass fermions, and reported results for three lattice spacings.  The nucleon mass with twisted mass fermions at maximal twist is expected to have very simple lattice spacing dependence~\cite{WalkerLoud:2005bt}, but more importantly, they do not find any significant lattice spacing corrections in their results~\cite{Alexandrou:2008tn}.  In Fig.~\ref{fig:MNcompare} we plot the resulting nucleon mass calculations from these groups along with our own straight-line in $m_\pi$ analysis of their data.  The only data for which the $SU(2)$ nucleon mass extrapolation formula performs significantly better than Eq.~\eqref{eq:MNstraightmpi} are those of the ETM collaboration.  Even in this case, the straight line is still a reasonable approximation as measured by the $\chi^2$.  

Within statistical error bars, our resulting nucleon mass calculations at $a\sim 0.124$~\texttt{fm} are consistent with those of the MILC superfine ($a\sim 0.06$~\texttt{fm}), RBC/UKQCD ($a\sim 0.114$~\texttt{fm}) and QCDSF/UKQCD ($a\sim 0.085-0.067$~\texttt{fm}) nucleon mass calculations, while the ETM nucleon mass results are systematically higher.  Our results are significantly different from the MILC calculations on the coarse lattices on which we performed our calculations.  The systematic difference with these coarse MILC results is highly suggestive of an expected overall additive $\mc{O}(a^2)$ difference in the nucleon mass from the two different actions.  Given the common method of scale setting between our calculation and that of MILC, and given the agreement of our results with the superfine MILC results,%
\footnote{We thank Doug Toussaint and other members of the MILC Collaboration for providing us with their preliminary superfine nucleon mass results, some of which are not yet published.} 
the DWF valence on the coarse MILC sea action seems to have significantly smaller discretization effects than the Asqtad action.

As is abundantly clear from this analysis, this linear behavior in $m_\pi$ is not unique to our mixed action calculation with domain-wall valence fermions on the MILC staggered sea configurations.  Is QCD conspiring to produce this remarkably straight line?  Or is this flattening of the chiral curvature of $M_N$ a lattice artifact?  Unfortunately, there are still too many variables to answer this question.  Half of the groups have used smaller lattice spacings, but only two flavors of light quarks.  The other half have used three dynamical light quarks, but employ larger lattice spacings (except for the superfine MILC results which have the smallest lattice spacing), albeit with very different lattice actions and expected lattice spacing dependence.  It is clear from Figs.~\ref{fig:MNExtrap}, 
\ref{fig:MN_LONLONNLO} and \ref{fig:MNcompare} that lighter pion masses are desirable, to determine if the lighter point would favor more curvature in the extrapolation function, although the lightest MILC superfine result has $m_\pi \sim 220$~\texttt{MeV}.  Also, it will be  important to use a second lattice spacing  with our mixed action as well as to perform complete domain wall calculations with domain wall sea quarks to explore the continuum extrapolation.  It is worth mentioning that each group has chosen an independent means to set the scale of their lattice calculations, so this phenomenon is not an accidental feature of a particular scale setting method.
To resolve this issue, a careful study with multiple lattice spacings, volumes and light pion masses will be needed.  Another feature that is now quite clear is that the nucleon mass is a poor observable to use to look for signatures of expected chiral non-analytic behavior, predicted from HB$\chi$PT, at least at the pion masses used in this work and other recent lattice calculations.%
\footnote{For an updated comparison and analysis of this phenomenon, see Ref.~\cite{WalkerLoud:2008pj}.}
%

\subsubsection{The (pion)-nucleon sigma term, $\s_N$}
We comment briefly on the pion-nucleon sigma-term, defined as the nucleon scalar form factor at zero momentum transfer.  From the Feynman-Hellmann theorem, we can relate this to the slope of the nucleon mass with respect to the quark mass,
\begin{equation}\label{eq:N_sigma}
	\s_N = m_q \frac{\partial}{\partial m_q} M_N\, .
\end{equation}
Up to NNLO differences, this is equal to
\begin{equation}
	\tilde{\s}_N = \frac{m_\pi}{2} \frac{\partial}{\partial m_\pi} M_N\, .
\end{equation}
While the explicit inclusion of the delta states has little impact upon the value of the chirally extrapolated nucleon mass, it has a significant effect upon the sigma-term.  For example, using the resulting NNLO HB$\chi$PT fits, Eq.~\eqref{eq:MN_NNLO}, we find
\begin{align}
	\tilde{\s}_N &= 84 \pm 17 \pm 20 \textrm{ MeV},\qquad \textrm{with } g_{\D N}=1.5(3)\, ,
\nonumber\\
	\tilde{\s}_N &= 42 \pm 14 \pm 9 \textrm{ MeV},\qquad \textrm{with } g_{\D N}=0\, ,
\end{align}
taking an average of the $m007$--$m040$ and $m007$--$m050$ fits.  While these fits are consistent at the 68\% confidence level, the explicit inclusion of the delta states increases the central value of the sigma-term by a factor of two.  With $g_{\D N}=1.5(3)$, the systematic error is dominated by this uncertainty in $g_{\D N}$.  Therefore, to really determine the nucleon sigma term from lattice QCD, one will need lattice data from which one can reliably determine $g_A$ and $g_{\D N}$ from the chiral extrapolation of the nucleon mass, or alternatively, lattice results in close proximity to the physical point on both sides.  Just for comparison purposes, we find, using the straight-line fit, a value of $\tilde{\s}_N = 67\pm 2$~\texttt{MeV}.

\subsubsection{The delta mass \label{sec:MDelta}}
The chiral expansion of the delta mass takes a similar form to the nucleon mass.  At NLO, the $SU(2)$ HB$\chi$PT extrapolation formula for the delta mass is given by~\cite{Tiburzi:2005na}
\begin{equation}\label{eq:MD_NLO}
M_\D = M_{0} + \D_0 + \g_M m_\pi^2
	-\frac{25 \pi g_{\D\D}^2}{27(4\pi f_\pi)^2}m_\pi^3 
	- \frac{2g_{\D N}^2}{3(4\pi f_\pi)^2} \mc{F}(m_\pi, -\D, \mu)\, .
\end{equation}
For $m_\pi < \D$, $\mc{F}(m_\pi, -\D, \mu)$ develops a branch cut, producing the imaginary part of the delta self-energy responsible for the strong decay to a p-wave nucleon-pion state.  In the finite box we work with, where the particle momenta are quantized in units of $p = 2\pi /L \sim 500$~\texttt{MeV}, this decay is kinematically forbidden.  In the center of mass frame, the nucleon and pion must each have about $p\sim 266$~\texttt{MeV}, and one might expect that as in the case of a delta that would be stable in infinite volume, the finite volume corrections to the delta mass should be exponential, $\textrm{exp}(-m_\pi L)$.  However, as noted in Ref.~\cite{Bernard:2007cm}, since the delta undergoes a strong decay and its width is therefore large, the plateau regions of the energy levels \textit{vs.} $L$ characterizing narrower resonances~\cite{Luscher:1991cf} are ``washed out" for the delta.  According to Ref.~\cite{Bernard:2007cm}, which uses an NLO covariant baryon $\chi$PT calculation, the energy levels receive power law in $L$ corrections, rather than exponential.  In this work, we perform two different extrapolations.  First, we perform the fit with the continuum, infinite volume extrapolation formula.  Second, we subtract from our lattice results the finite volume corrections calculated according to Ref.~\cite{Bernard:2007cm}, and then perform the chiral extrapolations.  Ultimately, a careful study with multiple volumes is required to test these predicted corrections and perform reliable extrapolations of the delta mass.

%
%
\begin{table}[t]
\caption{\label{tab:MD_SU2_LECs} Various two-flavor delta mass chiral extrapolation results.  We provide the minimization results for the NNLO fit to the delta mass using Eq.~\eqref{eq:MD_NNLO}, fitting directly to our lattice results as well as to the results with the predicted finite volume corrections subtracted, as discussed in the Sec.~\ref{sec:MDelta}.}
\begin{ruledtabular}
\begin{tabular}{ccccccccc}
NNLO, Eq.~\eqref{eq:MD_NNLO} & range &$M_{\D,0} [\textrm{GeV}]$ & $\g_M [\textrm{GeV}^{-1}]$ & $t_M [\textrm{GeV}^{-3}]$ & $t_A$ & $\chi^2$ & d.o.f. & $M_\D [\textrm{MeV}]$ \\
\hline
$f_0 = 121.9(8.8)$~\texttt{MeV} 
&007--040 & 1.59(13)(1) & -1.1(2.5)(0.7) & 4.5(1.8)(4.5) & -27(33)(6) & 0.34 & 1 & $1570\pm 92\pm58$  \\
$g_{\D N} = 1.5(3)$, $g_{\D\D}=2.2(6)$ 
&007--050 & 1.49(9)(2) & 1.1(1.6)(0.8) & 6.4(9)(4.7) & 3(20)(2) & 1.66 & 2 & $1500\pm92\pm78$\\
\\
\hline
NNLO, Eq.~\eqref{eq:MD_NNLO} with predicted $\d^\D_{FV}$ & range &$M_{\D,0} [\textrm{GeV}]$ & $\g_M [\textrm{GeV}^{-1}]$ & $t_M [\textrm{GeV}^{-3}]$ & $t_A$ & $\chi^2$ & d.o.f. & $M_\D [\textrm{MeV}]$ \\
\hline
$f_0 = 121.9(8.8)$~\texttt{MeV} 
&007--040 & 1.52(15)(1) & -1.0(2.9)(0.7) & 4.4(2.1)(4.9) & -28(39)(6) & 0.01 & 1 & $1509\pm92\pm6$  \\
$g_{\D N} = 1.5(3)$, $g_{\D\D}=2.2(6)$ 
&007--050 & 1.38(9)(10) & 2.0(1.8)(0.9) & 6.7(9)(4.8) & 13(22)(10) & 1.66 & 2 & $1412\pm92\pm9$ \\
\\
\hline
& 007 & 010 & 020 & 030 & & \\
\hline
$\d_{FV}^\D\equiv (M_\D[2.5\textrm{ fm}] - M_\D[\infty])$[MeV] 
& $42\pm18$ & $39\pm16$ & $4.9\pm2.2$ & $1.8\pm0.8$ & & 
\end{tabular}
\end{ruledtabular}
\end{table}

\begin{figure}
\begin{tabular}{cc}
\includegraphics[width=0.45\textwidth]{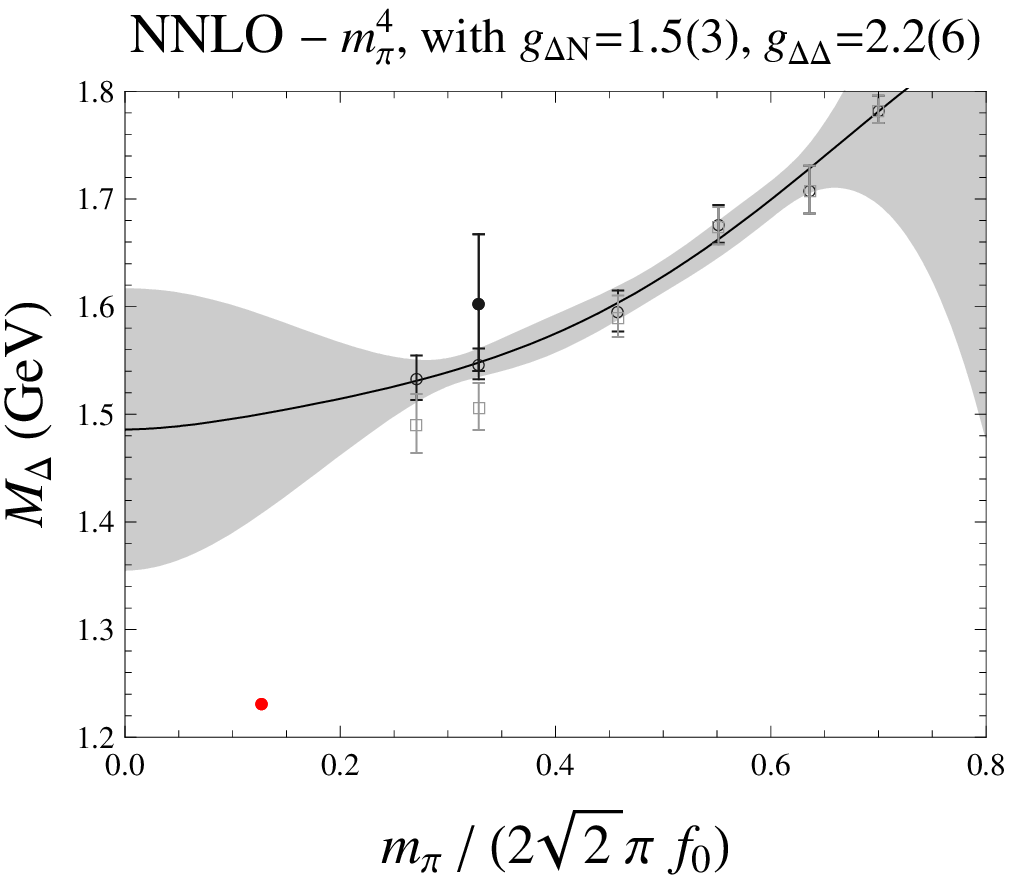}
&
\includegraphics[width=0.45\textwidth]{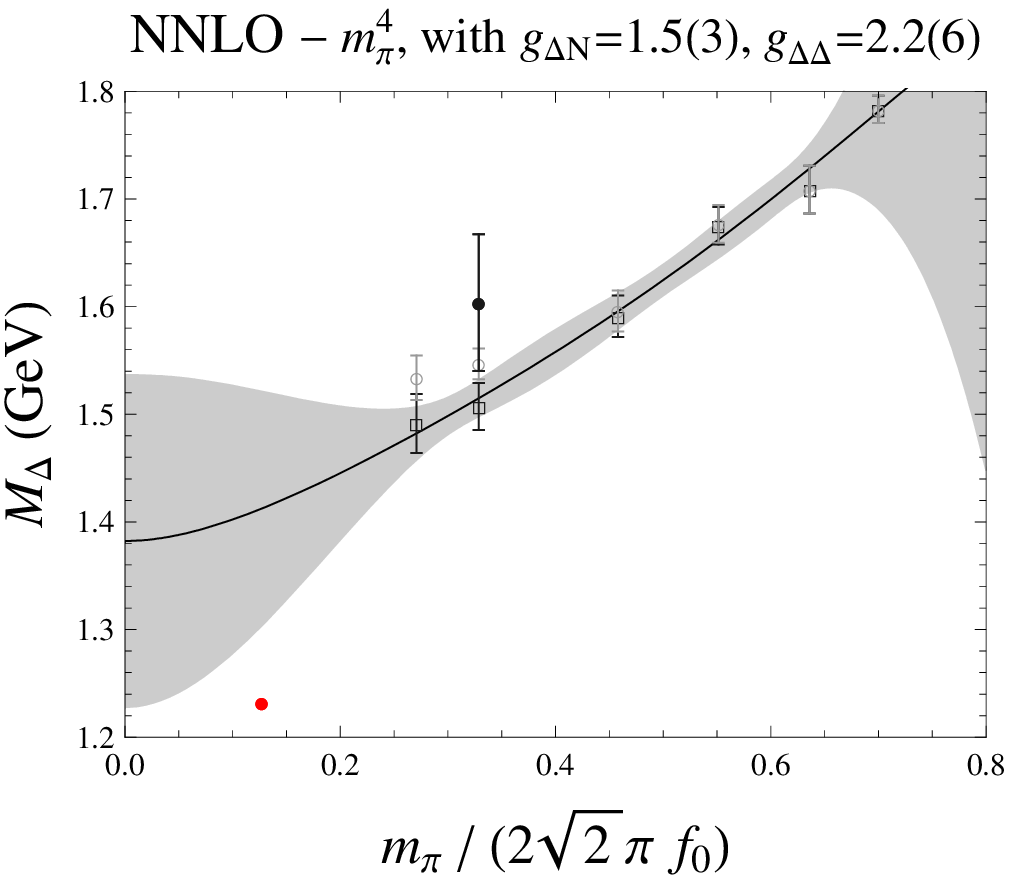}
\\
$(a)$ & $(b)$
\end{tabular}
\caption{\label{fig:MDeltaExtrap} Chiral extrapolations of the delta mass.  In $(a)$ we display the extrapolation of our lattice data using the continuum infinite volume formula.  In $(b)$, we display the extrapolation of our lattice data with the predicted finite volume corrections subtracted from our lattice results.  For comparison, both fits are done using all six mass points.  The open circles with error bars are our calculational results while the lower open squares are the predicted infinite volume extrapolations.  The filled circle with error bar represents our preliminary determination of the delta mass using the $m010$ ensemble on the $L\sim 3.5$~\texttt{fm} lattices.  Note that it is located on the opposite side compared to the predicted infinite volume extrapolation of Ref.~\cite{Bernard:2007cm}.  The filled (red) circle denotes the pole mass of the delta plotted at $m_\pi=137$~\texttt{MeV}.  A multiple volume study is required for the extrapolation of the delta mass, which is beyond the scope of this work.}
\end{figure}

\begin{figure}
\begin{tabular}{ccc}
\includegraphics[width=0.45\textwidth]{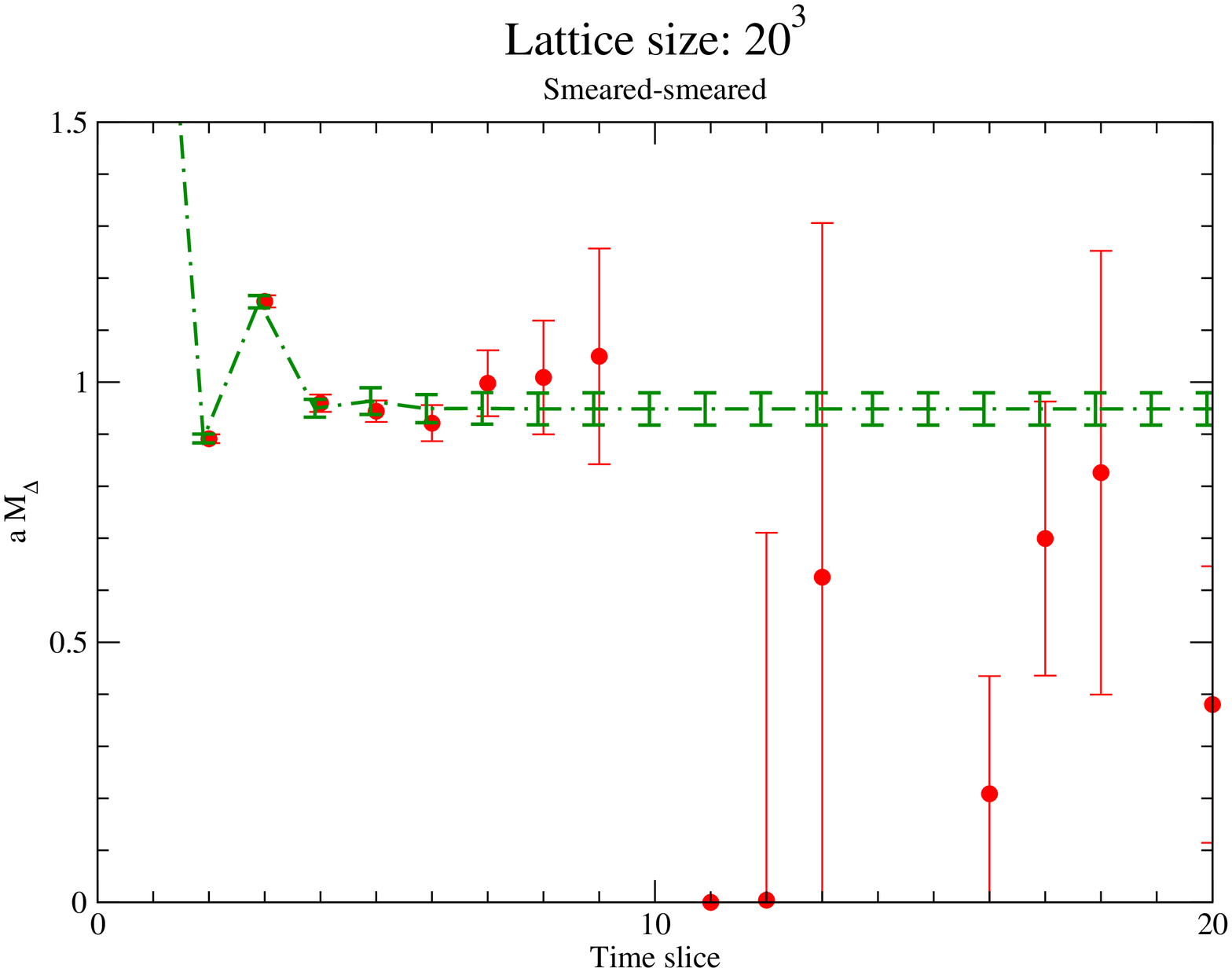}
&\phantom{sp}&
\includegraphics[width=0.45\textwidth]{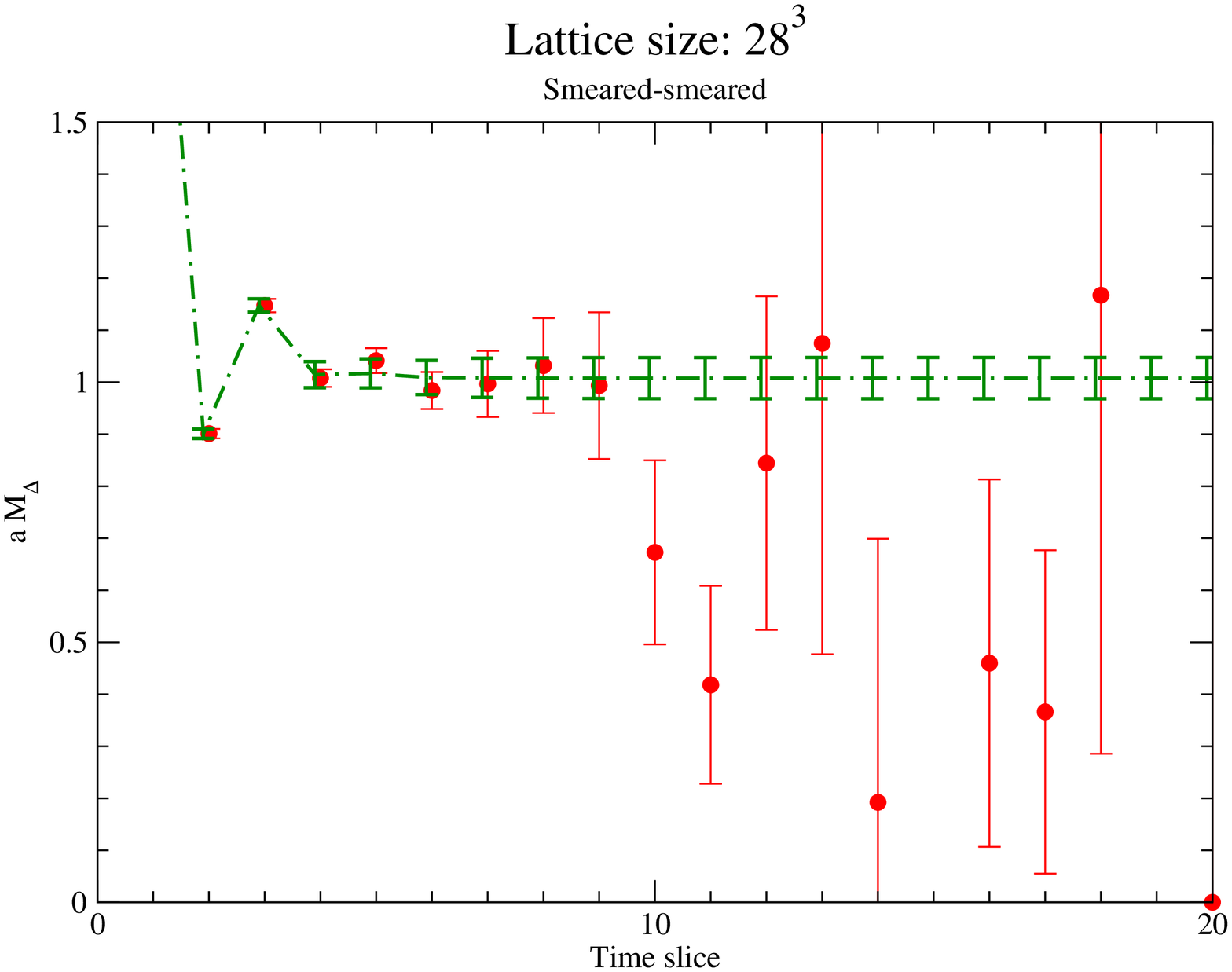}
\\
$(a)$ && $(b)$
\end{tabular}
\caption{\label{fig:MDelta_20vs28} Oscillating fits to the smeared-smeared delta correlation function, Eq.~\eqref{eq:osc} on the $m010$ $20^3\times64$, $(a)$ and the $28^3\times64$, $(b)$ lattices.  The delta mass on the larger volume is heavier than on the lighter volume, contrary to expectations.}
\end{figure}

We note that recently some groups have opted to perform the chiral extrapolation of the delta mass by setting the strong coupling of the delta to the nucleon-pion state to zero, by hand, accomplished by setting $g_{\D N}=0$ in Eq.~\eqref{eq:MD_NLO}~\cite{Gockeler:2007rx,Alexandrou:2008tn}.  We chose not to follow this approach.  To perform the chiral extrapolation of the delta mass, we use the continuum $SU(2)$ HB$\chi$PT extrapolation formula to NNLO~\cite{Tiburzi:2005na}, with a similarly modified renormalization prescription as discussed for the nucleon mass.  Taking the real part of the mass formula, we obtain
\begin{align}\label{eq:MD_NNLO}
\textrm{Re}[M_\D] =&\ M_{\D,0} + \g_M(\mu) m_\pi^2
	-\frac{25 \pi g_{\D\D}^2}{27(4\pi f_\pi)^2}m_\pi^3 
	- \frac{2g_{\D N}^2}{3(4\pi f_\pi)^2} \textrm{ Re}\Big[ \mc{F}(m_\pi, -\D, \mu) \Big]
	\nonumber\\&
	-\frac{3\g_M(\mu) m_\pi^4}{(4\pi f_\pi)^2} \ln \left( \frac{m_\pi^2}{\mu^2} \right)
	-\frac{25 g_{\D\D}^2 m_\pi^4}{48(4\pi f_\pi)^2 M_{\D,0}} \left( \ln \left( \frac{m_\pi^2}{\mu^2} \right) +\frac{19}{10} \right)
	\nonumber\\&
	-\frac{5 g_{\D N}^2 m_\pi^4}{8(4\pi f_\pi)^2 M_{\D,0}} \left( \ln \left( \frac{m_\pi^2}{\mu^2} \right) -\frac{1}{10} \right)
	-\frac{g_{\D N}^2 \g_M(\mu) m_\pi^2}{(4\pi f_\pi)^2} \textrm{ Re}\Big[ \mc{J}(m_\pi, -\D, \mu)\Big]
	\nonumber\\&
	+t_M(\mu) m_\pi^4 + t_A(\mu) \frac{m_\pi^4}{(4\pi f_\pi)^3}\ln \left( \frac{m_\pi^2}{\mu^2} \right)\, .
\end{align}
Some complex analysis shows
\begin{align}
\textrm{Re}\Big[\mc{F}(m,-\D,\mu)\Big] &= \left\{ \begin{array}{lc}
	-\mc{F}(m,\D,\mu), & m < |\D| \\
	-\mc{F}(m,\D,\mu) -2\pi (m^2 - \D^2)^{3/2}, & m > |\D|
	\end{array} \right.
\\
\textrm{Re}\Big[\mc{J}(m,-\D,\mu)\Big] &= \left\{ \begin{array}{lc}
	\mc{J}(m,\D,\mu), & m < |\D| \\
	\mc{J}(m,\D,\mu) + 2\pi \D \sqrt{m^2 - \D^2}, & m > |\D|
	\end{array}\right.
\end{align}
As with the nucleon mass, there are too many LECs to be determined from the lattice results alone.  We therefore fix $g_{\D N}=1.5(3)$ as above and $g_{\D\D}=2.2(6)$~\cite{Butler:1992pn,Savage:1996zd,FloresMendieta:1998ii}.  When performing our delta mass extrapolations, we take the results of the fits that include oscillatory terms.  Comparing Tables~\ref{tab:osc} and \ref{tab:decupletfit}, one can see that the resulting delta mass on the $m010$ ensemble is smaller using Eq.~\eqref{eq:osc} than with the single exponential fit (this discrepancy is negligible for all other ensembles).  Examining Fig.~\ref{fig:Deff}, it is clear that the oscillatory region provides constraints with small error bars, so that a fit including the oscillatory terms discussed in Sec.~\ref{sec:fit} provides a better description of the correlation function, which is confirmed by the $\chi^2$ minimization.  
We collect the results of our analysis in Table~\ref{tab:MD_SU2_LECs}, as well as the predicted finite volume corrections, defined as
\begin{equation}
	\d_{FV}^\D \equiv M_\D[2.5\textrm{ fm}] - M_\D[\infty]\, .
\end{equation}
These values have been obtained using the central values of $m_\pi$, $M_N$ and $M_\D$, listed in Tables~\ref{tab:meson_mass}, \ref{tab:octetfit} and \ref{tab:osc} respectively and by varying $g_{\D\D}$, $g_{\D N}$ and $f_\pi$ within the ranges specified in the present paper.
In Figure~\ref{fig:MDeltaExtrap} we present the resulting chiral extrapolations.  In the figure, the open circles with error bars are our lattice results.  The open squares that sit slightly below the circles are the lattice results after subtracting the predicted finite volume corrections, determined as described in Ref.~\cite{Bernard:2007cm}.  The filled circle with an error bar is our result for the delta mass using the $m010$ ensembles with $L\sim 3.5$~\texttt{fm} lattices, for which we display the correlation functions and oscillating fits in Fig.~\ref{fig:MDelta_20vs28}, and the resulting masses in Table~\ref{tab:MDelta_20vs28}.  Note, this lies in the opposite direction as that predicted from Ref.~\cite{Bernard:2007cm}.  However, the minimum momenta on these lattices are $p\sim 350$~\texttt{MeV}.  The pion-nucleon p-wave decay channel is therefore more accessible and this state may have much larger volume corrections.  Clearly, a multiple volume study will be required to study the delta mass and its chiral extrapolation.%
\footnote{A fit using the NNLO covariant baryon $\chi$PT delta mass extrapolation formula, as advocated in Ref.~\cite{Bernard:2005fy}, might be in more agreement with the pole position of the delta mass.  However, given our numerical results of the delta mass in the two different volumes, there are clearly more important systematic effects to be understood than which version of baryon EFT provides a better agreement with the physical pole mass.  We therefore do not make use here of the formula in Ref.~\cite{Bernard:2005fy}.} 
%

%
%
\begin{table}[t]
\caption{\label{tab:MDelta_20vs28} Resulting delta mass determinations on the $20^3\times64$ and $28^3\times64$ coarse MILC ensembles, using both the smeared--smeared (SS) as well as smeared-point (SP) correlation functions.  The results on the unchopped $20^3\times64$ ensembles are in agreement with our high statistics results on the chopped $20^3\times32$ ensembles.  Note the source on the $20^3\times32$ ensembles was placed at $t=10$.}
\begin{ruledtabular}
\begin{tabular}{ccccc}
&\multicolumn{2}{c}{SS} & \multicolumn{2}{c}{SP} \\
\cline{2-3}\cline{4-5}
Volume & Range & $M_\D$ & Range & $M_\D$ \\
$20^3\times32$ & 16--22 & 0.974(9) \\
$20^3\times64$ & 1--10 & 0.95(3) & 1--10 & 0.97(3) \\
$28^3\times64$ & 1--10 & 1.01(4) & 1--10 & 1.03(3) 
\end{tabular}
\end{ruledtabular}
\end{table}

%
%

%
%

%
%
\subsection{Three flavor chiral extrapolations \label{sec:SU3extrap}}

We now turn to the chiral extrapolations of the octet and decuplet baryon masses, making use of the mixed action generalization of $SU(3)$ heavy baryon $\chi$PT, as discussed above.%
\footnote{In Ref.~\cite{Frink:2005ru}, the chiral extrapolations of the octet baryon masses calculated by MILC on the coarse Asqtad ensembles was explored using the NNLO covariant formula determined in Ref.~\cite{Frink:2004ic}.} 
Before delving into the details of our chiral extrapolations, we first discuss our strategy.  In addition to fitting the baryon masses, we also fit the baryon mass splittings, both exclusively and in combination with the masses.  There are several advantages in studying the baryon mass splittings \textit{vs.} the masses, a technique first successfully exploited in Ref.~\cite{Bhattacharya:1995fz} in a quenched baryon spectroscopy calculation.  First, the leading lattice spacing corrections to the masses exactly cancel in the splittings, pushing these lattice artifacts to NNLO in the mixed action EFT~\cite{Tiburzi:2005is,Chen:2007ug}.%
\footnote{The virtual meson cloud contribution to the baryon masses, which enters the mass expressions at NLO, receives some contributions from mixed-mesons that have discretization errors in them.  However, the additive mixed meson mass splitting has been determined~\cite{Orginos:2007tw}, so this lattice artifact introduces no new unknown LECs, as opposed to those that enter at NNLO.} 
Second, the mass splittings must identically vanish both in the three-flavor chiral limit as well as the $SU(3)$ degenerate mass point.  This is reflected in the extrapolation formulae and leads to smaller theoretical error in the extrapolation to the physical point, as has been successfully employed for other quantities with similar limits calculated with chirally symmetric valence fermions~\cite{Beane:2005rj,Beane:2006kx,Beane:2006gj,Beane:2007xs,Beane:2007uh}.  Finally, fitting the mass splittings as well as the masses allows us to address in more detail the convergence of $SU(3)$ heavy baryon $\chi$PT, a topic that we address after detailing the mixed action Lagrangian.

The partially-quenched/mixed-action Lagrangian in the notation of Refs.~\cite{Chen:2001yi,Chen:2007ug}, is given by
\begin{align}\label{eq:BTLag}
\mathcal{L}_{MA} =&\ \bar{\mathcal{B}}\ i v\cdot D\ \mathcal{B} 
	+2 \a_M^{(PQ)} \left( \bar{\mc{B}} \mc{B} \mc{M}_+ \right)
	+2 \b_M^{(PQ)} \left( \bar{\mc{B}} \mc{M}_+ \mc{B} \right)
	+2 \s_M^{(PQ)} \left( \bar{\mc{B}} \mc{B} \right) \str (\mc{M}_+)
	+a^2 \s_a \left( \bar{\mc{B}} \mc{B} \right)
\nonumber\\&
	- \left( \bar{\mc{T}}^\mu \left[ i v \cdot D -\Delta \right] \mc{T}_\mu \right)
	+2\g_M^{(PQ)} \left( \bar{\mc{T}}^\mu \mc{M}_+ \mc{T}_\mu \right)
	-2\bar{\s}_M^{(PQ)} \left( \bar{\mc{T}}^\mu \mc{T}_\mu \right) \str (\mc{M}_+ )
	-a^2 \bar{\s}_a \left( \bar{\mc{T}}^\mu \mc{T}_\mu \right)
\nonumber\\&
	+2 \alpha^{(PQ)} \left( \bar{\mc{B}} S^\mu \mc{B} \mc{A}_\mu \right)
	+2 \beta^{(PQ)} \left( \bar{\mc{B}} S^\mu \mc{A}_\mu \mc{B} \right)
	+2 \mc{H}^{(PQ)} \left( \bar{\mc{T}}^\mu S^\nu \mc{A}_\nu \mc{T}_\mu \right)
\nonumber\\&
	+\sqrt{\frac{3}{2}} \mc{C}^{(PQ)} \left( \bar{\mc{T}}^\mu \mc{A}_\mu \mc{B} 
		+ \bar{\mc{B}} \mc{A}_\mu \mc{T}^\mu \right)\, .
\end{align}
The last term in each of the first two lines provides the leading lattice spacing correction to the baryon masses.  As can be seen, these terms treat all the octet and decuplet baryon masses the same, and therefore, as mentioned above, this leading lattice spacing dependence drops out of the mass splittings.  For our purposes, we only need the leading term of the mass spurion field,
\begin{equation}
	\mc{M}_+ = m_q +\dots\, ,
\end{equation}
where the dots represent terms with higher powers of the meson fields, and $m_q$ is the quark mass matrix.  Here, we have suppressed all flavor indices, and the braces $(\ )$ represent the flavor traces as defined in Ref.~\cite{Chen:2001yi}, and $\mathrm{str}()$ stands for a super-trace over the graded flavor algebra.  The relation between the LECs of this Lagrangian and the standard $SU(3)$ Lagrangian~\cite{Jenkins:1990jv,Jenkins:1991ne} can be determined by matching this theory onto the valence sector~\cite{Chen:2001yi}, for which one finds
\begin{align}\label{eq:MN_MANLO}
	&D = \frac{1}{4} \left( \a^{(6|3)} -2\b^{(6|3)} \right)\, ,&
	&F = \frac{1}{12} \left( 5\a^{(6|3)} +2\b^{(6|3)} \right)\, ,&
	\nonumber\\
	&\mc{H} = \mc{H}^{(6|3)}\, , &
	&\mc{C} = -\mc{C}^{(6|3)}\, .&
	\nonumber\\
	&b_D = \frac{1}{4} \left( \a_M^{(6|3)} -2\b_M^{(6|3)} \right)\, ,&
	&b_F = \frac{1}{12} \left( 5\a_M^{(6|3)} +2\b_M^{(6|3)} \right)\, ,&
	\nonumber\\
	&b_0 = \s_M^{(6|3)} +\frac{1}{6}\a_M^{(6|3)} +\frac{2}{3}\b_M^{(6|3)}\, ,&
	&\g_M^{(3)} = \g_M^{(6|3)}\, ,& 
	\nonumber\\
	&\ol{\s}_M^{(3)} = \ol{\s}_M^{(6|3)}\, ,&
\end{align}
From this Lagrangian, for example, the nucleon mass is given to NLO by~\cite{Chen:2001yi}
\begin{align}
M_N =&\ M_0(a,\mu)
	-m_\pi^2 \Big[ \a_M^\prime(\mu) +\b_M^\prime(\mu) +\s_M^\prime(\mu) \Big]
	-m_K^2\, 2\s_M^\prime(\mu)
	\nonumber\\ &\ 
	+\frac{\mc{F}(m_\pi,0,\mu)}{(4\pi f)^2} \left[ \frac{1}{3}(D-3F)(11D-9F)
		-\frac{2}{3} \frac{\D_{PQ}^2(D-3F)^2}{\tilde{m}_X^2 -m_\pi^2} 
		+\frac{\D_{PQ}^4 (D-3F)^2}{(\tilde{m}_X^2 - m_\pi^2)^2} \right]
	\nonumber\\&
	-\frac{2}{3}(5D^2 -6DF+9F^2) \left[ 
		\frac{2 \mc{F}(\tilde{m}_{ju},0,\mu)}{(4\pi f)^2} 
		+\frac{\mc{F}(\tilde{m}_{ru},0,\mu)}{(4\pi f)^2} \right]
	\nonumber\\&
	-\frac{2\mc{C}^2}{3} \left[ \frac{2 \mc{F}(m_\pi,\D,\mu)}{(4\pi f)^2}
		+\frac{2 \mc{F}(\tilde{m}_{ju},\D,\mu)}{(4\pi f)^2}
		+\frac{\mc{F}(\tilde{m}_{ru},\D,\mu)}{(4\pi f)^2} \right]
	\nonumber\\&
	-\frac{\mc{F}(\tilde{m}_{X},0,\mu)}{(4\pi f)^2} \frac{(D-3F)^2}{3} \left[ 1
		-\frac{2\D_{PQ}^2}{\tilde{m}_X^2 -m_\pi^2}
		+\frac{\D_{PQ}^4}{(\tilde{m}_X^2 -m_\pi^2)^2} \right]
	\nonumber\\&
	-\frac{\partial_{m_\pi^2} \mc{F}(m_\pi,0,\mu)}{(4\pi f)^2} \D_{PQ}^2 (D-3F)^2 \left[ 1
		-\frac{\D_{PQ}^2}{3(\tilde{m}_X^2 -m_\pi^2)} \right]\, ,
\end{align}
where $\mc{F}(m_\phi,\D,\mu)$ is defined in Eq.~\eqref{eq:F} and $\mc{F}(m_\phi,0,\mu)=\pi m_\phi^3$.  We use modified LECs
\begin{equation}
(\a_M^\prime, \b_M^\prime, \s_M^\prime) = 
	\Big( \frac{\a_M}{B_0}, \frac{\b_M}{B_0}, \frac{\s_M}{B_0} \Big)\, ,
\end{equation}
that have mass-dimension $-1$, and also we subsequently drop the \textit{primes}.  The parameter $B_0$ is related to the chiral condensate that appears in the Gell-Mann--Oakes--Renner relation~\cite{GellMann:1968rz} and the meson chiral Lagrangian.  We further have
\begin{align}
	&\tilde{m}_X^2 = m_\eta^2 +\D_{PQ}^2\, ,&
	&\D_{PQ}^2 = a^2 \D_\mathrm{I}&
	\nonumber\\
	&\tilde{m}_{ju}^2 = m_\pi^2 +a^2\D_\mathrm{Mix}\, ,&
	&\tilde{m}_{ru}^2 = m_K^2 +a^2\D_\mathrm{Mix}\, ,&
\end{align}
where $a^2 \D_\mathrm{I}$ and $a^2\D_\mathrm{Mix}$ can be found in Ref.~\cite{Aubin:2004fs} and \cite{Orginos:2007tw} respectively.  Similar expressions for all the octet and decuplet baryon masses are determined using Refs.~\cite{Tiburzi:2005is,Chen:2001yi,Beane:2002vq,WalkerLoud:2004hf,Tiburzi:2004rh,Tiburzi:2005na,WalkerLoud:2006sa}.  We use these formulae to perform the chiral extrapolation analysis of our baryon mass results, that we turn to after a brief digression on the expected convergence of $SU(3)$ HB$\chi$PT.

%
%
\subsubsection{On the convergence of $SU(3)$ heavy baryon $\chi$PT}
The expansion parameter of $\chi$PT~\cite{Weinberg:1978kz,Gasser:1983yg,Gasser:1984gg} in the meson sector is given in terms of generic meson masses in the set $\{m_\phi: m_\pi, m_K, m_\eta\}$, as
\begin{equation}
	\varepsilon_\phi \sim \frac{m_\phi^2}{\Lambda_\chi^2}\, .
 \end{equation}
The inclusion of heavy matter fields in the effective theory, such as baryons or heavy mesons, leads to the expansion parameter
\begin{equation}
	\varepsilon_H \sim \frac{m_\phi}{\Lambda_\chi}\, ,
\end{equation}
for which the convergence of the theory becomes worse and/or questionable.  In fact, using general knowledge of asymptotic series and taking the ratio of physical parameters, $m_\eta/\Lambda_\chi \sim 1/2$, one may expect that $SU(3)$ heavy baryon $\chi$PT would only have a chance of converging for 2-3 orders before the asymptotic nature of the theory became apparent.  This problem is exacerbated with current lattice calculations, in which the meson masses are heavier than those in nature, casting doubt on the ability to reliably extrapolate observable quantities computed on the lattice, using three-flavor heavy baryon $\chi$PT, or any of its generalizations.%
\footnote{Those generalizations include explicit inclusion of the decuplet degrees of freedom, partially quenched or mixed action generalizations for the lattice, covariant baryon $\chi$PT with infrared regularization~\cite{Becher:1999he}, or some other regulator like finite range regularization~\cite{Donoghue:1998rp,Leinweber:1999ig,Young:2002ib,Leinweber:2003dg}.} 

Recently, the first calculation of the hyperon axial couplings from the lattice was performed, in which it was found that $SU(3)$ heavy baryon $\chi$PT at NLO failed to describe the lattice results~\cite{Lin:2007ap}.  Additionally, the RBC/UKQCD Collaborations have advocated the use of two-flavor chiral extrapolations for their recent calculations of meson quantities including the decay constants as well as the kaon bag parameter~\cite{Allton:2008pn}.  In this work, we find that both $SU(3)$ heavy baryon $\chi$PT as well as its mixed action generalization describe the lattice results of the baryon masses and mass splittings, with reasonable $\chi^2$ per d.o.f.  However, these fits (presented in Sec.~\ref{sec:SU3extraps}) return values of the axial coupling LECs, $D,F$ and $C$ that are far from the recent lattice calculations~\cite{Alexandrou:2007zz,Lin:2007ap} as well as the  phenomenologically determined values~\cite{Butler:1992pn,Savage:1996zd,FloresMendieta:1998ii,Cabibbo:2003cu,Ratcliffe:2004jt}, which are in agreement with each other.  For example, matching $SU(3)$ onto $SU(2)$ for the nucleon mass, one expects $D+F = g_A + \mc{O}(m_s)$.  Furthermore, when the resulting formulae are used to extrapolate to the physical point, they are in disagreement with the experimentally measured values of the baryon mass splittings.  By comparing the predicted NLO contributions to the mass splittings of the octet baryons with their experimentally determined values, we demonstrate that a fit that only includes $\mathcal{O}(m_\phi^3)$, or NLO terms, is doomed to fail to reproduce these LECs, and also it is most probable that the $SU(3)$ heavy baryon $\chi$PT determination of the mass splittings as well as the masses fails to converge.  In Table~\ref{tab:predNLOmassSplit}, we present the experimental knowledge of the octet baryon masses and mass splittings, as well as the predicted NLO contribution to the masses (at the physical point), \textit{i.e.} we have set $\alpha_M = \beta_M = \sigma_M = 0$ (see Eq.~\eqref{eq:BTLag}).  We have chosen the charge neutral baryon masses, as our lattice calculations in which electromagnetism is turned off are most similar to these.  To determine the central values and errors of the various $\d M_B^{(3/2)}$, we have used the values of the axial couplings $D=0.715(50)$ and $F=0.453(50)$, which were taken from Ref.~\cite{Lin:2007ap} (we have inflated the uncertainties to be consistent with those in Ref.~\cite{Jenkins:1991es,Butler:1992pn,Savage:1996zd,FloresMendieta:1998ii,Cabibbo:2003cu,Ratcliffe:2004jt}) and we have used $C=1.2(2)$ that is consistent with Refs.~\cite{Butler:1992pn,Alexandrou:2007zz}.  We have performed this exercise with heavy baryon $\chi$PT both with and without $(\Delta \hskip-0.6em\slash)$ explicit decuplet degrees of freedom.  
%
%
\begin{table}[t]
\caption{\label{tab:predNLOmassSplit} NLO, $\delta M_B^{(3/2)}$, contributions to octet baryon masses and mass splittings in heavy baryon $\chi$PT both with decuplet degrees of freedom and without $(\Delta \hskip-0.6em\slash)$.  We use the charge-neutral baryons in these relations as the lattice computations without electromagnetism are most similar to these.  To determine the predicted NLO mass contributions and their corresponding errors, we use $D=0.715(50)$, $F=0.453(50)$ and $C=1.2(2)$.}
\begin{ruledtabular}
\begin{tabular}{|c|c|cc||c|c|cc|}
Quantity & Experimental & HB$\chi$PT& $\Delta \hskip-0.6em\slash \ (C=0)$& Quantity& Experimental & HB$\chi$PT& $\Delta \hskip-0.6em\slash \ (C=0)$ \\
& & $\d M_{B^\prime}^{(3/2)} - \d M_B^{(3/2)} $ & $\d M_{B^\prime}^{(3/2)} - \d M_B^{(3/2)} $ &&&
$\delta M_B^{(3/2)}$ & $\delta M_B^{(3/2)}$ \\
\hline
$M_\Lambda -M_N$ (MeV) & 176 & -285(65) & -326(65) 
& $M_N$ (MeV)& 940 & -195(38) & -278(38)  \\
$M_ \Sigma - M_N$ (MeV)& 253 & -152(60) & -287(46) 
& $M_\Lambda$ (MeV)& 1116 & -480(84) & -604(84) \\
$M_\Xi -M_N$ (MeV)& 375 & -516(120) & -637(120) 
& $M_\Sigma$ (MeV)& 1193 & -347(79) & -565(66) \\
$M_\Sigma - M_ \Lambda $ (MeV)& 77 & 133(76) & 39(76)
& $M_\Xi$ (MeV)& 1315 & -711(124) & -915(124) \\
$M_\Xi - M_ \Lambda $ (MeV)& 199 & -232(39) & -311(39) &&&& \\
$M_\Xi - M_ \Sigma $ (MeV)& 122 & -367(81) & -351(81) &&&& \\ 
\hline
$\Delta_{GMO}$ (MeV)& 10 & 9(4) & 3(4) &&&&
\end{tabular}
\end{ruledtabular}
\end{table}

\begin{figure}
\includegraphics[width=0.48\textwidth]{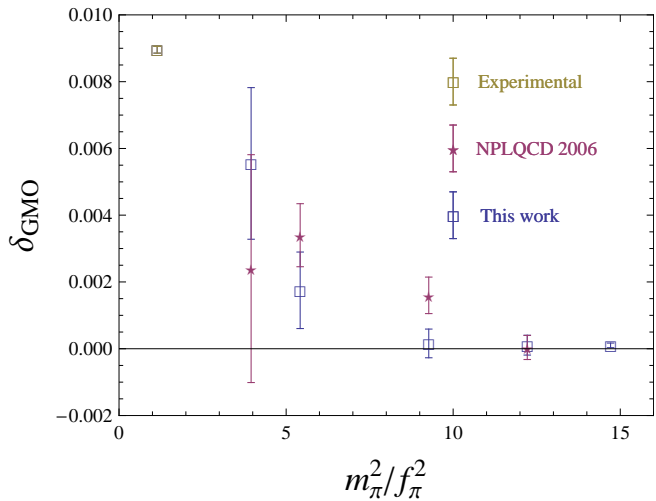}
\caption{\label{fig:GMO} The Gell-Mann--Okubo (GMO) mass ratio with $\delta_{\rm GMO}=\frac{\D_{GMO}}{M_B}$, where $\D_{GMO}$ is defined in Eq.~\eqref{eq:DGMO} and the centroid mass is $M_B = \frac{1}{8}M_\L+\frac{3}{8}M_\S+\frac{1}{4}M_N+\frac{1}{4}M_\Xi$.  Here we collect the results of this work, as well as those of NPLQCD~\cite{Beane:2006pt}.  We also put the experimental number in the figure, using the charge-neutral baryon masses.}
\end{figure}

As can be seen in Table~\ref{tab:predNLOmassSplit}, the predicted NLO contributions to the various mass splittings are in all but one case larger in magnitude than the actual splitting itself, and generally opposite in sign.  To accommodate these NLO mass corrections and the experimental results, the LO contributions would have to be larger still and they would not lead to the expected hierarchy of higher order contributions, $\d M_{B^\prime}^{(3/2)} - \d M_{B}^{(3/2)} \ll \d M_{B^\prime}^{(1)} - \d M_B^{(1)}$ essential to the convergence of the expansion.  The explicit inclusion of the decuplet degrees of freedom generally improves the situation but not enough to alleviate concerns of convergence.%
\footnote{There is a subtle issue that arises with the inclusion of the decuplet states, requiring the inclusion of an extra chiral-singlet parameter in the theory, the mass splitting of the decuplet and octet baryons in the chiral limit.  Because of this extra parameter, one can never completely disentangle the LO and NLO contributions to the baryon masses as the new parameter leads to an arbitrary finite renormalization of all the existing LECs.  One can therefore only make rigorous statements about the complete mass calculation to a given order and not the absolute size of a given order in $m_q$~\cite{WalkerLoud:2004hf,Tiburzi:2004rh,Tiburzi:2005na}.} 
This was discussed in some detail in Ref.~\cite{Bernard:1993nj}.  
One exception to this picture is the Gell-Mann--Okubo (GMO) formula, where we have defined the mass splittings,
\begin{equation}\label{eq:DGMO}
	\Delta_{GMO} = M_\L +\frac{1}{3}M_\S -\frac{2}{3}M_N -\frac{2}{3}M_\Xi\, .
\end{equation}
The NLO contributions to the octet baryon masses provide the leading correction to GMO formula, and as can be seen in Table~\ref{tab:predNLOmassSplit}, the $SU(3)$ HB$\chi$PT formula is in remarkable agreement with the experimental measurements.  This can be understood in part from the fact that the quark mass matrix transforms as an $\mathbf{8}\oplus\mathbf{1}$ under the $SU(3)$ flavor symmetry.  Therefore, a single insertion of the quark mass matrix in the Lagrangian, such as the operators in Eq.~\eqref{eq:BTLag} with coefficients $\a_M$ and $\b_M$, will automatically satisfy the GMO relation.  Violations of the GMO relation must come from higher dimensional representations, the first of which is the $\mathbf{27}$ and appears at NLO in the HB$\chi$PT expansion.  In Ref.~\cite{Beane:2006pt}, the NPLQCD Collaboration provides a more detailed discussion of this topic and presents a detailed calculation of the violations of the GMO relation, finding the agreement between the NLO HB$\chi$PT formula and the lattice calculation holds for heavier quark masses as well.  In Figure~\ref{fig:GMO} we display our resulting calculations of the deviation of the GMO relation, divided by the centroid octet mass, $M_B = \frac{1}{8}M_\L+\frac{3}{8}M_\S+\frac{1}{4}M_N+\frac{1}{4}M_\Xi$, following NPLQCD, whose results we find agreement with and are plotted along side ours.

%
%

One must be careful in making this convergence analysis.  The leading axial couplings in the Lagrangian, $D$, $F$, and $C$ are not physical observables.  However, with sufficiently light \textit{up}, \textit{down} and \textit{strange} quark masses, one would expect the fits to give values for these LECs which are fairly stable to the inclusion of higher orders in the chiral expansion.  These axial couplings provide the leading order contribution to the axial matrix elements that give, $g_A=D+F$, $g_{\L \S}=2D$, $g_{\S\S}=2F$ and $g_{\Xi\Xi}=D-F$ (following the normalization of Ref.~\cite{Jiang:2008aqa}).  To fully test issues of convergence for $SU(3)$ HB$\chi$PT, one should perform a combined analysis of the octet baryon axial charges and masses, using the NNLO expressions for both.  This is beyond the scope of this work.  We now proceed to perform fits to our octet mass results using the HB$\chi$PT formula.

%
%
\subsubsection{Baryon mass extrapolations \label{sec:SU3extraps}}

In this section, we perform a large variety of chiral extrapolation analyses to our calculated octet and decuplet baryon masses.  We find that, whereas the resulting minimizations are statistically consistent with the lattice results, as measured by the $\chi^2$ per d.o.f., in general, the resulting chiral extrapolations are in disagreement with the physical masses and mass splittings.  We perform both the continuum $SU(3)$ analysis as well as the mixed-action analysis, using the mass formulae that can be determined from Eq.~\eqref{eq:BTLag}, for example Eq.~\eqref{eq:MN_MANLO}.  Before presenting our results, we need to note that it is known the strange quark mass on these MILC lattices is too large~\cite{Aubin:2004ck,Bernard:2007ps}.  As we have used only one value for the strange quark mass, we can not control the strange quark mass extrapolation.  Attempting to correct for this leads to a reduction in all the baryon masses containing strange quarks, which provides for a larger disagreement between our extrapolated baryon mass splittings and the physical mass splittings as evidenced from the resulting fits presented in Tables~\ref{tab:SU2_LOMBmnsMN} to \ref{tab:SU63_NLOMT}.

\begin{figure}[t]
\begin{tabular}{cc}
\includegraphics[width=0.48\textwidth]{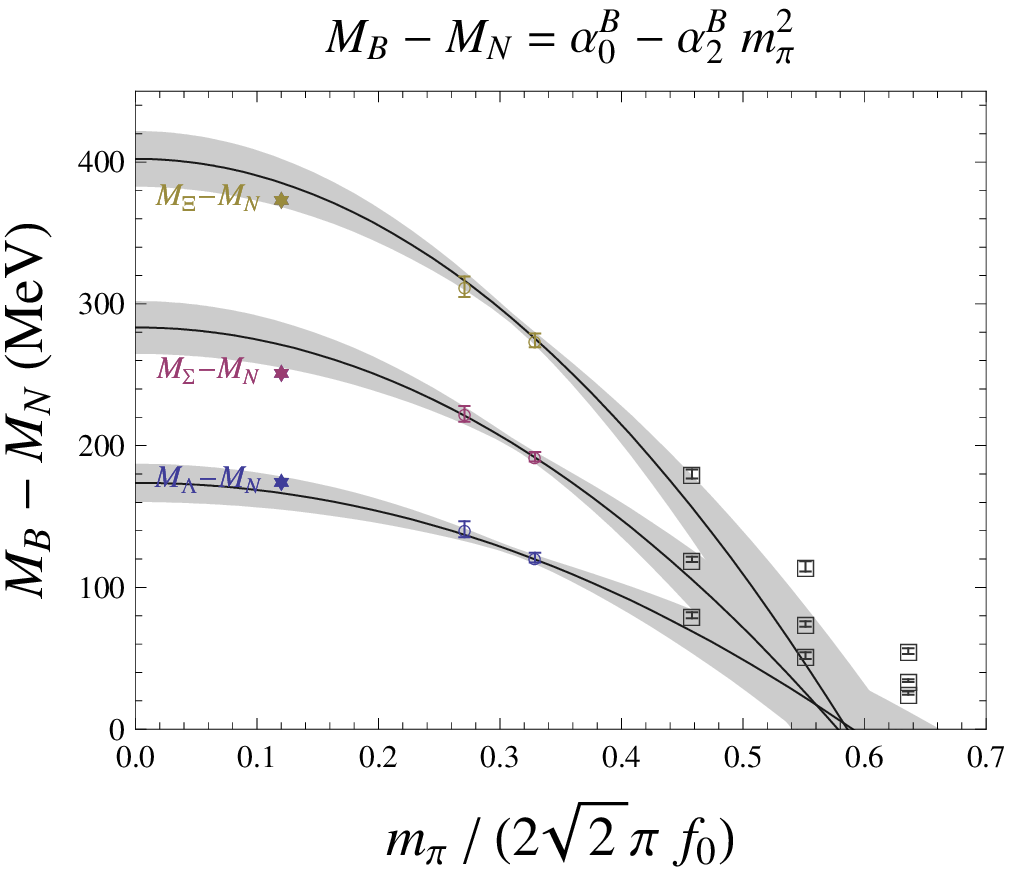}
&\includegraphics[width=0.48\textwidth]{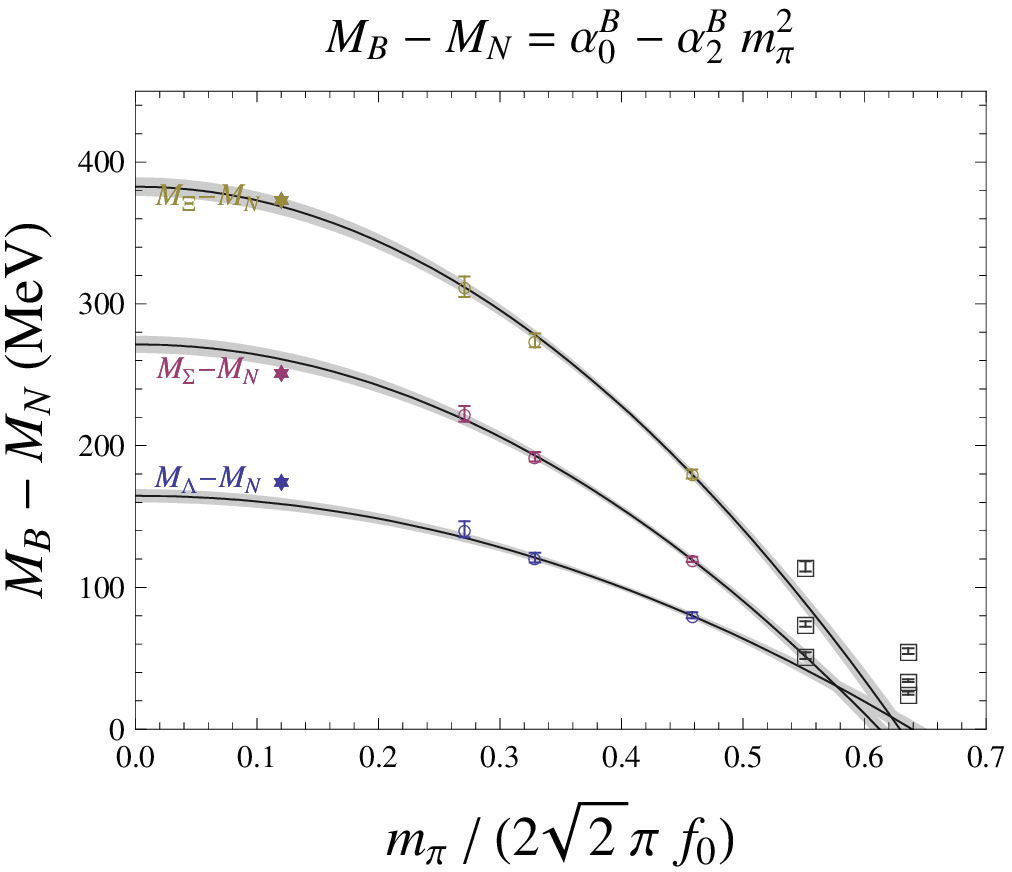}
\\
$(a)$ & $(b)$
\end{tabular}
\caption{\label{fig:MBmnsMNSU2fits} We display the LO $SU(2)$ fits to the mass splittings.  The left plot, $(a)$ is the result of fitting only to the lightest two mass points plus the 68\% confidence bands.  The right plot, $(b)$ is the result of fitting to the lightest three mass points.  The stars represent the physical baryon mass splittings and are not included in the analysis.}
\end{figure}

We begin with a LO two-flavor extrapolation of the octet baryon mass splittings, in which the strange quark is integrated out, but we enforce $SU(3)$ symmetry.  We perform the fit to $M_\L-M_N$, $M_\S-M_N$ and $M_\Xi-M_N$ using the following formula
\begin{align}\label{eq:SU2_BMasses}
	M_\L-M_N &= \tilde{\a}_s -\frac{1}{2}\tilde{\a}_u m_\pi^2\, ,
	\nonumber\\
	M_\S-M_N &= \frac{1}{3}\tilde{\a}_s +\frac{4}{3}\tilde{\b}_s 
		-\left(\frac{1}{6}\tilde{\a}_u +\frac{2}{3}\tilde{\b}_u \right) m_\pi^2\, ,
	\nonumber\\
	M_\Xi-M_N &= \frac{5}{3}\tilde{\a}_s +\frac{2}{3}\tilde{\b}_s 
		-\left(\frac{5}{6}\tilde{\a}_u +\frac{1}{3}\tilde{\b}_u \right) m_\pi^2\, ,
\end{align}
where in terms of the LECs of Eq.~\eqref{eq:BTLag}, we have
\begin{align}\label{eq:SU23_LECs}
	&\tilde{\a}_s = -\a_M m_s\, ,&
	&\tilde{\b}_s = -\b_M m_s\, ,&
	&\tilde{\a}_u = -\frac{\a_M}{B_0}\, ,&
	&\tilde{\b}_u = -\frac{\b_M}{B_0}\, .&
\end{align}
The results of this analysis are collected in Table~\ref{tab:SU2_LOMBmnsMN} and in Figure~\ref{fig:MBmnsMNSU2fits} we display the resulting chiral extrapolations from the analysis using the lightest two and three quark mass ensembles.
%
%
\begin{table}[b]
\caption{\label{tab:SU2_LOMBmnsMN} Results from LO $SU(2)$ extrapolations with $SU(3)$ symmetry, Eqs.~\eqref{eq:SU2_BMasses} and \eqref{eq:SU23_LECs}.}
\begin{ruledtabular}
\begin{tabular}{clcccccc}
FIT: LO $SU(2)$& range & $\tilde{\a}_s$ & $\tilde{\b}_s$ & $\tilde{\a}_u$ & $\tilde{\b}_u$ & $\chi^2$ & d.o.f. \\
&& $[\textrm{MeV}]$&$[\textrm{MeV}]$&$[\textrm{MeV}^{-1}]$&$[\textrm{MeV}^{-1}]$ \\
\hline
$M_B - M_N$ 
&007--010 & 174(13)& 169(15)& 0.00085(22)& 0.00087(25)& 1.1& 2 \\
&007--020 & 164(04)& 162(04)& 0.00069(04)& 0.00075(04)& 2.5& 5 \\
&007--030 & 158(03)& 150(03)& 0.00061(02)& 0.00061(02)& 42.4& 8 \\
\end{tabular}
\begin{tabular}{lcccccc}
``Predictions" & $M_\L -M_N$ & $M_\S - M_N$ & $M_\Xi -M_N$ 
& $M_\S - M_\L$ & $M_\Xi - M_\L$ & $M_\Xi - M_\S$  \\
Phys. MeV & 176 & 254 & 376 & 77 & 199 & 122 \\ 
\hline
007--010 & 166(11)& 270(15)& 384(16)& 104(21)& 218(8)& 114(20)  \\
007--020 & 158(04)& 260(05)& 367(05)& 102(07)& 209(3)& 107(07) \\
007--030 & 152(03)& 243(04)& 350(04)& 90(05)& 197(2)& 107(05) \\
\end{tabular}
\end{ruledtabular}
\end{table}

%
%
\begin{figure}
\begin{tabular}{cc}
\includegraphics[width=0.45\textwidth]{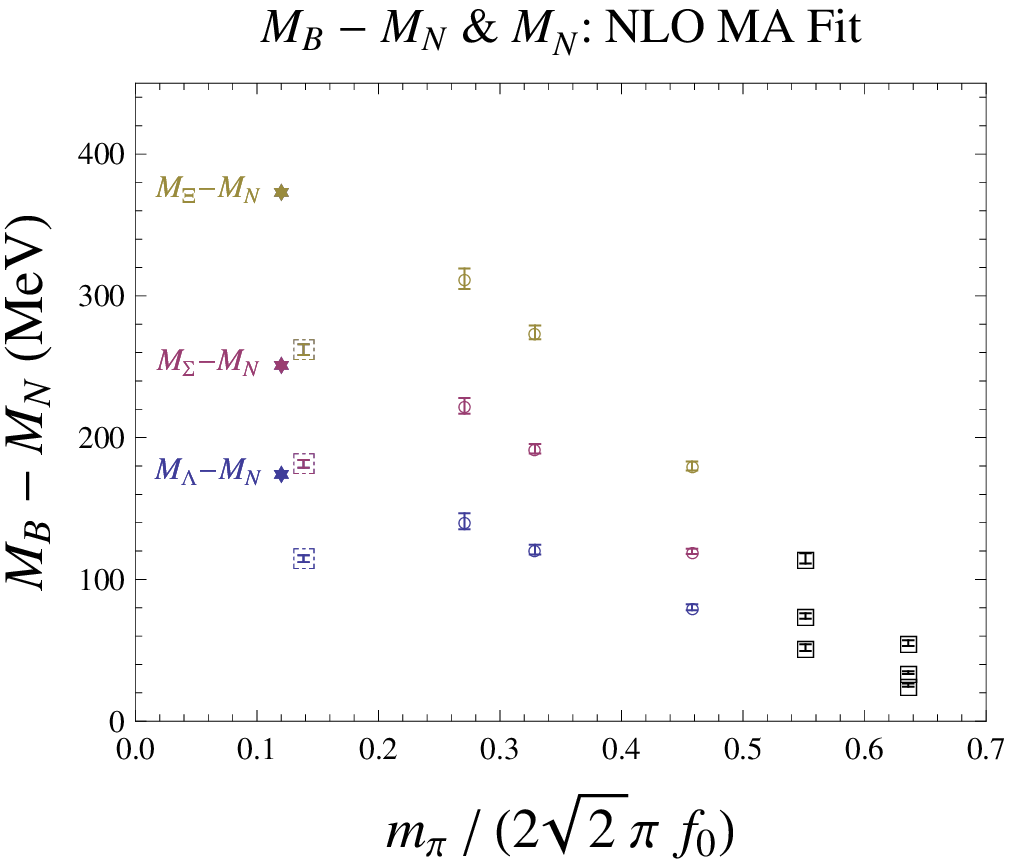}
&
\includegraphics[width=0.45\textwidth]{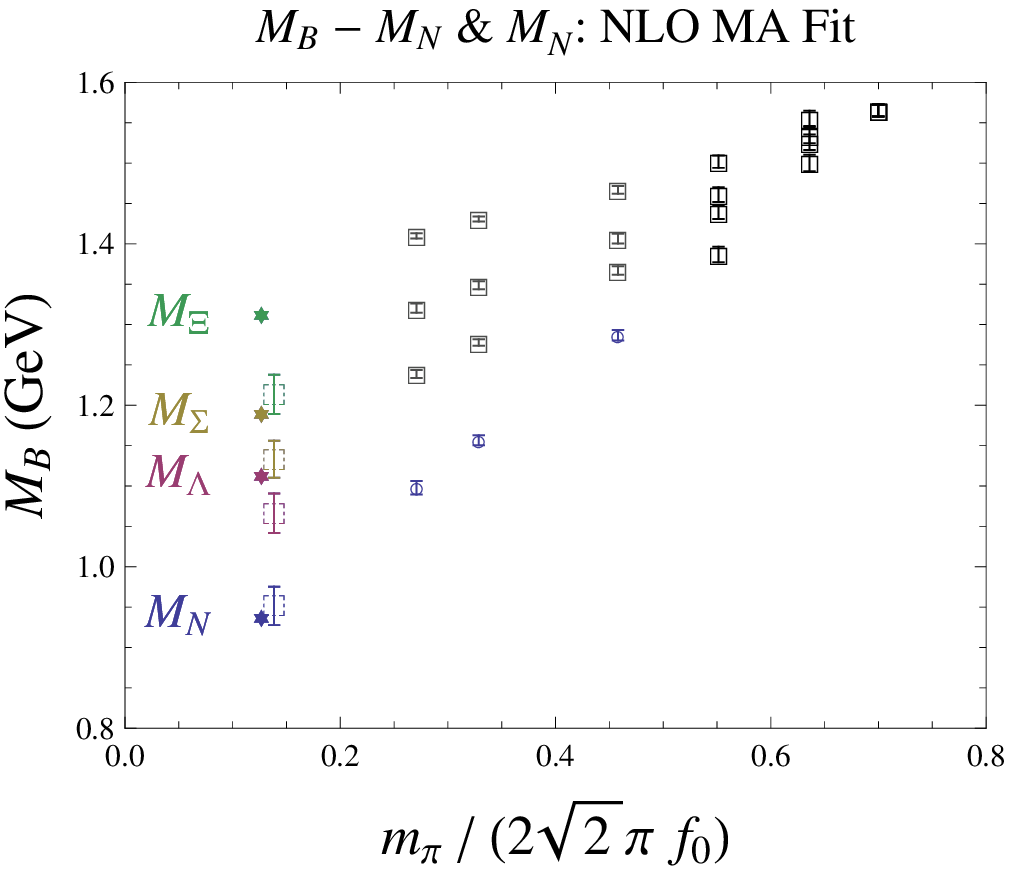}
\\
$(a)$ & $(b)$
\\ \\
\includegraphics[width=0.45\textwidth]{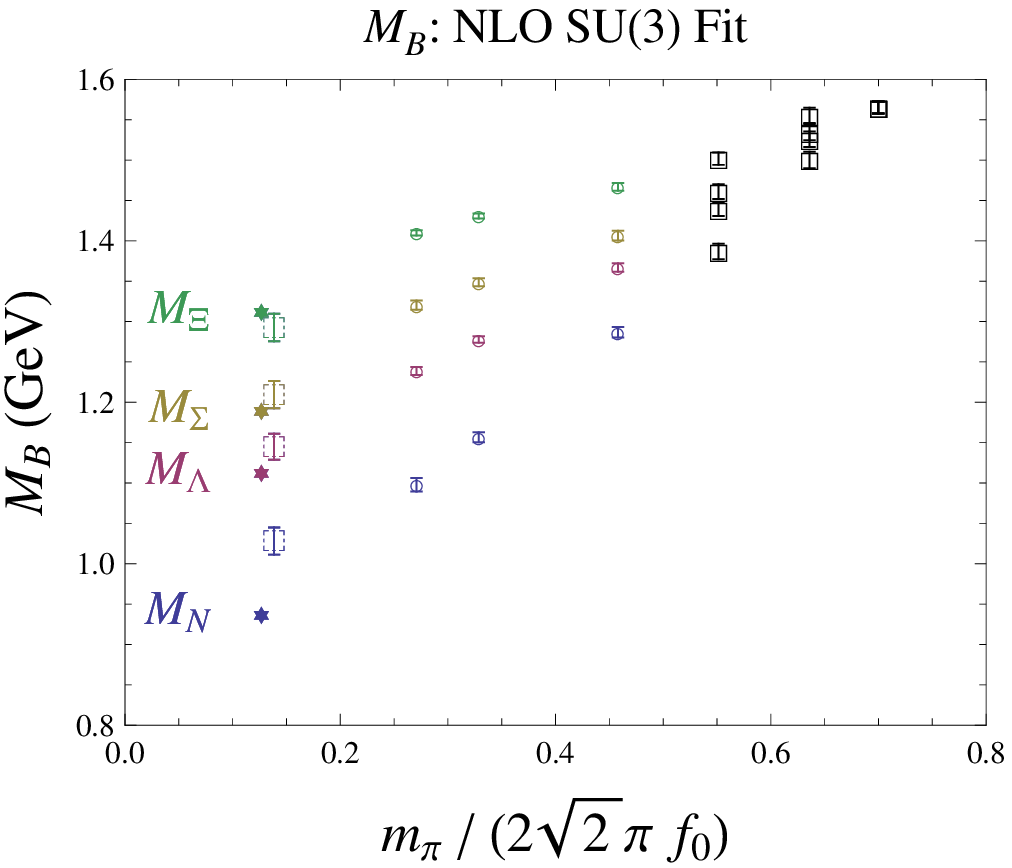}
&\includegraphics[width=0.45\textwidth]{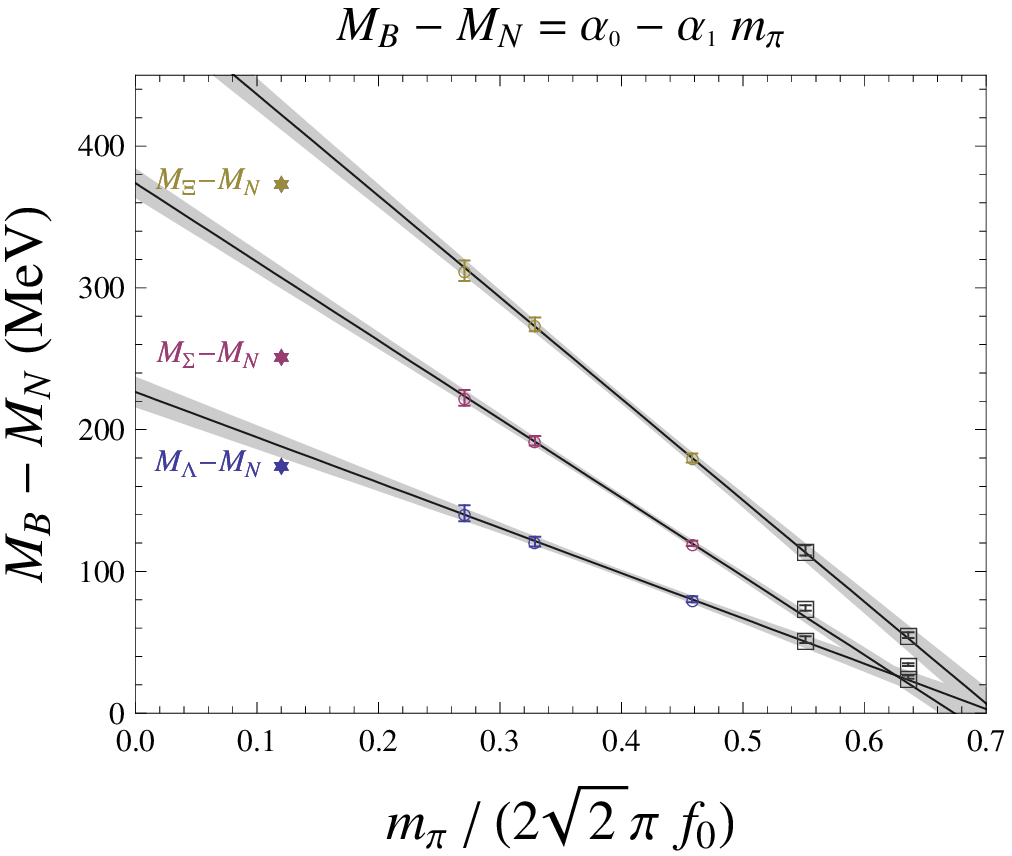}
\\
$(c)$ & $(d)$
\end{tabular}
\caption{\label{fig:MBmnsMNSU63fits} We display various chiral extrapolations of the octet baryon masses and mass splittings.  In all fits, the (gray) squares are points not included in the analysis, the (colored) open circles with error bars are the points that are included and the (colored) stars are the physical masses/splittings, which are never included in the minimization.  The dashed (colored) squares with error bars are the resulting predictions, slightly displaced horizontally for clarity.  Figs.~$(a)$ and $(b)$ are from a combined fit of $M_B-M_N$ and $M_N$, fit to the lightest three quark mass ensembles, using the NLO mixed-action extrapolation formulae.  Fig.~$(c)$ is the result of the NLO mixed-action fit to the octet baryon masses.  Fig.~$(d)$ is the result of the straight line analysis.  As can be seen, the straight line fit reproduces the lattice results not included in the minimization but conflicts dramatically with experiment.  The error bars/bands represent the 68\% confidence interval from the statistical uncertainty.}
\end{figure}

We see that the fits to the lightest two and three quark mass ensembles have a good $\chi^2$ per d.o.f., while the inclusion of the fourth lightest point, the $m030$ ensemble, results in a poor fit.  Furthermore, these two-flavor extrapolations give mass splittings at the physical pion mass that are reasonably close to the physical octet mass splittings.  Next, we perform a variety of NLO analysis using both the continuum $SU(3)$ and the mixed-action formulae;
\begin{itemize}
\item
$M_B-M_N$, Table~\ref{tab:SU63_NLOMBmnsMN}

\item
$\frac{M_B-M_N}{f_\pi}$, Table~\ref{tab:SU63_NLOMBmnsMNfU}.  Dividing by $f_\pi$ removes scale setting ambiguities.

\item
$\frac{M_B-M_N}{f_K}$.  Dividing by $f_K$ also removes scale setting ambiguities, but more importantly $f_K$ has significantly milder chiral corrections than $f_\pi$.  Therefore, since the chiral corrections to $f$ appear at NNLO in the mass splittings, beyond the order of this fit, using $f_K$ instead of $f_\pi$ provides more stable fits, as seen in Table~\ref{tab:SU63_NLOMBmnsMNfU}.

\item
$M_B-M_N$ and $M_N$, Table~\ref{tab:SU63_NLOMBmnsMNMN}.

\item
$M_N$, $M_\L$, $M_\S$ and $M_\Xi$, Table~\ref{tab:SU63_NLOMB}.

\item
$M_B-M_N$ and $M_N$ at NNLO.  We do not find a stable minimization for this fit, as there are 18 unknown LECs that must be determined.

\item
$M_B = \a_0^B +\a_1^B m_\pi$, the straight line analysis as for the nucleon, Table~\ref{tab:MB_mpiFit}.

\end{itemize}
In Figure~\ref{fig:MBmnsMNSU63fits} we display some of the resulting extrapolations.  In all cases, the (gray) boxes denote points that were not included in the minimization.  The (colored) circles with error bars are points that were included, and the dashed boxes with error bars are the resulting predictions at the physical pion mass, slightly displaced horizontally for clarity.  The (colored) stars are the physical masses (mass splittings) that are never included in the minimization analysis.  We perform a similar analysis for the decuplet baryon masses, with the same caveat mentioned for the delta mass in Sec.~\ref{sec:MDelta}. Aside from the $\O^-$, the other decuplet states have similar volume issues as the delta discussed in Sec.~\ref{sec:MDelta}.  The resulting fits are collected in Tables~\ref{tab:SU63_NLOMTmnsMD} to \ref{tab:SU63_NLOMT}.

%
%

There are a few points worth mentioning.  First, as measured by the $\chi^2$ per d.o.f., most of the fits presented in Tables~\ref{tab:SU2_LOMBmnsMN} to \ref{tab:SU63_NLOMT} provide a good description of our lattice results, and are in disagreement with the physical baryon masses and mass splittings.  One should not place much emphasis on the disagreement with the physical masses.  First, as mentioned above, the strange quark mass is known to be too large~\cite{Aubin:2004ck,Bernard:2007ps}.  Furthermore, the issues of convergence of $SU(3)$ HB$\chi$PT may lead to large NNLO corrections, or worse, indicate a lack of convergence.  This problem has been discussed in some detail in Ref.~\cite{Donoghue:1998rp,Borasoy:1996bx}.  Support of this statement is found in the LO analysis presented in Table~\ref{tab:SU2_LOMBmnsMN} (and displayed in Fig.~\ref{fig:MBmnsMNSU2fits}).  The resulting LECs and predicted baryon masses (mass splittings) from the various fits are all in reasonable agreement with each other, despite providing axial couplings, $D$, $F$, $C$ and $H$ that are in stark disagreement with their known phenomenological values.  We should mention that the inclusion of the NNLO terms of $\mc{O}(m_\phi^4)$, may provide stability to the fits, such that the values of these couplings are in closer agreement with phenomenology.  However, performing the NNLO analysis with $D$, $F$, $C$ and $H$ fixed, we also do not find a stable minimization.%
\footnote{Holding $H$ fixed may not be ideal regardless, as in a recent investigation of $H$ in $SU(2)$ $\chi$PT (matching provides $H=g_{\D\D}$), it was found that $H$ has a large expected quark mass dependence~\cite{Jiang:2008we}.} 
Examining Tables~\ref{tab:SU63_NLOMTmnsMD} through \ref{tab:SU63_NLOMT}, it is clear that the decuplet extrapolation is in far worse condition than the octet masses.  Understanding the decuplet masses will require a multiple volume study, which is beyond the scope of this work.
Lastly, we note that for the straight line in $m_\pi$ analysis, using $M_B = \a_0^B +\a_1^B m_\pi$, we find (see Table~\ref{tab:MB_mpiFit}) that the resulting values of the parameters $\a_1^B$ are approximately simple fractions: $\a_1^N\sim 1$, $\a_1^\L\sim 2/3$, $\a_1^\S\sim 1/2$ and $\a_1^\Xi\sim 1/3$.  Determining whether this is a phenomenon of QCD or perhaps a combined finite-volume--lattice-spacing artifact will require further investigation with multiple lattice spacings and volumes.

%
%
\begin{table}[h]
\caption{\label{tab:MB_mpiFit} Results from straight-line in $m_\pi$ fit of the octet masses.  A noteworthy feature is the coefficient of the $m_\pi$ term for each of the masses.}
\begin{ruledtabular}
\begin{tabular}{ccccccccc}
$M_B = \a_0^B +\a_1^B m_\pi$ & $M_N$ & $M_\L$ & $M_\S$ & $M_\Xi$ & $M_N$ & $M_\L$ & $M_\S$ & $M_\Xi$ \\
\cline{2-5}\cline{6-9}
&\multicolumn{4}{c}{$\a_0^B [\textrm{GeV}]$}&\multicolumn{4}{c}{$\a_1^B$} \\
\cline{1-1}\cline{2-5}\cline{6-9}
007--030 & 0.82(2) & 1.05(1)& 1.19(1)& 1.33(1)& 0.94(4) & 0.65(3)& 0.45(3)& 0.29(2) \\
007--040 & 0.80(1) & 1.03(1)& 1.17(1)& 1.32(1)& 0.99(3) & 0.69(2)& 0.50(2)& 0.32(2) \\
007--050 & 0.79(1) & 1.03(1)& 1.16(1)& 1.31(1)& 1.01(2) & 0.70(2)&0.52(2)& 0.33(1)
\end{tabular}
\end{ruledtabular}
\end{table}

%
%
\begin{table}[h]
\caption{\label{tab:SU63_NLOMBmnsMN} Results from NLO  bootstrap $\chi$-extrapolations of the octet mass splittings, using mixed action (MA) and $SU(3)$ heavy baryon $\chi$PT.}
\begin{ruledtabular}
\begin{tabular}{clccccccc}
FIT: NLO & range & $\a_M$ & $\b_M$ & C & D & F & $\chi^2$ & d.o.f. \\
&& $[\textrm{GeV}^{-1}]$&$[\textrm{GeV}^{-1}]$&& \\
\hline
$M_B - M_N$ &007--020: MA &-0.49(04)& -0.43(03)& 0.38(3)& 0.07(07)& 0.03(3)& 10.1& 4 \\
& 007--020: $SU(3)$ & -0.87(38)& -0.40(18)& 0.36(3)& 0.23(19) & 0.27(9)& 3.1& 4\\
& 007--030: MA & -0.48(03)& -0.43(03)& 0.38(2)& 0.06(07)& 0.02(3) & 11.4& 7 \\
& 007--030: $SU(3)$ & -0.77(32)& -0.32(19)& 0.37(2)& 0.15(19)& 0.27(8)& 4.8& 7 \\
& 007--040: MA & -0.48(04)& -0.44(04)& 0.38(2)& 0.10(08)& 0.03(3)& 11.6& 10 \\
& 007--040: $SU(3)$ & -0.86(36)& -0.38(20)& 0.37(2)& 0.21(20)& 0.28(8) & 8.8& 10 \\
\end{tabular}
\begin{tabular}{lcccccc}
``Predictions" & $M_\L -M_N$ & $M_\S - M_N$ & $M_\Xi -M_N$ 
& $M_\S - M_\L$ & $M_\Xi - M_\L$ & $M_\Xi - M_\S$  \\
Phys. MeV & 176 & 254 & 376 & 77 & 199 & 122 \\ 
\hline
007--020: MA & 115(3)& 182(4)& 262(4)& 68(5)& 148(3)& 80(4) \\
007--020: $SU(3)$ & 131(13)& 195(17)& 286(27)& 64(4)& 154(14)& 90(11) \\
007--030: MA & 115(3)& 182(4)& 262(4)& 68(4)& 147(3)& 79(4) \\
007--030: $SU(3)$ & 127(9)& 193(14)& 276(19)& 66(7)& 149(11)& 83(5) \\
\end{tabular}
\end{ruledtabular}
\end{table}

%
%
\begin{table}[h]
\caption{\label{tab:SU63_NLOMBmnsMNfU} Results from NLO extrapolations of the octet mass splittings in $f_\pi$ and $f_K$ units, using mixed action (MA) and $SU(3)$ heavy baryon $\chi$PT.}
\begin{ruledtabular}
\begin{tabular}{clccccccc}
FIT: NLO & range & $\a_M$ & $\b_M$ & C & D & F & $\chi^2$ & d.o.f. \\
&& $[\textrm{GeV}^{-1}]$&$[\textrm{GeV}^{-1}]$&& \\
\hline
$\frac{M_B - M_N}{f_\pi}$ &007--020: MA 
& -1.7(4)& -0.8(2)& 0.00(41)& 0.50(05)& 0.33(09)& 2.2& 4 \\
& 007--020: $SU(3)$ & -0.6(2)& -0.5(3)& 0.27(08)& 0.09(80)& 0.04(26)& 3.1& 4\\
& 007--030: MA & -0.6(8)& -0.5(2)& 0.25(10)& 0.08(51)& 0.10(70)& 5.1& 7 \\
& 007--030: $SU(3)$ & -0.6(2)& -0.6(2)& 0.25(06)& 0.20(24)& 0.00(16)& 4.8& 7 \\
& 007--040: MA & -0.6(3)& -0.5(4)& 0.23(08)& 0.1(1.1)& 0.02(35)& 10.6& 10 \\
& 007--040: $SU(3)$ & -0.6(2)& -0.6(2)& 0.21(07)& 0.32(15)& 0.00(10)& 8.8& 10 \\
\end{tabular}
\begin{tabular}{lcccccc}
``Predictions" & $M_\L -M_N$ & $M_\S - M_N$ & $M_\Xi -M_N$ 
& $M_\S - M_\L$ & $M_\Xi - M_\L$ & $M_\Xi - M_\S$  \\
Phys. MeV & 176 & 254 & 376 & 77 & 199 & 122 \\ 
\hline
007--020: MA & 220(43)& 233(09)& 443(55)& 13(48)& 223(13)& 210(58) \\
007--020: $SU(3)$ & 134(8)& 209(16)& 306(21)& 74(14)& 171(16)& 97(09) \\
007--030: MA & 139(61)& 210(14)& 313(93)& 71(50)& 174(33)& 103(81) \\
007--030: $SU(3)$ & 133(8)& 210(09)& 306(14)& 78(10)& 173(09)& 95(09) \\
\end{tabular}
\begin{tabular}{clccccccc}
FIT: NLO & range & $\a_M$ & $\b_M$ & C & D & F & $\chi^2$ & d.o.f. \\
&& $[\textrm{GeV}^{-1}]$&$[\textrm{GeV}^{-1}]$&& \\
\hline
$\frac{M_B - M_N}{f_K}$ &007--020: MA 
& -1.1(2)& -0.8(2)& 0.14(19) & 0.57(13)& 0.24(06)& 4.0& 4 \\
& 007--020: $SU(3)$ & -0.8(1)& -0.8(2)& 0.19(14)& 0.53(13)& 0.18(04)& 5.4& 4\\
& 007--030: MA & -0.9(2)& -0.8(2)& 0.24(10)& 0.48(14)& 0.18(05)& 7.4& 7 \\
& 007--030: $SU(3)$ & -0.7(1)& -0.7(2)& 0.26(10)& 0.45(13)& 0.14(04)& 8.0& 7 \\
& 007--040: MA & -0.8(1)& -0.9(2)& 0.19(12)& 0.52(14)& 0.16(04)& 12.4& 10 \\
& 007--040: $SU(3)$ & -0.7(1)& -0.8(2)& 0.20(11)& 0.48(12)& 0.12(03)& 12.9& 10 \\
\end{tabular}
\begin{tabular}{lcccccc}
``Predictions" & $M_\L -M_N$ & $M_\S - M_N$ & $M_\Xi -M_N$ 
& $M_\S - M_\L$ & $M_\Xi - M_\L$ & $M_\Xi - M_\S$  \\
Phys. MeV & 176 & 254 & 376 & 77 & 199 & 122 \\ 
\hline
007--020: MA & 168(20)& 234(25)& 373(32)& 66(33)& 205(19)& 139(29) \\
007--020: $SU(3)$ & 136(07)& 221(14)& 321(17)& 85(13)& 185(13)& 100(07) \\
007--030: MA & 114(15)& 223(23)& 338(26)& 74(27)& 189(17)& 115(22) \\
007--030: $SU(3)$ & 131(06)& 211(12)& 307(14)& 80(12)& 175(11)& 95(06) \\
\end{tabular}
\end{ruledtabular}
\end{table}

%
%
\begin{table}[h]
\caption{\label{tab:SU63_NLOMBmnsMNMN} Results from NLO bootstrap $\chi$-extrapolations of the octet mass splittings combined with the nucleon mass, using mixed action (MA) and $SU(3)$ heavy baryon $\chi$PT.}
\begin{ruledtabular}
\begin{tabular}{clccccccccc}
FIT: NLO & range & $M_0$ & $\s_M$ & $\a_M$ & $\b_M$ & C & D & F & $\chi^2$ & d.o.f. \\
&&$[\textrm{GeV}]$ &$[\textrm{GeV}^{-1}]$&$[\textrm{GeV}^{-1}]$&$[\textrm{GeV}^{-1}]$& \\
\hline
$M_B - M_N$&007--020: MA & 0.807(50) & -0.23(5) & -0.49(3) & -0.43(2) & 0.37(3) & 0.07(6) & 0.03(3)& 10.9 & 5 \\
$M_N$ &007--020: $SU(3)$ & 0.895(43) & -0.25(6) & -0.80(26) & -0.37(11) & 0.36(2) & 0.21(13) & 0.26(7)&9.2 & 5 \\
& 007--030: MA & 0.848(38)& -0.19(3)& -0.49(3)& -0.43(2)& 0.38(2)& 0.09(6)& 0.03(3)&16.8& 9 \\
& 007--030: $SU(3)$ & 0.952(66)& -0.21(7)& -0.94(38)& -0.42(22)& 0.36(2)& 0.27(20)& 0.29(8)& 13.1& 9 \\
& 007--040: MA & 0.864(30) & -0.18(3) & -0.50(4) & -0.45(3) & 0.38(2)& 0.14(5)& 0.04(3)&17.0& 13 \\
& 007--040: $SU(3)$ &0.946(55) &-0.22(6)&-0.92(36)&-0.41(20)&0.36(2)& 0.26(20)& 0.29(8)& 13.3& 13 \\
\end{tabular}
\begin{tabular}{lcccccccccc}
``Predictions" & $M_\L -M_N$ & $M_\S - M_N$ & $M_\Xi -M_N$ 
& $M_\S - M_\L$ & $M_\Xi - M_\L$ & $M_\Xi - M_\S$ 
& $M_N$ & $M_\L$ & $M_\S$ & $M_\Xi$ \\
Phys. MeV & 176 & 254 & 376 & 77 & 199 & 122 & 938 & 1116 & 1193 & 1314 \\ 
\hline
007--020: MA & 115(02) & 182(03) & 262(04) & 67(4)& 147(2)& 81(3)& 952(24) & 1066(24) & 1133(23) & 1214(24) \\
007--020: $SU(3)$ & 129(10) & 185(07)& 271(14)& 60(2)& 147(7)& 86(7)& 980(20)& 1105(23)& 1165(23)& 1251(29) \\
007--030: MA & 115(02) & 182(03) & 262(04) & 67(3)& 148(2)& 80(3)& 968(20) & 1083(19) & 1150(18) & 1231(19) \\
007--030: $SU(3)$ & 125(07) & 196(18)& 281(23)& 68(7)& 153(13)& 85(6)& 1004(25)& 1133(33)&1200(40)& 1285(45) 
\end{tabular}
\end{ruledtabular}
\end{table}

%
%
\begin{table}[h]
\caption{\label{tab:SU63_NLOMB} Results from NLO bootstrap $\chi$-extrapolations of the octet baryon masses, using mixed action (MA) and $SU(3)$ heavy baryon $\chi$PT.}
\begin{ruledtabular}
\begin{tabular}{clccccccccc}
FIT: NLO & range & $M_0$ & $\s_M$ & $\a_M$ & $\b_M$ & C & D & F & $\chi^2$ & d.o.f. \\
&&$[\textrm{GeV}]$ &$[\textrm{GeV}^{-1}]$&$[\textrm{GeV}^{-1}]$&$[\textrm{GeV}^{-1}]$& \\
\hline
$M_N$, $M_\L$,& 007--020: MA 
& 1.087(51) & -0.03(5) & -0.72(8) & -0.62(4) & 0.15(9) & 0.33(4) & 0.14(3) & 6.0 & 5 \\
$M_\S$, $M_\Xi$& 007--020: $SU(3)$ 
& 1.014(32) & -0.07(4) & -0.77(10) & -0.56(5) & 0.18(9) & 0.30(6) & 0.19(4) & 5.5& 5 \\
& 007--030: MA 
& 1.149(57) & 0.01(4) & -0.79(11) & -0.67(7) & 0.12(9) & 0.38(6) & 0.16(3) & 14.4& 9 \\
& 007--030: $SU(3)$ 
& 1.091(66) & -0.04(3) & -0.99(28) & -0.73(19) & 0.1(1) & 0.44(14) & 0.24(7) & 11.9& 9 \\
& 007--040: MA 
& 1.147(52) & 0.01(3) & -0.78(10) & -0.68(6) & 0.13(9) & 0.39(6) & 0.16(3) & 14.9& 13 \\
& 007--040: $SU(3)$ 
& 1.090(61) & -0.04(3) & -0.99(26) & -0.73(18) & 0.1(1) & 0.45(13) & 0.25(6) & 12.5& 13 \\
\end{tabular}
\begin{tabular}{lcccccccccc}
``Predictions" & $M_\L -M_N$ & $M_\S - M_N$ & $M_\Xi -M_N$ 
& $M_\S - M_\L$ & $M_\Xi - M_\L$ & $M_\Xi - M_\S$ 
& $M_N$ & $M_\L$ & $M_\S$ & $M_\Xi$ \\
Phys. MeV & 176 & 254 & 376 & 77 & 199 & 122 & 938 & 1116 & 1193 & 1314 \\
\hline
007--020: MA & 122(4) & 183(3) & 277(6) & 61(6)& 155(2)& 94(7)& 1082(23)&1204(26)&1266(23)&1360(27)  \\
007--020: $SU(3)$ & 117(3)& 181(2)& 265(4)& 65(3)& 148(2)& 83(3)& 1028(17)& 1145(16)& 1240(23)& 1293(17) \\
007--030: MA & 125(5) &184(3)&283(8)& 59(6)& 158(4)& 99(8)&1112(27)&1237(32)& 1296(28)&1395(34) \\
007--030: $SU(3)$ & 121(8)& 189(9)& 277(16)& 68(3)& 156(8)& 88(7)& 1051(18)& 1172(22)& 1237(21)& 1328(29) \\
\end{tabular}
\end{ruledtabular}
\end{table}

%
%
\begin{table}[h]
\caption{\label{tab:SU63_NLOMTmnsMD} Results from NLO bootstrap $\chi$-extrapolations of the decuplet mass splittings using mixed action (MA) and $SU(3)$ heavy baryon $\chi$PT.}
\begin{ruledtabular}
\begin{tabular}{clccccc}
FIT: NLO & range & $\g_M$ & $C$ & $H$ & $\chi^2$ & d.o.f. \\
&& $[\textrm{GeV}]$&$[\textrm{GeV}^{-1}]$&$[\textrm{GeV}^{-1}]$& \\
\hline
$M_T - M_\D$ &007--020: MA 
	& 1.2(2)& 0.00(14)& 1.19(32)& 8.4 & 6 \\
& 007--020: $SU(3)$ 
	& 0.5(6)& 0.83(14)& 0.00(93)& 3.4 & 6 \\
& 007--030: MA 
	& 0.48(01)& 0.00(18)& 0.75(35)& 11.7 & 9 \\
& 007--030: $SU(3)$ 
	& 0.5(5)& 0.65(15)& 0.00(82)& 8.1& 9 \\
& 007--040: MA 
	& 0.95(13)& 0.00(15)& 0.99(21)& 13.2 & 12 \\
& 007--040: $SU(3)$ 
	& 0.5(5)& 0.70(11)& 0.00(86)& 8.3& 12 \\
\end{tabular}
\begin{tabular}{lccc}
``Predictions" & $M_{\S^*} -M_\D$ & $M_{\Xi^*} -M_\D$ & $M_{\O^-} -M_\D$ \\
Phys. MeV & 152 & 300 & 440 \\ 
\hline
007--020: MA & 110(20)& 207(32)& 292(38) \\
007--020: $SU(3)$ & 73(03)& 142(05)& 207(08) \\
007--030: MA & 87(13)& 169(21)& 246(25) \\
007--030: $SU(3)$ & 71(02)& 140(05)& 206(08) \\
\end{tabular}
\end{ruledtabular}
\end{table}

%
%
\begin{table}[h]
\caption{\label{tab:SU63_NLOMTmnsMDMD} Results from NLO bootstrap $\chi$-extrapolations of the decuplet mass splittings combined with the delta mass, using mixed action (MA) and $SU(3)$ heavy baryon $\chi$PT.}
\begin{ruledtabular}
\begin{tabular}{clccccccc}
FIT: NLO & range & $M_{T,0}$ & $\bar{\s}_M$ & $\g_M$ & $C$ & $H$ & $\chi^2$ & d.o.f. \\
&& $[\textrm{GeV}]$&$[\textrm{GeV}^{-1}]$&$[\textrm{GeV}^{-1}]$& \\
\hline
$M_T - M_\D$ &007--020: MA 
	&1.74(14)& -0.01(08)& 1.2(4)& 0.00(13)& 1.22(31)& 11.4& 7 \\
$M_\D$& 007--020: $SU(3)$ 
	&1.69(12)& 0.18(11)& 0.50(02)& 0.71(16)& 0.00(1.02)& 10.3& 7\\
& 007--030: MA 
	&1.55(10)& -0.05(05)& 0.8(3)& 0.00(16)& 0.80(33)& 15.5& 11 \\
& 007--030: $SU(3)$ 
	&1.54(08)& 0.03(07)& 0.48(01)& 0.52(18)& 0.00(80)& 14.7& 11 \\
& 007--040: MA 
	&1.66(08)& 0.00(03)& 0.9(2)& 0.00(30)& 0.99(20)& 18.8& 15  \\
& 007--040: $SU(3)$ 
	&1.62(07)& 0.10(05)& 0.49(01)& 0.57(12)& 0.00(1.04)& 19.2& 15  \\
\end{tabular}
\begin{tabular}{lcccccccc}
``Predictions" & $M_{\S^*} -M_\D$ & $M_{\Xi^*} -M_\D$ & $M_{\O^-} -M_\D$
& $M_\D$ & $M_{\S^*}$ & $M_{\Xi^*}$ & $M_{\O^-}$  \\
Phys. MeV & 152 & 300 & 440 & 1232 & 1384 & 1532& 1672 \\ 
\hline
007--020: MA 
	& 112(20)& 210(32)& 295(38)& 1627(75)& 1739(92)& 1837(103)& 1923(109) \\
007--020: $SU(3)$ 
	&73(02)& 143(05)& 211(8)& 1590(56)& 1663(56)& 1734(55)& 1801(53) \\
007--030: MA 
	& 88(13)& 171(21)& 248(25)& 1533(55)& 1622(66)& 1705(74)& 1782(77) \\
007--030: $SU(3)$ 
	&72(02)& 142(05)& 211(8)& 1520(40)& 1592(40)& 1662(39)& 1731(37)  \\
\end{tabular}
\end{ruledtabular}
\end{table}

%
%
\begin{table}[h]
\caption{\label{tab:SU63_NLOMT} Results from NLO bootstrap $\chi$-extrapolations of the decuplet masses, using mixed action (MA) and $SU(3)$ heavy baryon $\chi$PT.}
\begin{ruledtabular}
\begin{tabular}{clccccccc}
FIT: NLO & range & $M_{T,0}$ & $\bar{\s}_M$ & $\g_M$ & $C$ & $H$ & $\chi^2$ & d.o.f. \\
&& $[\textrm{GeV}]$&$[\textrm{GeV}^{-1}]$&$[\textrm{GeV}^{-1}]$& \\
\hline
$M_\D$, $M_{\S^*}$, &007--020: MA 
	&1.68(10) & -0.04(3)& 1.2(3)& 0.00(07)& 1.2(2)& 18.9& 7 \\
$M_{\Xi^*}$, $M_{\O^-}$& 007--020: $SU(3)$ 
	& 1.52(05) &-0.20(4)& 1.3(3)& 0.00(15)& 1.4(3)& 20.3& 7\\
& 007--030: MA 
	& 1.64(08)& -0.05(2)& 1.1(2)& 0.00(07)& 1.1(2)& 21.0& 11 \\
& 007--030: $SU(3)$ 
	& 1.52(04)& -0.19(4)& 1.3(3)& 0.00(15)& 1.4(3)& 21.1& 11 \\
& 007--040: MA 
	&1.73(08)& -0.01(1)& 1.2(2)& 0.00(06)& 1.2(2)& 32.8& 15 \\
& 007--040: $SU(3)$ 
	& 1.57(04)& -0.18(4)& 1.4(3)& 0.00(14)& 1.6(2)& 34.8& 15\\
\end{tabular}
\begin{tabular}{lcccccccc}
``Predictions" & $M_{\S^*} -M_\D$ & $M_{\Xi^*} -M_\D$ & $M_{\O^-} -M_\D$
& $M_\D$ & $M_{\S^*}$ & $M_{\Xi^*}$ & $M_{\O^-}$  \\
Phys. MeV & 152 & 300 & 440 & 1232 & 1384 & 1532& 1672 \\ 
\hline
007--020: MA 
	& 115(16)& 217(27)& 307(34)& 1591(48)& 1706(63)& 1808(74)& 1897(80) \\
007--020: $SU(3)$ 
	& 94(09)& 170(12)& 227(10)& 1442(13)& 1536(11)& 1612(10)& 1669(7)  \\
007--030: MA 
	&107(14)& 204(24)& 290(30)& 1572(43)& 1680(56)& 1777(65)& 1863(71) \\
007--030: $SU(3)$ 
	& 94(09)& 169(12)& 225(10)& 1445(12)& 1538(09)& 1613(8)& 1670(06)  \\
\end{tabular}
\end{ruledtabular}
\end{table}

\section{Conclusions \label{sec:conclusions}}
In this work, we have presented a detailed study of the light hadron spectrum in the mixed action framework that consists of computing domain-wall valence fermion propagators on the background of the Asqtad improved, rooted, staggered MILC sea configurations.  These results assume the validity of the rooting procedure used in the staggered sea sector.  With the domain wall pion masses tuned to within a few percent of the staggered Goldstone pion masses, we found that the other meson masses had no systematic trend compared to the masses determined by MILC.  
However, as discussed in detail in Sec.~\ref{sec:mesons}, there are additional complications which must be addressed to compare the vector meson masses computed in this work to those of MILC. 
In contrast, we found that the baryon masses were systematically lighter (and therefore in better agreement with the experimental values) than those computed on the coarse MILC lattices, suggesting the mixed action has smaller lattice spacing corrections.  This is supported in the case of the nucleon mass by comparing to the (preliminary) super-fine MILC results displayed in Fig.~\ref{fig:MNcompare}$(a)$.  We also find our nucleon mass results are consistent with those calculated by the RBC/UKQCD group using a domain wall valence on domain wall sea action, as can be inferred from Fig.~\ref{fig:MNcompare}$(a)$ and $(c)$, as well as with the results of the QCDSF/UKQCD group using two-flavors of $\mc{O}(a)$ improved Wilson fermions, inferred from Fig.~\ref{fig:MNcompare}$(b)$.

In addition to the spectroscopy calculations, we have performed a detailed chiral extrapolation analysis of the octet and decuplet baryon masses, using both the continuum $SU(3)$ heavy baryon $\chi$PT as well as its mixed action generalization.  
We have performed extensive three flavor chiral extrapolations, the results of which are collected in Tables~\ref{tab:SU63_NLOMBmnsMN}--\ref{tab:SU63_NLOMT}.  In most cases, the extrapolation analysis is in good agreement with our calculational  results, as measured by the $\chi^2$ per degree of freedom.  However, in all cases, the resulting values of the axial couplings, $D$ and $F$ are different from those determined phenomenologically.  This is suggestive that the three flavor heavy baryon $\chi$PT is not converging for these masses.  Furthermore, the analysis presented in Table~\ref{tab:predNLOmassSplit} is suggestive that the theory is not converging even at the physical point, due to the large kaon mass, except for special observables like the Gell-Mann--Okubo relation.  To form a conclusive analysis of the convergence $SU(3)$ heavy baryon $\chi$PT, one would need to perform a chiral extrapolation of the octet masses (or mass splittings) simultaneously with the hyperon axial charges, with enough lattice-data points to perform the NNLO analysis.  This is beyond the scope of this work.  Alternatively, one can perform $SU(2)$ chiral extrapolations of the masses, as advocated in Ref.~\cite{Tiburzi:2008bk}, which will have much better convergence properties.  We have performed the leading order $SU(2)$ extrapolation of the hyperon mass splittings, while enforcing $SU(3)$ symmetry, the results of which are collected in Table~\ref{tab:SU2_LOMBmnsMN}.  We found good agreement with both the our numerical results as well as the physical mass splittings. 

We have also performed the continuum $SU(2)$ chiral extrapolation of the nucleon and delta masses using the $\mc{O}(m_\pi^4)$ mass formulae.  These extrapolations required the input of the nucleon and delta axial couplings, $g_A$, $g_{\D\D}$ and $g_{\D N}$.  Even after fixing these parameters, the mixed action formula has too many unknown parameters to be determined from our lattice results for either the nucleon or delta, and so we did not perform this mixed action analysis.  In the case of the delta mass, the extrapolation does not agree with the known pole position of the $\D(1232)$, even with the inclusion of the predicted large volume corrections from Ref.~\cite{Bernard:2007cm}.  Furthermore, the delta mass on the one larger volume ($L\sim3.5$~\texttt{fm}) ensemble we use is heavier than on the smaller volume ($L\sim2.5$~\texttt{fm}), yielding a stronger discrepancy with the physical pole mass.  For the lightest two mass points, on one of which we have the two volumes, the delta would decay in infinite volume but due to the restricted values of allowed momenta, this decay is kinematically forbidden.  From our analysis, it is clear that an understanding of the delta mass as well as its chiral extrapolation will require a multi-volume study where the finite volume systematics can be carefully explored.  We expect similar issues for the $\S^*$ and $\Xi^*$ decuplet baryon resonances.

We found that the $\mc{O}(m_\pi^4)$ $SU(2)$ chiral extrapolation of the nucleon mass is in statistical agreement with the lattice data as well as the physical nucleon mass.  An examination of resulting contributions from the first three orders in the expansion, however, revealed that only for the lightest one or two of six mass points does one trust the resulting fits to the nucleon mass in the strict sense of an order by order convergence.  The most surprising result of our analysis is that a linear fit in the pion mass, $M_N = \a_0^N +\a_1^N m_\pi$, Eq.~\eqref{eq:MNstraightmpi}, provides a remarkable agreement with both the lattice data as well as the physical nucleon mass.  To explore our uncontrolled systematics, the lattice spacing and finite volume artifacts, we compared our results to the recent nucleon mass calculations of the MILC~\cite{Bernard:2007ux}, QCDSF/UKQCD~\cite{Gockeler:2007rx}, RBC/UKQCD~\cite{Yamazaki:2007mk} and ETM~\cite{Alexandrou:2008tn} Collaborations.  We found that Eq.~\eqref{eq:MNstraightmpi} provides a statistically good description of the nucleon mass lattice data of all these groups, although the heavy baryon extrapolation was noticeably better in the case of the ETM calculation.  This straight line in $m_\pi$ analysis is not based upon any understanding of low energy QCD we currently have; it is a lattice phenomenological form.  At this point, we are unable to determine if this phenomenon is a conspiracy of QCD or whether it arises from lattice artifacts.  To resolve this issue, more lattice calculations are needed at lighter pion masses, as well as multiple lattice spacings and volumes.  The RBC/UKQCD Collaborations, which use $2+1$ flavor domain-wall valence and sea fermions with $a\sim0.114$~\texttt{fm}, have nucleon mass results that are consistent with ours.  The MILC Collaboration has performed high statistics calculations with three lattice spacings ranging from the coarse lattice, $a\sim 0.124$~\texttt{fm}, to the super-fine lattices with $a\sim0.06$~\texttt{fm}.  While the coarse MILC nucleon mass results are systematically higher than ours, the super-fine nucleon mass results, with a pion mass as light as $m_\pi\sim220$~\texttt{MeV}, lie within the statistical errors on the straight line drawn through our mass results, as can be seen in Fig.~\ref{fig:MNcompare}.  As is clear from this analysis, a reliable chiral extrapolation of the nucleon mass will be much more challenging and demanding than perhaps previously thought.  The nucleon mass displays too little structure to determine all the LECs from an extrapolation of the nucleon mass alone.  To make progress, one will need to perform a global extrapolation analysis with a sufficient set of observables designed to strongly constrain the values of the LECs, notedly the axial couplings, $g_A$, $g_{\D N}$ and $g_{\D\D}$.  We leave this to future work.

\begin{acknowledgments}
We would like to thank the NPLQCD collaborations for providing us with the strange quark propagators needed for a study of the hyperon spectrum as well as many of the light quark propagators used for the high statistics computations.  We thank Doug Toussaint for providing us with the nucleon mass results from the MILC Collaboration, including preliminary numbers calculated on the superfine lattices that were not presented in Ref.~\cite{Bernard:2007ux}.  A.W-L. would like to thank Brian Tiburzi for many useful discussions.  J. N  and M. P would like to thank Ulf-G. Mei{\ss}ner and Akaki Rusetsky for useful discussion and correspondence.  H.-W. Lin would like to thank Christopher Aubin for discussions on the $a_0$ meson and mixed $\chi$PT.
W.S. acknowledges support by the National Science Council of Taiwan under the grant numbers NSC96-2112-M002-020-MY3 and NSC96-2811-M002-026.
K.O. acknowledges support from the Jeffress Memorial Trust grant J-813.
These calculations were performed using the Chroma software suite~\cite{Edwards:2004sx} on clusters at Jefferson Laboratory using time awarded under the SciDAC Initiative.
This work was supported in part by U.S. DOE Contract No. DE-AC05-06OR23177 under which JSA operates Jefferson Laboratory, by the DOE Office of Nuclear Physics under grants DE-FG02-94ER40818, DE-FG02-04ER41302, DE-FG02-96ER40965, DE-FG02-93ER40762, DE-FG02-05ER25681, DE-FG02-07ER41527, the DFG (Forschergruppe Gitter-Hadronen-Ph\"{a}nomenologie) and in part by the EU Integrated Infrastructure Initiative Hadron Physics (I3HP) under contract number RII3-CT-2004-506078. Ph. H. and B. M. acknowledge support by the DFG Emmy-Noether program and the cluster of excellence `Origin and Structure of the Universe'. Ph. H., M.P. and W. S. would like to thank the A.v. Humboldt-foundation for support by the Feodor-Lynen program.  It is a pleasure to acknowledge the use of computer resources provided by the DOE through the USQCD project at Jefferson Lab and through its support of the MIT Blue Gene/L.  We also acknowledge the use of computing resources provided by the William and Mary Cyclades Cluster. We are indebted to members of the MILC Collaboration for providing the dynamical quark configurations that made our full QCD calculations possible.
\end{acknowledgments}

\bibliography{LHPC_spectrum}

\end{document}